\documentclass[11pt]{cernyrep}
\usepackage{graphicx}
\usepackage{epsfig}
\usepackage{axodraw}
\usepackage{pst-coil}
\usepackage{pstricks}
\usepackage{cite}

\let\vec\mathbf
\providecommand{\Rule}[2][0mm]{\rule[#1]{0mm}{#2}} % zero width rule, #1 lift, #2 height
\begin{document}

\title{Neutrino physics}
\author{P.~Hern\'andez\\
IFIC, 
Universidad de Val\`encia and CSIC, 
E-46071 Valencia, Spain}
\maketitle

\begin{abstract}
This is the writeup of the lectures on neutrino physics delivered at various schools: TASI and Trieste in 2013 and the CERN-Latin American School  in 2015. The topics discussed in this lecture include: general properties of
neutrinos in the SM, the theory of neutrino masses and mixings (Dirac
and Majorana), neutrino oscillations both in vacuum and in matter, as well as an
overview of the experimental evidence for neutrino masses and of the
prospects in neutrino oscillation physics. We also briefly review the
relevance of neutrinos in leptogenesis and in
beyond-the-Standard-Model physics.\\\\
{\bfseries Keywords}\\
Neutrino; particle physics; lectures; neutrino oscillations; mixing; standard model.
\end{abstract}

\section{Introduction}

The history of neutrinos is tightly linked to that of the Standard Model. The discovery of neutrinos and the measurement of their tiny masses has been a scientific {\it tour de force}.  

\begin{figure}[htbp]
\begin{center}
 \vspace{0.5cm}
 \includegraphics[width=6cm]{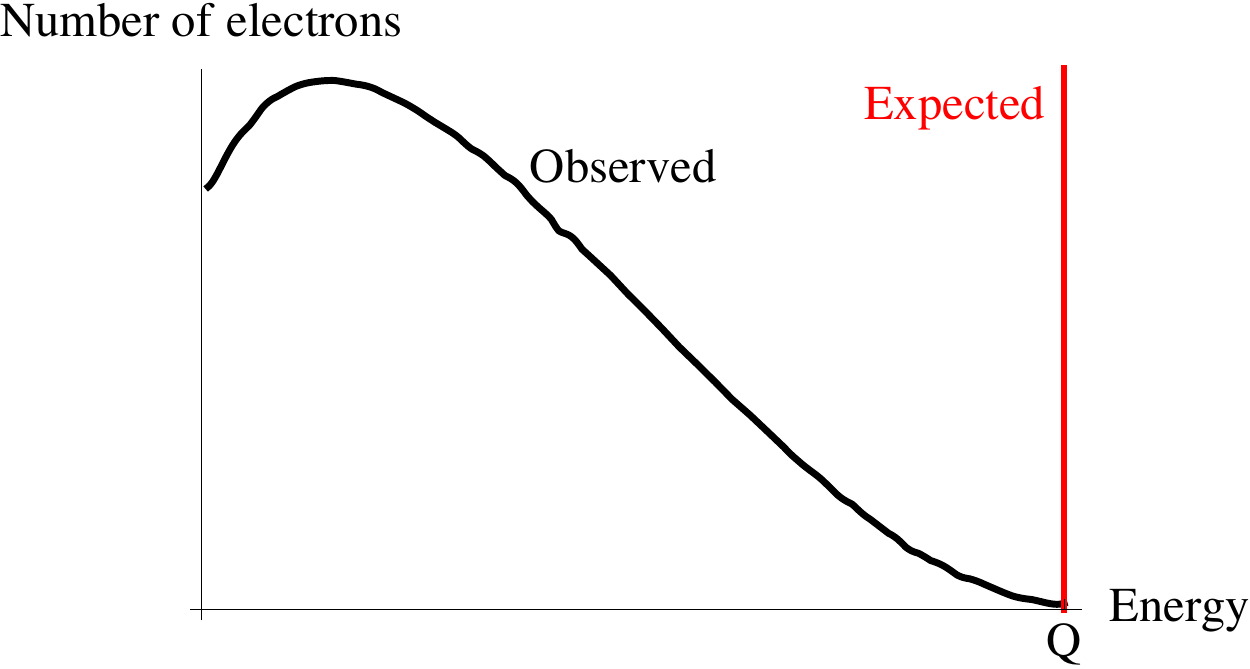}
\caption{Electron spectrum of $\beta$-decay.}
\label{fig:mybeta}
\end{center}
\end{figure}

Neutrinos made their appearance at the beginning of the 20th century as {\it dark} particles in radioactive $\beta$-decay. In this process a nucleus undergoes a transition
\begin{eqnarray}
^A_Z X\rightarrow ^A_{Z+1}X' + e^- 
\end{eqnarray}
emitting
 an electron, which, by  energy conservation, should have an energy equal to the difference of the parent and daughter nuclear masses, $Q$, see Fig.~\ref{fig:mybeta}.

\begin{figure}[htbp]
\begin{center}
\includegraphics[width=5cm]{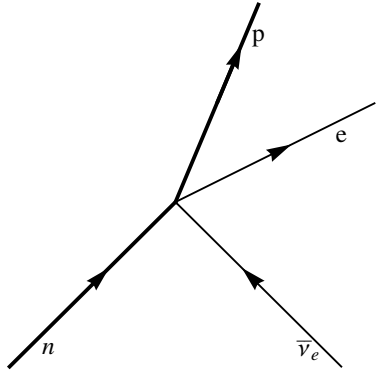}
\caption{Fermi four-fermion coupling responsible for $\beta$-decay.}
\label{fig:fermi}
\end{center}
\end{figure}

The spectrum of the electrons
 was measured to be instead continuous with an end-point at $Q$. It took almost 20 years to come up with an explanation to this apparent violation of energy conservation. 
 W.~Pauli called for a {\it desperate remedy}, proposing that in the decay, a neutral and relatively light particle was being emitted together with the electron and escaped undetected.  In that  case the spectrum of the electron would indeed be  continuous since only the sum of the energy of the electron and the phantom particle should equal 
 $Q$. The dark particle got an italian name: {\it neutrino} in honour of E.~Fermi, who was among the first to take seriously Pauli's hypothesis, from which he constructed the famous theory of $\beta$-decay \cite{Fermi:1934sk}. In this theory, the interaction responsible for $\beta$-decay can be depicted as in Fig.~\ref{fig:fermi}, a four-fermion interaction with strength
  given by $G_F$, the fermi constant. 
   Such interaction implies that neutrinos should also scatter
  off  matter through the inverse beta process, $\bar{\nu} ~p \rightarrow n e^+$.   Bethe and Pearls \cite{Bethe:1934qn} estimated the cross section for such process to be
  \begin{eqnarray}
  \sigma_{\bar{\nu}} \leq 10^{-44} ~\Ucm^2,  \;\;\; E_{\bar \nu} \simeq 2~ \UMeV
  \end{eqnarray}
  and concluded that {\it "it is absolutely impossible to observe processes of this kind"}. Indeed this tiny cross section implies that 
  a neutrino has  a mean free path of thousands of light-years in water. 
 
\begin{figure}[htbp]
\begin{center}
\includegraphics[width=6cm]{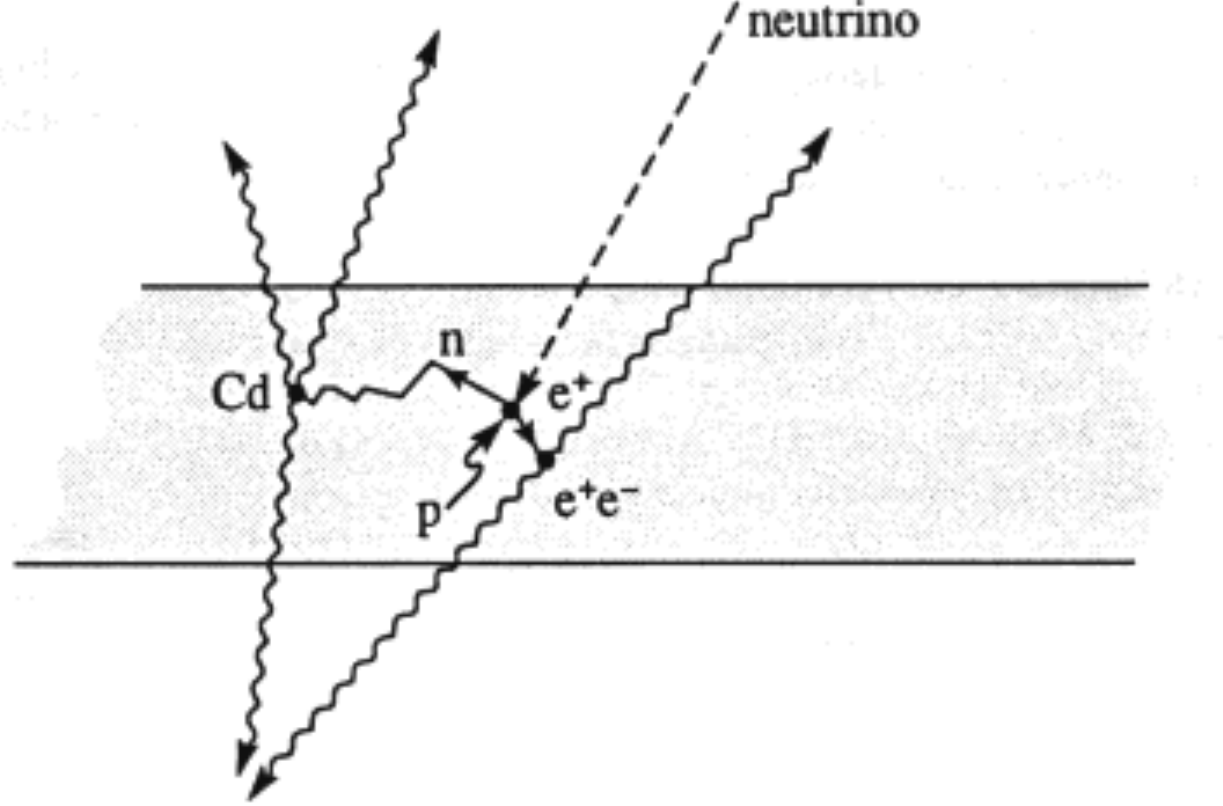}
\caption{Detection technique in  the Reines-Cowan experiment.}
\label{fig:RC}
\end{center}
\end{figure}

  Pontecorvo \cite{Pontecorvo:1946mv} however was among the first to realise that it was not so hopeless. One could get a few events per day in a ton-mass scale detector with a neutrino flux 
  of $10^{11} \nu$/cm$^2$/s.  Such is the neutrino flux of a typical nuclear reactor at a few tens of meters. Reines and Cowen (RC) succeeded   \cite{Reines:1956rs,Cowan:1992xc}. They were able to detect neutrinos via  inverse beta decay in a very massive detector thanks to the extremely robust and clean signal which combines the detection of the positron and the neutron in delayed coincidence, see Fig.~\ref{fig:RC}. 
This experiment not only lead 
  to the discovery of anti-neutrinos, but introduced a detection technique that is still being used today in state-of-the-art reactor neutrino experiments, that continue to make discoveries in neutrino physics.      

  Soon after anti-neutrinos were discovered, it was realised that they come in flavours. The muon had been discovered in cosmic rays much earlier, but it took a long time 
  to understand that this particle was a heavier version of the electron and not the pion. The analogous of the $\beta$-process involving muons is pion decay
  \begin{eqnarray}
  \pi^- \rightarrow \mu^- {\bar \nu}_\mu.
  \end{eqnarray}
  It was understood that also in this case a neutrino was being emitted but that such neutrino, accompanying the $\mu$, had a different identity to that in $\beta$-decay.  Since the energies involved in this process are higher than in $\beta$-decay and neutrino cross-sections grow fast with energy in the Fermi theory, it would actually be easier
  to detect this new type of neutrinos. 
  
\begin{figure}[htbp]
\begin{center}
\vspace{0.5cm}
\includegraphics[width=8cm]{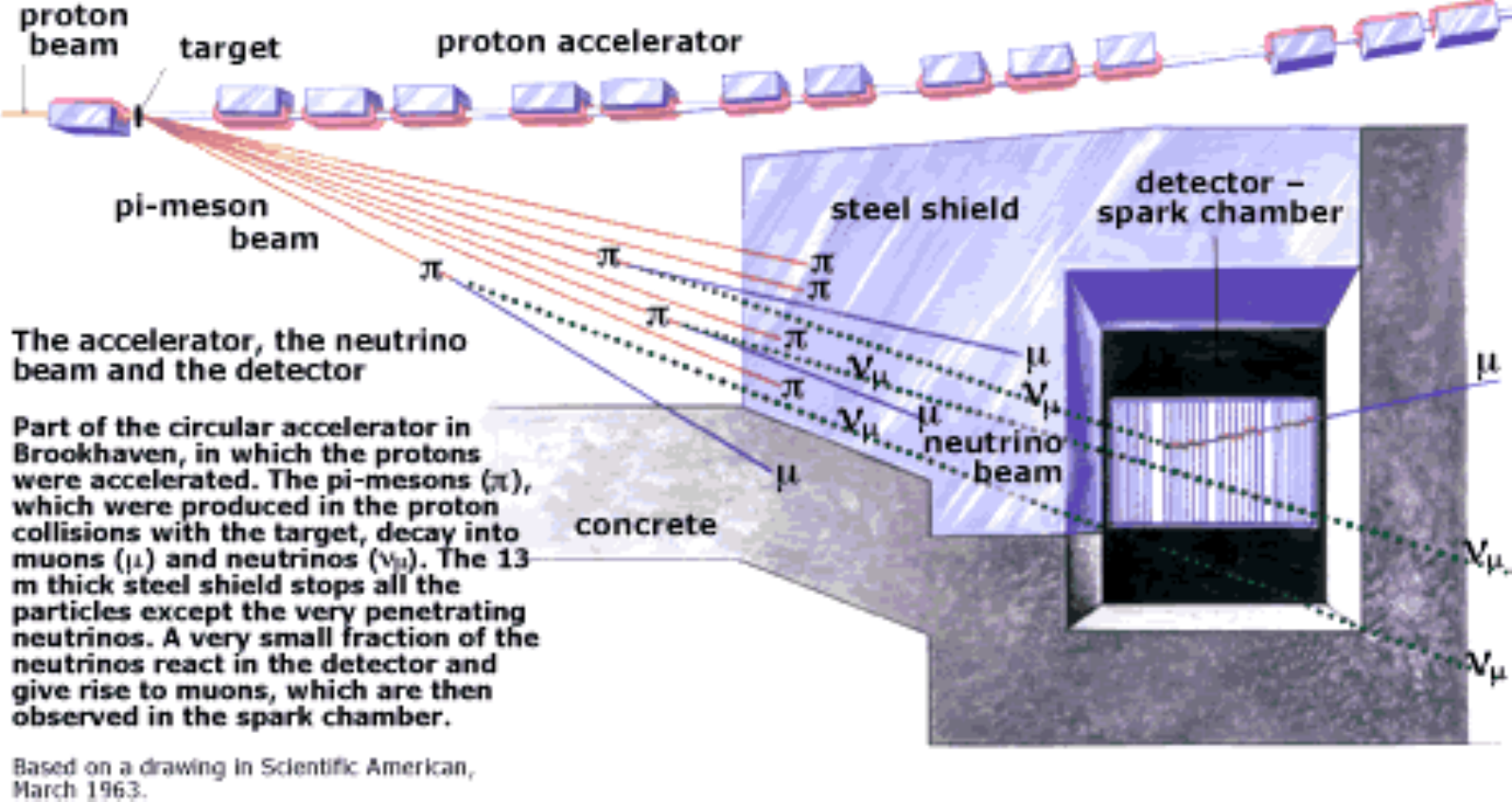}
\caption{Lederman, Schwartz, Steinberger experiment.}
\label{fig:LSS}
\end{center}
\end{figure}

  In 1962 Lederman, Schwartz and Steinberger (LSS)  achieved this goal by creating the first accelerator neutrino beam \cite{Danby:1962nd}. In such a beam, an boosted proton beam hits a target
 producing pions and other hadrons that decay into neutrinos and other particles, mimicking what happens in cosmic rays. If a thick shield intercepts the beam, all particles except the neutrinos can be 
 stopped, see Fig.~\ref{fig:LSS}. A neutrino detector is located behind the shield. A neutrino event will be seen from the appearance of a muon in the detector. 
 Again this was such a great idea that we are still making discoveries with the modern versions of the LSS experiment, in the so-called conventional accelerator neutrino beams. 

\begin{figure}
\centering
\includegraphics[width=.5\linewidth]{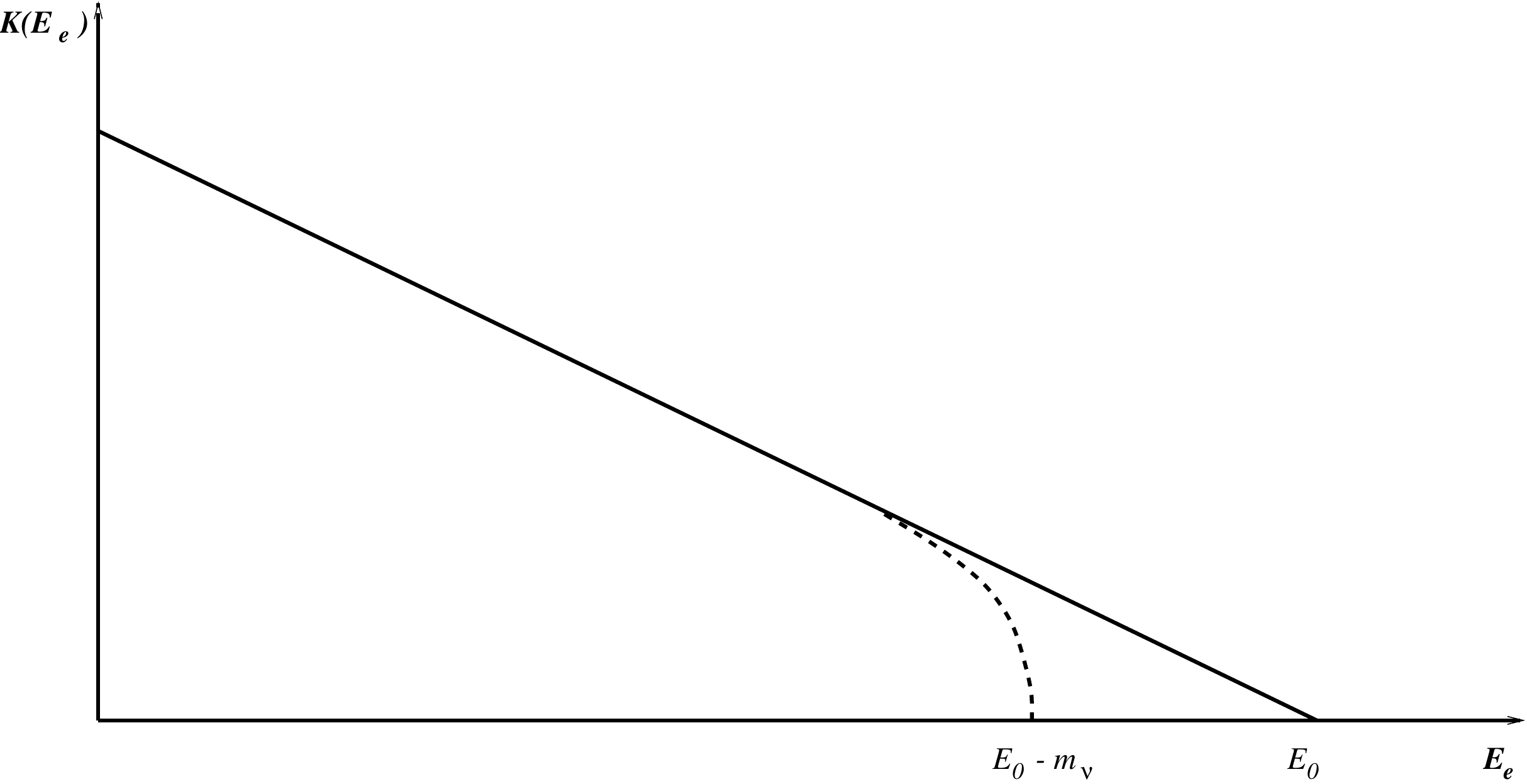}
\caption[]{Effect of a neutrino mass in the end-point of the lepton
           energy spectrum in $\beta$ decay.}
\label{fig:kurie}
\end{figure}

 Kinematical effects of neutrino masses were searched for by
  measuring very precisely  the end-point of the lepton energy spectrum in weak decays, that gets modified if neutrinos are massive.  In particular the most stringent limit is
obtained from tritium $\beta$-decay for the "electron" neutrino:
\begin{eqnarray}
^3H \rightarrow ^3{\rm He} + e^- + \bar{\nu}_e .
\end{eqnarray}
Fig.~\ref{fig:kurie} shows the effect of a neutrino mass in the
end-point electron energy spectrum in this decay. 
%The functional form
%of this curve is $K(E_e) \propto \sqrt{(E_0 - E_e)((E_0 - E_e)^2 -
%m_\nu^2)^{1/2}}$.
The best limit has been obtained by the Mainz and Troitsk experiments. The PDG combination 
 gives
\cite{Agashe:2014kda}:
\begin{eqnarray}
m_{\nu_e} < 2 \UeV (95\% {\rm CL}) \SPp . 
\label{mnue}
\end{eqnarray}
  The direct limits from processes involving $\mu, \tau$ leptons 
are much weaker. The best limit on the $\nu_\mu$ mass
($m_{\nu_\mu} < 170 \UkeV$ \cite{Assamagan:1995wb}) was obtained from the
end-point spectrum of the decay $\pi^+ \rightarrow \mu^+ \nu_\mu$,
while that on the $\nu_\tau$ mass was obtained at LEP
($m_{\nu_\tau} < 18.2\UMeV$ \cite{Barate:1997zg}) from the decay $\tau
\rightarrow 5 \pi \nu_\tau$. Neutrinos in the SM where therefore 
conjectured to be massless.   
  
\section{Neutrinos in the Standard Model}

The  Standard Model (SM)  is a gauge theory based on the gauge group $SU(3)\times SU(2)\times U_{Y}(1)$. All elementary particles 
arrange in irreducible representations of this gauge group. The quantum numbers of the fermions $(d_{SU(3)}, d_{SU(2)})_Y$ are listed in table \ref{tab:reps}. 

\begin{table}
\[
\begin{array}{@{}cc|ccc@{}}                          \hline\hline
\Rule[-1em]{2.5em}
(\vec{1},\vec{2})_ {-\frac{1}{2}} 
 & (\vec{3},\vec{2})_{-\frac{1}{6}} 
  & (\vec{1},\vec{1})_{-1} 
   & (\vec{3},\vec{1})_{-\frac{2}{3}} 
    & (\vec{3},\vec{1})_{-\frac{1}{3}}             \\\hline  
\Rule[-2em]{4em}
\begin{pmatrix}\nu_e  \\ e \end{pmatrix}_{_L} 
 & \begin{pmatrix}u^i \\ d^i \end{pmatrix}_{_L} 
  & e_R 
   & u^i_R 
    & d^i_R                                        \\ 
\Rule[-2em]{4em}
\begin{pmatrix}\nu_\mu \\ \mu\end{pmatrix}_{_L} 
 & \begin{pmatrix}c^i  \\ s^i\end{pmatrix}_{_L} 
  & \mu_R 
   & c^i_R 
    & s^i_R                                        \\ 
\Rule[-2em]{4em}
\begin{pmatrix} \nu_\tau \\ \tau\end{pmatrix}_{_L} 
 & \begin{pmatrix}t^i \\ b^i\end{pmatrix}_{_L} 
  & \tau_R 
   & t^i_R 
    & b^i_R                                        \\\hline\hline
\end{array}
\]
\caption[]{Irreducible fermionic representations in the Standard Model: ($d_{SU(3)}, d_{SU(2)})_{Y}$.}
\label{tab:reps}
\end{table}

Under gauge transformations neutrinos transform as doublets of $SU(2)$, they are singlets under $SU(3)$ and their hypercharge is $-1/2$. The electric charge, given by $Q=T_3+Y$, vanishes.  They are therefore the only particles in the SM that carry no conserved charge. 

The two most intriguing features of table \ref{tab:reps} are its left-right or chiral asymmetry,  and the three-fold repetition of family structures. Neutrinos have been essential in establishing both features. 

\subsection{Chiral structure of the weak interactions}

The left and right entries in table~\ref{tab:reps} have well defined chirality, negative and positive respectively. They are two-component spinors or Weyl fermions, that is the smallest irreducible representation of the Lorentz group representing spin $1/2$ particles.   
%Concerning their transformation properties under the Lorentz group, $\nu_{\alpha}$ are Weyl fermions (two component spinors), that is the smallest irreducible representation of the Lorentz group with spin $1/2$.
Only fields with negative chirality  (i.e. eigenvalue of $\gamma_5$ minus one) carry the $SU(2)$ charge.
%\begin{eqnarray}
%\Psi = \Psi_R + \Psi_L = \underbrace{\left(\frac{1 + \gamma_5}{2}\right)}_{P_R}  
%       \Psi + \underbrace{\left(\frac{ 1 - \gamma_5}{2}\right)}_{P_L}  \Psi\SPp.
%\end{eqnarray}
For free fermions moving at the speed of light (\ie massless), it is easy to see that the
chiral projectors are equivalent to the projectors
on helicity components:
\begin{eqnarray}
P_{R,L} \equiv {1\pm \gamma_5 \over 2} = \frac{1}{2} \left( 1 \pm \frac{\vec{s}\cdot\vec{p}}{|p|} \right) 
        + O\left(\frac{m_i}{E}\right),
\end{eqnarray}
where the helicity operator $\vec\Sigma = \frac{\vec{s}\cdot\vec{p}}{|p|}$ measures
the component of the spin in the direction of the momentum. Therefore
for massless fermions only the left-handed states (with the spin pointing 
in the opposite direction to the momentum) carry $SU(2)$ charge. This is
not inconsistent with Lorentz invariance, since for a fermion travelling 
at the speed of light, the helicity is the same in any reference frame. In 
other words, the helicity operator commutes with the Hamiltonian for a massless
fermion and is thus a good quantum number. 

The discrete symmetry under CPT (charge conjugation, parity, and time
reversal), which is a basic building block of any Lorentz invariant
and unitary quantum field theory (QFT), requires that for any left-handed particle,
there exists a right-handed antiparticle, with opposite charge, but
the right-handed particle state may not exist. A Weyl fermion field represents
therefore a particle of negative helicity and an antiparticle with positive one. 

Parity however transforms left and right fields into each other, thus the left-handedness of the weak
interactions implies that parity
is maximally broken in the SM. The breaking  is nowhere more obvious 
than for neutrinos where the parity partner of the neutrino does not exist. All the 
remaining fermions in the SM come in parity pairs, albeit with different 
$SU(2) \times U(1)$ charges. Since this gauge symmetry is spontaneously broken, the 
left and right fields combine into massive Dirac fermions, that is a four component 
representation of the Lorentz group and parity, which represents a particle 
and an antiparticle with either helicity. 

\begin{figure}[htbp]
\begin{center}
\includegraphics[width=5cm]{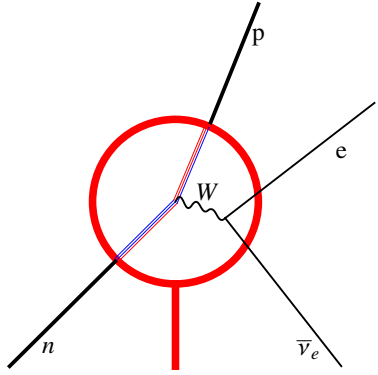}
\caption{$\beta$-decay process in the SM.}
\label{fig:smnint}
\end{center}
\end{figure}

The SM resolved the Fermi interaction as being the result of the exchange of the $SU(2)$ massive $W$ boson
as in Fig.~\ref{fig:smnint}.

Neutrinos interact in the SM via charged and neutral currents:
\begin{eqnarray}
{\mathcal L}_{SM} -{g \over \sqrt{2}} \sum_\alpha \bar{\nu}_{\alpha} \gamma_\mu P_L l_\alpha W^+_\mu - {g \over 2 \cos \theta_W}  \sum_\alpha \bar{\nu}_{\alpha} \gamma_\mu P_L \nu_\alpha Z^+_\mu +h.c.
\end{eqnarray}

\begin{figure}
\centering
\includegraphics[width=.6\linewidth]{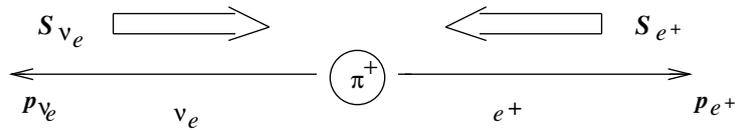}
\caption[]{Kinematics of pion decay.}
\label{fig:pidecay}
\end{figure}

The weak current is therefore $V$--$A$ since it only couples to the left
fields: $\gamma_\mu P_L \propto \gamma_\mu$--$\gamma_\mu \gamma_5$.
This structure is clearly seen in the kinematics
of weak decays involving neutrinos, such as the classic example of
pion decay to $e ~{\bar\nu}_e$ or $\mu
 {\bar\nu}_\mu$. In the limit of vanishing
electron or muon mass, this decay is forbidden, because the spin of
the initial state is zero and thus it is impossible to conserve
simultaneously momentum and angular momentum if the two recoiling
particles must have opposite helicities, as shown in
\Fref{fig:pidecay}. Thus the ratio of the decay rates to electrons and
muons, in spite of the larger phase space in the former, is strongly
suppressed by the factor $\left(\frac{m_e}{m_\mu}\right)^2 \sim
2\times 10^{-5}$.

Another profound consequence of the chiral nature of the weak
interaction is anomaly cancellation. The chiral coupling of fermions
to gauge fields leads generically to inconsistent gauge theories due
to chiral anomalies: if any of the diagrams depicted in \Fref{fig:ano}
is non-vanishing, the weak current is conserved at tree level but not
at one loop, implying a catastrophic breaking of gauge
invariance. Anomaly cancellation is the requirement that all these
diagrams vanish, which imposes strong constraints on the hypercharge
assignments of the fermions in the SM, which are \emph{miraculously}
satisfied:
\begin{figure}
\begin{center}
%\begin{picture}(280,70)(-55,15)
%
% \Line(27,33)(33,27) \Line(27,27)(33,33)
%  \ArrowLine(30,30)(120,60) \ArrowLine(120,60)(120,0)
 % \ArrowLine(120,0)(30,30)
 % \Text(-25,30)[]{$W,B,G,g \;\gamma_5$}
 % \Photon(120,60)(150,60){4}{4} \Text(195,60)[]{$W,B,G,g$}
 % \Photon(120,0)(150,0){4}{4} \Text(195,0)[]{$W,B,G,g$}
 % \Vertex(120,60)1 \Vertex(120,0)1
%\end{picture}
\includegraphics[width=.6\linewidth]{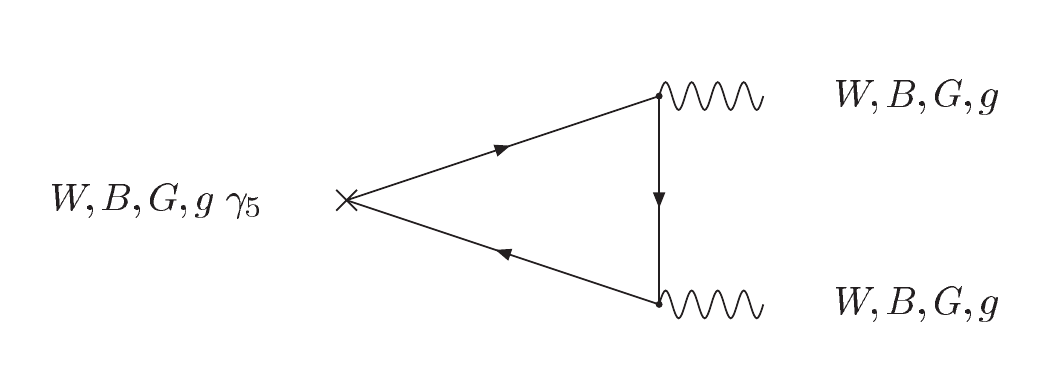}
\end{center}
\caption[]{Triangle diagrams that can give rise to anomalies. $W, B,
           G$ are the gauge bosons associated to the $SU(2), U_Y(1),
           SU(3)$ gauge groups, respectively, and $g$ is the
           graviton.}
\label{fig:ano}
\end{figure}
\begin{eqnarray}
\overbrace{\sum_{i=\text{quarks}} Y^L_i- Y^R_i}^{GGB} = 
\overbrace{\sum_{i=\text{doublets}} Y^L_i}^{WWB} = 
\overbrace{\sum_{i} Y^L_i-Y^R_i}^{Bgg} = 
\overbrace{\sum_i (Y_i^L)^3-(Y_i^R)^3}^{B^3} = 0, 
\label{eq:ano}
\end{eqnarray}
where $Y^{L/R}_i$ are the hypercharges of the left/right components of the 
fermionic field $i$, and the triangle diagram corresponding to each of the 
sums is indicated above the bracket. 

\subsection{Family structure}

\begin{figure}
\centering
\includegraphics[width=.5\linewidth]{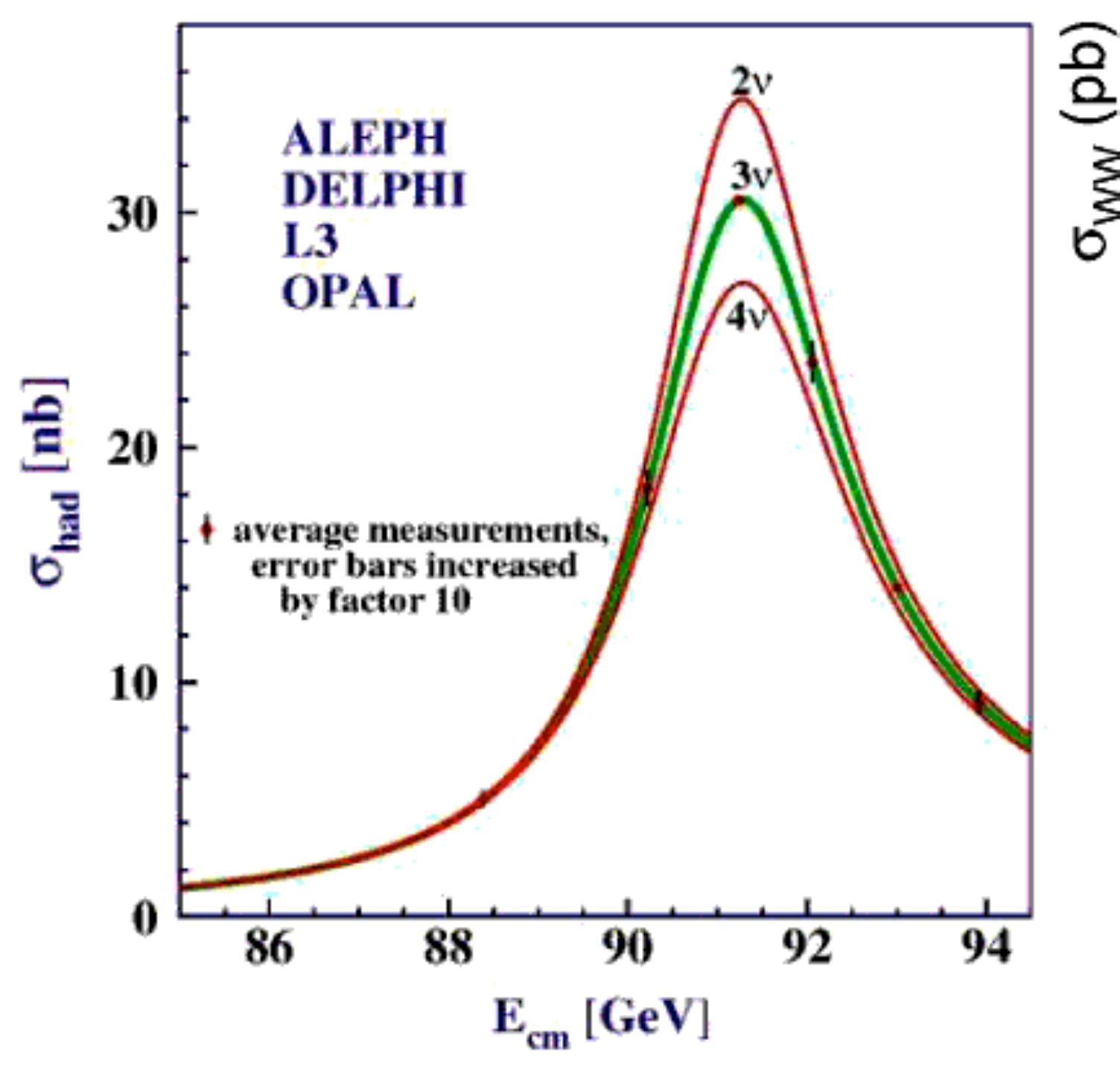}
\caption[]{$Z^0$ resonance from the LEP experiments. Data are
           compared to the case of $N_\nu=2,3$ and 4.}
\label{fig:z0} 
\end{figure}

Concerning the family structure, we know, thanks to neutrinos, that
there are exactly three families in the SM. An extra SM family with quarks
and charged leptons so heavy that cannot be produced, would also
have massless neutrinos that would contribute to the invisible $Z^0$ decay:
\begin{eqnarray}
Z^0 \rightarrow \bar{\nu}_\alpha\nu_\alpha.
\end{eqnarray}
The invisible width of the $Z^0$  has been measured at LEP with an impressive
precision, as shown in \Fref{fig:z0}. This measurement excludes any
number of standard families different from three \cite{Agashe:2014kda}:
\begin{eqnarray}
N_\nu = {\Gamma_{\rm inv} \over \Gamma_{\bar{\nu}\nu}} = 
 2.984 \pm 0.008.
\end{eqnarray}

\section{Massive Neutrinos}

Neutrinos are ubiquitous in our surroundings. If we open our hand, it will be crossed each second by about ${\mathcal O}(10^{12})$ neutrinos from the sun, about ${\mathcal O}(10)$ from the atmosphere, about ${\mathcal O}(10^{9})$ from natural radioactivity in the Earth and even ${\mathcal O}(10^{12})$ relic neutrinos from the Big Bang. 
In 1987, the Kamiokande detector in Japan observed the neutrino burst from a SuperNova that exploded in the Large Magellanic Cloud, at a distance of 168 thousand light years from Earth. For a few seconds, the supernova flux was of the same order of magnitude as the flux of solar neutrinos! 

Using many of these sources as well as others man-made, a decade of revolutionary neutrino experiments have demonstrated that, for the time being, neutrinos are the less standard of the SM particles. They have tiny masses and this necessarily requires new degrees of freedom
with respect to those in table~\ref{tab:reps}.

  A massive
fermion necessarily has two states of helicity, since it is always
possible to reverse the helicity of a state that moves at a slower
speed than light by looking at it from a boosted reference frame. 
What is the right-handed state of the neutrino ?
% In
%fact a mass can be thought of as the strength of the coupling between
%the two helicity states:
%\begin{eqnarray}
%m \;\overline{\psi_L} \psi_R + \text{h.c.}
%\end{eqnarray}
%In order to include such a coupling in the SM for the neutrinos we need
%to identify the neutrino right-handed state, which in the SM is
%absent.
It turns out there are two ways to proceed.

%\subsubsection*{Free case}
Let us consider the case of free fermions. A four-component Dirac fermion can be made massive adding the following mass term to the Lagrangian:
\begin{eqnarray}
-{\mathcal L}_m^{Dirac}=m\bar{\psi}\psi=m(\overline{\psi_L+\psi_R})(\psi_L+\psi_R)=m(\overline{\psi_L}\psi_R+\overline{\psi_R}\psi_L).
\label{eq:dirac}
\end{eqnarray}
A Dirac mass term couples the left-handed and right-handed chiral components of the fermion field, and therefore this coupling vanishes 
identically in the case of  a Weyl fermion. 

Can one give a mass to a  two-component Weyl fermion ?
%The answer to these questions is: \textit{Yes}. 
As first realized by Majorana, this indeed can be done with the following mass term:
\begin{eqnarray}
-{\mathcal L}_{m}^{Majorana}=\frac{m}{2}\overline{\psi^{c}}\psi+\frac{m}{2}\overline{\psi}\psi^{c}=\frac{m}{2}\psi^{T}C\psi+\frac{m}{2}\bar{\psi}C\bar{\psi}^{T},
\label{eq:majorana}
\end{eqnarray}
where 
\begin{eqnarray}
\psi^c\equiv C\bar{\psi}^{T}=C\gamma_0\psi^{*}.
\end{eqnarray}
It is easy to check that the Majorana mass term satisfies the required properties:
\begin{itemize}
 \item[1)] It can be constructed with a two-component spinor or Weyl fermion:
 if  $\psi=P_L \psi \equiv (\psi_L,0)$
\begin{eqnarray}
\psi^{T}C\psi=\psi_L^{T}i\sigma_2\psi_L,
\end{eqnarray}
and it does not vanish in the absence of the right chiral component. 
  \item[2)] It is  Lorentz invariant. It is easy to show, using the properties of the gamma matrices that under a Lorentz transformation $\psi$ and $\psi^c$ transform in the same way, 
\begin{eqnarray}
\psi\rightarrow e^{-\frac{i}{4}\omega_{\mu\nu}\sigma^{\mu\nu}}\psi\equiv S(\Lambda)\psi, \;\; \psi^c\rightarrow S(\Lambda)\psi^c, \;\; 
\end{eqnarray}
with $\sigma_{\mu\nu}\equiv = {i\over 4} [\gamma_\mu, \gamma_\nu]$, and therefore the bilinear $\overline{\psi^c} \psi$ is Lorentz invariant. 
 \item[3)] The equation of motion derived from eq.~(\ref{eq:majorana}) for a free majorana fermion has  plane wave solutions satisfying the relativistic relation for a massive fermion:
\begin{eqnarray*}
E^2-\vec{p}^2=m^2. 
\end{eqnarray*}
\end{itemize}

In the SM none of the mass terms of eqs.~(\ref{eq:dirac}) and (\ref{eq:majorana}) are gauge invariant. Spontaneous symmetry breaking 
allows to generate the Dirac mass term from Yukawa couplings for all fermions in the SM, while the Majorana mass term can only be generated for neutrinos. 

\subsection{Massive Dirac neutrinos}
\label{sec:dirac}

\begin{figure}
\centering
\includegraphics[width=.3\linewidth]{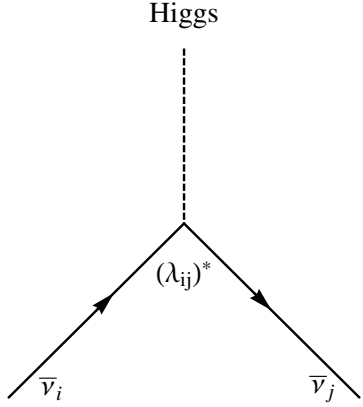}
\caption[]{Neutrino Yukawa coupling.}
\label{fig:dirac} 
\end{figure}

\begin{figure}
\centering
\includegraphics[width= 0.8\linewidth]{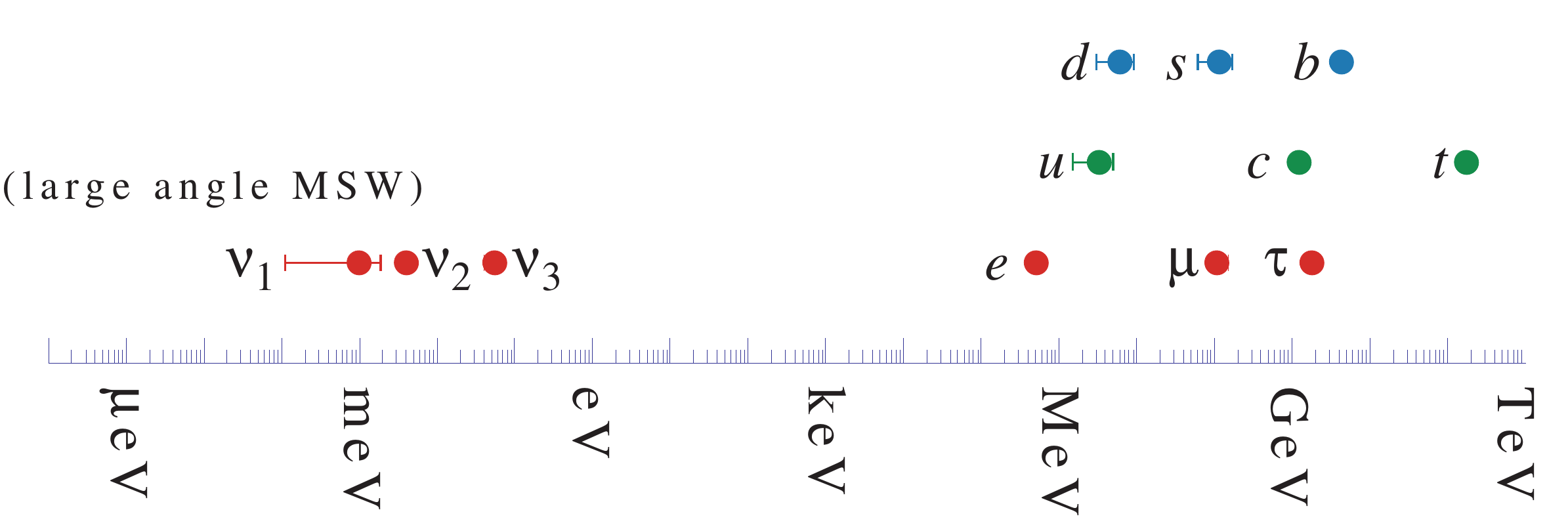}
\caption[]{Fermion spectrum in the Standard Model.}
\label{fig:hierar}
\end{figure}

We can enlarge the SM by adding a set of three right-handed neutrino, $\nu_R$
states, with quantum numbers $(1,1)_0
$, ie 
 singlets under all the gauge groups.  A new Yukawa (Fig.~\ref{fig:dirac}) coupling of these new states with the lepton doublet 
 is exactly gauge invariant and therefore can be added to the SM:
\begin{eqnarray}
-{\mathcal L}_m^{Dirac}= \;\overline{L}\; \lambda \tilde\Phi\; \nu_{R} + \text{ h.c. }  
%  \rightarrow m_\nu = \lambda_\nu\; v,
\label{eq:yukawadirac}
\end{eqnarray}
where $L= (\nu\,\, l)$ is the lepton doublet, $\tilde\Phi\equiv i \sigma_2 \phi^*$
and $\phi$ is the Higgs field, with quantum numbers $(\mathbf{1,2})_{-{1\over 2}}$.
Upon spontaneous symmetry breaking 
the scalar doublet gets a vacuum expectation value $\langle
\tilde\Phi \rangle = ({v\over \sqrt{2}} \,\,0)$, and therefore a neutrino Dirac mass term is generated 
\begin{eqnarray}
-{\mathcal L}_m^{Dirac} \rightarrow  -\;\overline{\nu_L}\; \lambda {v \over \sqrt 2} \nu_{R} + \text{ h.c. } .
%  \rightarrow m_\nu = \lambda_\nu\; v,
\end{eqnarray}
The neutrino mass matrix is proportional to the Higgs vev, in complete analogy to the remaining fermions:
\begin{eqnarray}
m_\nu = \lambda {v \over \sqrt{2}}.
\end{eqnarray}
 There are two important consequences of Dirac neutrinos.  First, there is a new hierarchy problem
in the SM to be explained: why are neutrinos so much lighter than the
remaining leptons, even those in the same family (see
\Fref{fig:hierar}), if they get the mass in the same way ? Secondly, an accidental global symmetry, lepton number ${\rm L}$,  that counts the number of leptons minus
that of antilepton, remains  exactly conserved  at
the classical level \footnote{As usual ${\rm B+L}$ is broken by the anomaly
and only ${\rm B-L}$ remains exact at all orders.}, just as baryon number,
${\rm B}$, is.

\subsection{Massive Majorana neutrinos}

\begin{figure}
\centering
\includegraphics[width=.3\linewidth]{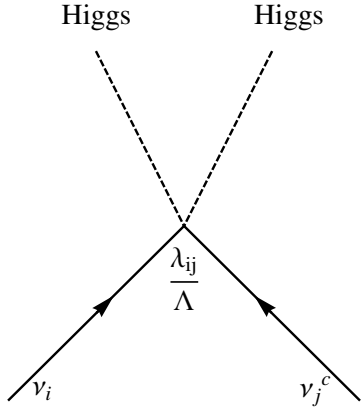}
\caption[]{Weinberg operator.}
\label{fig:majorana} 
\end{figure}

Since the combination ${\bar L}\tilde{\phi}$ is a singlet under all gauge groups, the Majorana-type 
contraction (see Fig.~\ref{fig:majorana}):
\begin{eqnarray}
-{\mathcal L}_{m}^{Majorana}=~ {\bar L}\tilde{\phi}~ \alpha C \tilde{\phi}^T \bar{L}^T + h.c.,
\label{eq:weinberg}
\end{eqnarray}
 is gauge invariant. This term, first writen down by Weinberg \cite{Weinberg:1979sa}, gives rise to a Majorana mass term for neutrinos upon spontaneous symmetry breaking:
 \begin{eqnarray}
-{\mathcal L}_{m}^{Majorana}\rightarrow {\bar \nu_L} \alpha {v^2\over 2} C  \bar{\nu}_L^T + h.c.,
\end{eqnarray}
The neutrino mass matrix in this case is given by:
\begin{eqnarray}
m_\nu = \alpha {v^2}.
\end{eqnarray}
The Weinberg operator has dimension 5, and therefore the coupling $[\alpha]=-1$. We can write
it  in terms of a dimensionless coupling as
\begin{eqnarray}
\alpha= {\lambda\over \Lambda},
\end{eqnarray}
where $\Lambda$ is a new physics scale, in principle unrelated to the electroweak scale. 

The consequences of the SM neutrinos being massive Majorana particles
are profound. 

If the scale $\Lambda$ is much higher than the electroweak scale $v$, a
strong hierarchy between the neutrino and the charged lepton masses
arises naturally. If all dimensionless couplings $\lambda$ are of the same order, neutrino masses
are suppressed by a factor $v/\Lambda$ with respect to the charged fermions.
On the other hand, Weinberg's operator violates lepton number ${\rm L}$ and provides a new seed for generating 
the matter/antimatter asymmetry in the Universe as we will see.   

Even though the Majorana mechanism to generate neutrino masses does not involve any extra degree of freedom with respect to those in the SM, the existence of the Weinberg coupling implies that cross sections involving for example the scattering of neutrinos and the higgs will grow with energy, ultimately violating unitarity. The situation is analogous to that of the Fermi interaction of Fig.~\ref{fig:fermi}. The SM resolved 
this interaction at higher energies as being the result of the interchange of a heavy vector boson, Fig.~\ref{fig:smnint}. The Majorana coupling, if it exists, should also  represent the effect at low-energies of the exchange of one or more unknown massive  states. What those states
are remains one of the most interesting open questions in neutrino physics.

Finally, it is interesting to note that the anomaly cancellation conditions fix all the hypercharges in this case
      (\ie there is only one possible choice for the hypercharges that
      satisfies \Erefs{eq:ano}), which implies that
      electromagnetic charge quantization is the only possibility in a
      field theory with the same matter content as the SM.
      
      \subsection{Neutrino masses and physics beyond-the-standard-model}

Any new physics beyond the standard model  (BSM) characterized by a high scale, $\Lambda$,  will induce effects at low energies $E \ll \Lambda$ that  can be described by an effective field theory \cite{Weinberg:1978kz,Buchmuller:1985jz}:
\begin{eqnarray}
{\mathcal L}_{\rm eff} = {\mathcal L}_{\rm SM} + \sum_i {\alpha_i \over \Lambda} O_i^{d=5} + \sum_i {\beta_i\over \Lambda^2} O_i^{d=6} + ...
\label{eq:eft}
\end{eqnarray}
It is the most general Lagrangian which includes the SM and an infinite tower of  operators constructed out of the SM fields respecting Lorentz and gauge symmetries. In principle such a theory depends on infinite new couplings, one per new independent operator, and it is therefore not predictive. However, if the energy we are interested in effects at a given finite order, $n$,  in $\left
({E\over \Lambda}\right)^n$, we can truncate the series to include only operators of dimension $d\leq n+4$. The operators of lowest dimension are the most relevant at low energies. 

It turns out that there is only one such operator of  the lowest possible dimension, $d=5$, which is precisely the Weinberg operator of eq.~(\ref{eq:weinberg}). In this perspective, it is natural to expect that  the first indication of BSM physics is precisely Majorana neutrino masses, and while
many types of BSM theories can give rise to neutrino masses,  generically they will induce other new physics effects represented by the operators of $d=6$ or higher. 

\section{Neutrino masses and lepton mixing}

\begin{figure}[h]
\centering
\includegraphics[angle=270,width=.475\linewidth]{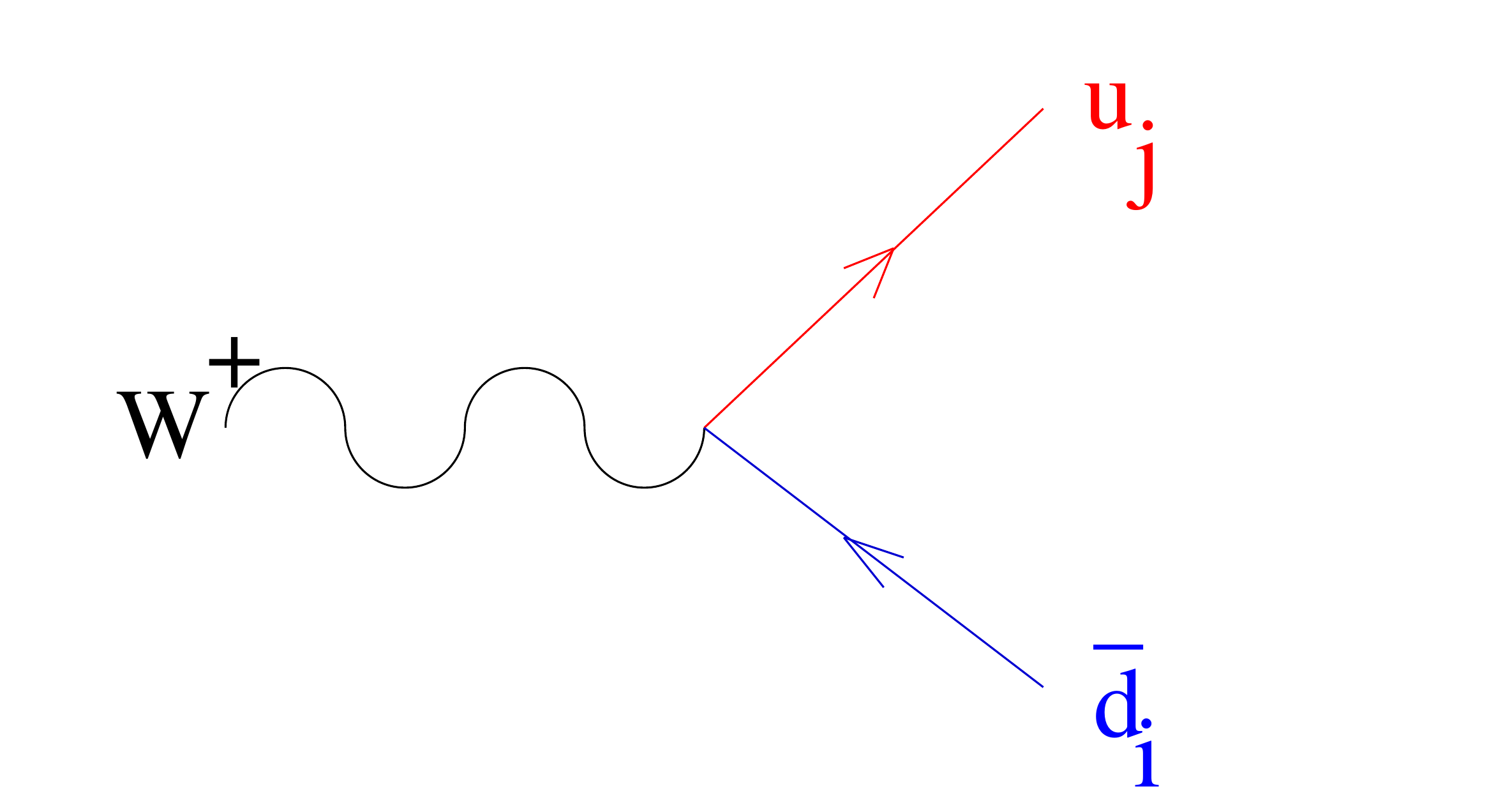}\qquad
\includegraphics[angle=270,width=.475\linewidth]{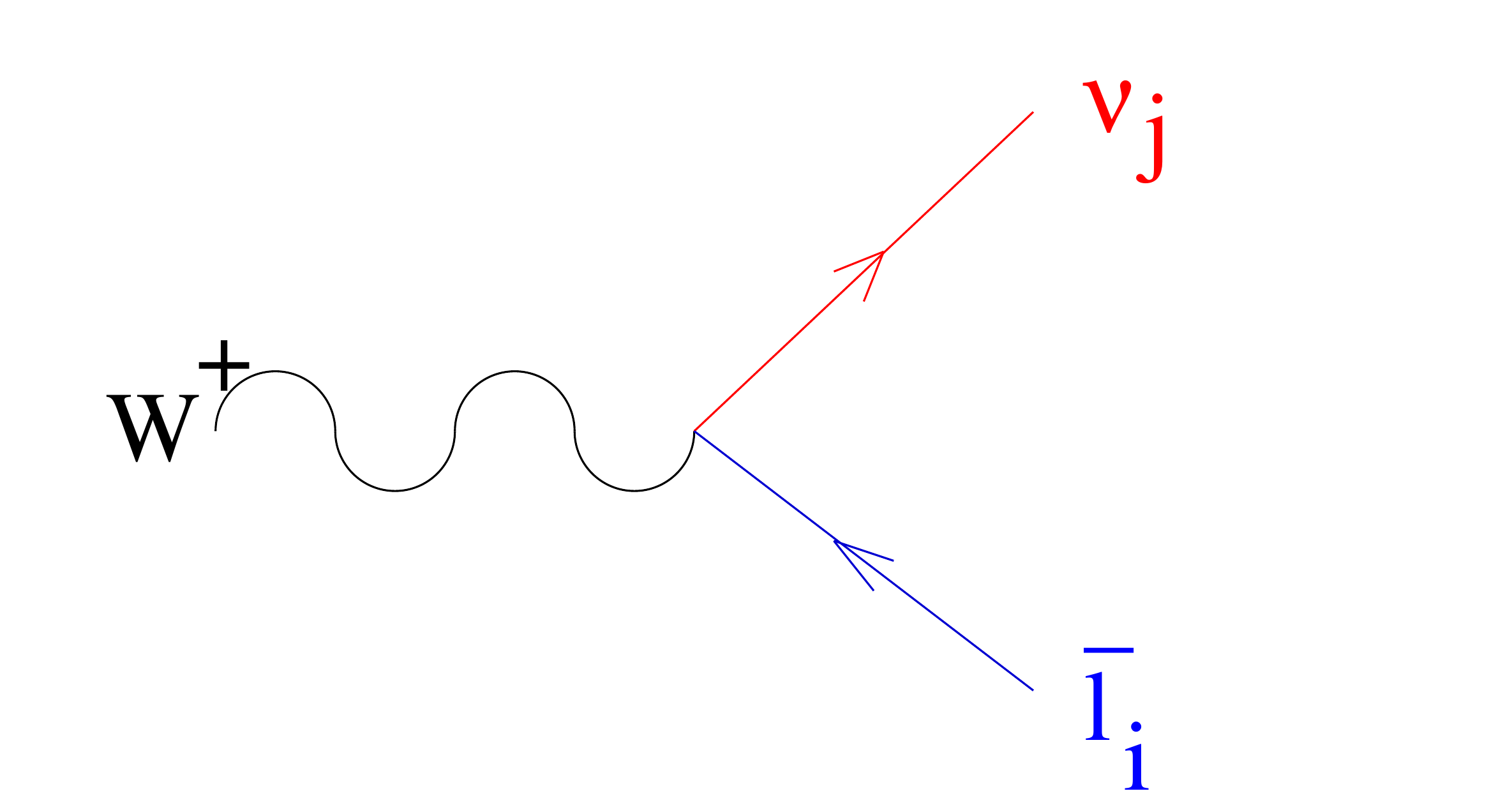}
\caption[]{Quark and lepton mixing.}
\label{fig:mix}
\end{figure}

Neutrino masses, whether Dirac or Majorana, imply lepton mixing \cite{Maki:1962mu,Pontecorvo:1967fh}. The Yukawa couplings in eq.~(\ref{eq:yukawadirac}) is a generic complex matrix in flavour space, while that in eq.~(\ref{eq:weinberg}) is a  generic complex symmetric matrix, and therefore the same holds for 
the corresponding mass matrices:
\begin{eqnarray}
-\mathcal{L}_m^{\rm Dirac} &=& \overline{\nu^i_L} \left(M_\nu\right)_{ij} \nu^j_R  + \overline{l^i_L} \left(M_l\right)_{ij} l^j_R+\text{ h.c. } \\
-\mathcal{L}_m^{\rm Majorana} &=& \frac{1}{2}  \overline{\nu^i_L}  \left(M_\nu\right)_{ij}  \nu^{c j} _L + \overline{l^i_L} \left(M_l\right)_{ij} l^j_R+ \text{ h.c. } .
\end{eqnarray}
In the Dirac case, the two mass matrices can be diagonalized by a bi-unitary rotation:
\begin{eqnarray}
M_\nu = U_\nu^\dagger {\rm Diag}(m_1,m_2,m_3) V_\nu, ~~M_l = U_l^\dagger {\rm Diag}(m_e,m_\mu,m_\tau) V_l,
\end{eqnarray}
while in the Majorana case, the neutrino mass matrix, being symmetric, can be taken to a diagonal form by
\begin{eqnarray}
M_\nu = U_\nu^\dagger {\rm Diag}(m_1,m_2,m_3) U^*_\nu. 
\end{eqnarray}
We can go to the mass basis by rotating the fields as:
\begin{eqnarray}
\nu'_R = V_\nu \nu_R,~\nu'_L = U_\nu \nu_L, ~~l'_R =V_l l_R,~l'_L = U_l l_L. ~~
\end{eqnarray}
In this basis  the charged current interactions are no longer diagonal, in complete analogy 
with the quark sector (see \Fref{fig:mix}): 
\begin{eqnarray}
\mathcal{L}^{\rm lepton}_{CC} = -{g \over \sqrt{2}} {\bar l'_{i}} \gamma_\mu P_L W_\mu^+
 \underbrace{(U^\dagger_l U_\nu)_{ij}}_{U_{\rm PMNS}} \nu'_{j} + \text{h.c.}  
\end{eqnarray}
The  mixing matrix in the lepton sector is referred to as the Pontecorvo-Maki-Nakagawa-Sakata (PMNS) matrix, analogous to the CKM one in 
the quark sector. 

The number of  physical parameters  in the
lepton mixing matrix, $U_\text{PMNS}$,  can easily be computed by counting the
number of independent real and imaginary elements of the Yukawa
matrices and eliminating those that can be absorbed in field
redefinitions. The allowed field redefinitions are the unitary
rotations of the fields that leave the rest of the Lagrangian invariant, only those that are not symmetries of the full
Lagrangian when lepton masses are included.
\begin{table}[h]
\centering
\begin{tabular}{@{}l|l|l|l|l|l@{}} \hline\hline
 & Yukawas 
  & Field redefinitions 
   & $No.~�m$ 
    & $No.~ \theta$ 
     & $No.~ \phi$                 \\ \hline 
 &&&&&                           \\ 
Dirac 
 & $\lambda_l,~\lambda_\nu$ 
  & $U(n)^3/U(1)$   
   &&&                           \\  
 & $4 n^2$   
  &  $\dfrac{3(n^2-n)}{2},~\dfrac{3(n^2+n)-1}{2}$ 
   & $2 n$ 
    & $\dfrac{n^2-n}{2}$ 
     & $\dfrac{(n-2)(n-1)}{2}$   \\
 &&&&&                           \\ \hline
 &&&&&                           \\
Majorana 
 & $\lambda_l,~\alpha_\nu^T=\alpha_\nu$ 
  & $U(n)^2$ 
   &&&                           \\  
 & $3 n^2 + n$ 
  & $n^2-n,~n^2+n$ 
   & $2 n$ 
    & $\dfrac{n^2-n}{2}$ 
     &  $\dfrac{n^2-n}{2}$       \\ 
 &&&&&                           \\\hline\hline
\end{tabular}
\caption[]{Number of real and imaginary parameters in the Yukawa
           matrices, of those that can be absorbed in field
           redefinitions. The difference between the two is the number
           of observable parameters: the lepton masses ($m$), mixing
           angles ($\theta$), and phases ($\phi$).}
\label{table:mix}
\end{table}

In the Dirac case, it is possible to rotate independently the
left-handed lepton doublet, together with the right-handed charged
leptons and neutrinos, that is $U(n)^3$, for a generic number of
families $n$. However, this includes total lepton number which remains
a symmetry of the massive theory and thus cannot be used to reduce the
number of physical parameters in the mass matrix. The
parameters that can be absorbed in field redefinitions are thus the
parameters of the group $U(n)^3/U(1)$ (that is $\frac{3 (n^2-n)}{2}$
real, $\frac{3(n^2+n)-1}{2}$ imaginary).

In the case of Majorana neutrinos, there is no independent
right-handed neutrino field, nor is lepton number a good
symmetry. Therefore the number of field redefinitions is the number of
parameters of the elements in $U(n)^2$ (that is $n^2-n$ real and
$n^2+n$ imaginary).

The resulting real physical parameters are the mass eigenstates and
the mixing angles, while the resulting imaginary parameters are
CP-violating phases. All this is summarized in
\Tref{table:mix}. Dirac and Majorana neutrinos differ only in the
number of observables phases. For three families ($n=3$), there is
just one Dirac phase and three in the Majorana case.\\

A standard parametrization of the mixing matrices for  Dirac, $U_{\rm PMNS}$, and Majorana, ${\tilde U}_{\rm PMNS}$, is given by
\begin{eqnarray}
U_\text{PMNS} &=& 
%V^+_l V_\nu \equiv 
\begin{pmatrix}
1 & 0 & 0           \\
0 & c_{23} & s_{23} \\
0 & -s_{23} & c_{23} 
\end{pmatrix}
\begin{pmatrix}
c_{13} & 0 & s_{13} e^{-i \delta} \\
0 & 1 & 0           \\
-s_{13} e^{i \delta} & 0  & c_{13}
\end{pmatrix}
\begin{pmatrix}
c_{12} & s_{12} & 0  \\
-s_{12} & c_{12}  & 0 \\\
0 & 0 & 1 
\end{pmatrix},\nonumber\\
{\tilde U}_\text{PMNS} &=& 
%V^\dagger_l O_\nu \equiv 
U_\text{PMNS}(\theta_{12},\theta_{13}, \theta_{23}, \delta) 
\begin{pmatrix}
  1 & 0 & 0             \\
  0 & e^{i \alpha_1} & 0\\
0 & 0 & e^{i \alpha_2}
\end{pmatrix}\SPp,
\label{mns}
\end{eqnarray}
where in all generality $\theta_{ij} \in [0, \pi/2]$ and $\delta,\alpha_1,\alpha_2 \in [0, 2 \pi]$. 

\section{Majorana versus Dirac}

\begin{figure}[h]
\centering
\includegraphics[angle=270,width=.75\linewidth]{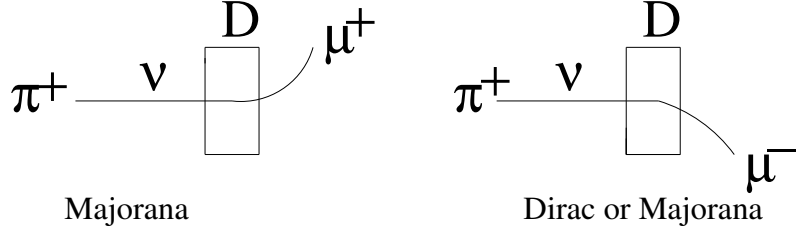}
\caption[]{A neutrino beam from $\pi^+$ decay ($\nu_\mu$) could
           interact in the magnetized detector producing a $\mu^+$
           only if neutrinos are Majorana.}
\label{fig:signature}
\end{figure}

It is clear that establishing the Majorana nature of neutrinos is of
great importance, since it would imply the existence of a new physics scale.  
In principle there are very clear signatures, such
as the one depicted in \Fref{fig:signature}, where a $\nu_\mu$ beam
from $\pi^+$ decay is intercepted by a detector. In the Dirac case,
the interaction of neutrinos on the detector via a charged current
interaction will produce only a $\mu^-$ in the final state. If neutrinos
are Majorana, a wrong-sign muon in the final state is also possible.
Unfortunately the rate for $\mu^+$ production is suppressed by
$m_\nu/E$ in amplitude with respect to the $\mu^-$. For example, for
$E_\nu = \mathcal{O}(1)\UGeV$ and $m_\nu \sim \mathcal{O}(1)\UeV$ the cross--section for this process will be roughly $10^{-18}$ times the usual CC
neutrino cross-section.

\begin{figure}[h]
\centering
\includegraphics[width=.75\linewidth]{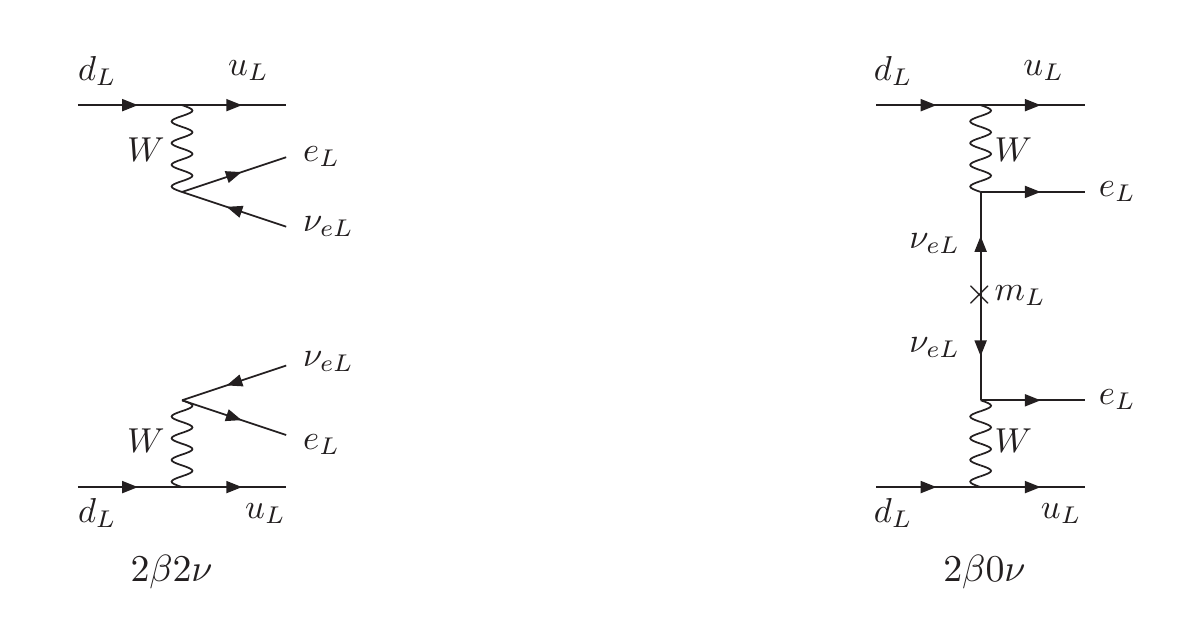}
\caption[]{$2\beta$ decay: normal (left) and neutrinoless (right).}
\label{fig:dbd}
\end{figure}

The best hope of observing a rare process of this type seems to be the
search for neutrinoless double--beta decay ($2\beta 0\nu$), the right
diagram of \Fref{fig:dbd}. The background to this process is the
standard double--beta decay depicted on the left of \Fref{fig:dbd},
which has been observed to take place for various isotopes with a lifetime of $T_{2\beta
2\nu} > 10^{19}$--$10^{21}$ years.

If the source of this process is just the Majorana $\nu$ mass, the
inverse lifetime for this process is given by
\begin{eqnarray}
T^{-1}_{2\beta 0\nu} \simeq 
  \underbrace{G^{0\nu}}_\text{Phase}
\underbrace{\left|M^{0\nu}\right|^2}_{\rm Nuclear M.E.}
\underbrace{\left|\sum_i \left({\tilde U}^{ei}_\text{PMNS}\right)^2 m_i
\right|^2}_{|m_{ee}|^2} .
\end{eqnarray}

In spite of the suppression in the neutrino mass (over
the energy of this process), the neutrinoless mode has a phase
factor orders of magnitude larger than the $2\nu$ mode, and as a result 
present experiments searching for this rare process have already set bounds on 
neutrino masses in the $\UeVZ$
range as shown in \Tref{tab:2b0n}.

\begin{table}[h]
\centering
\begin{tabular}{@{}lll@{}}                         \hline\hline
%\Rule[-0.5em]{1.5em}
Experiment 
 & Nucleus 
  & $|m_{ee}|$                                     \\\hline
 EXO-200  
 & \Isotope[136]{Xe} 
 & $<0.19$--$0.45\UeV $\\
 %\Rule{1.2em}
%\Rule[-0.5em]{1.5em}
NEMO-3 
 & \Isotope[100]{Mo} 
 & $<0.33$--$0.87\UeV $         \\
 GERDA 
 & \Isotope[76]{Ge} 
  & $<0.2$--$0.4\UeV$                    \\
%\Rule{1.2em}
 KamLAND-Zen
 & \Isotope[136]{Xe} 
 & $<0.12$--$0.25\UeV $   \\
CUORICINO 
 & \Isotope[130]{Te} 
  &  $<0.2$--$0.7\UeV $ \\
 \hline\hline
\end{tabular}
\caption[]{Present bounds at $90\%$CL from some
 recent neutrinoless double-beta-decay
           experiments \cite{Agashe:2014kda}.}
\label{tab:2b0n} 
\end{table}

\section{Neutrino Oscillations}

The most spectacular implication of neutrino masses and mixings is the macroscopic quantum phenomenom of neutrino oscillations, first introduced by 
B.~Pontecorvo \cite{Pontecorvo:1957cp}.  The Nobel prize in 2015 has been awarded to T.~ Kajita (from the SuperKakiokande collaboration) and A. B.~ McDonald (from the SNO collaboration) for the {\it discovery of neutrino oscillations, which shows that neutrinos have a mass}. 

\begin{figure}
\centering
\includegraphics[
width=1\linewidth]{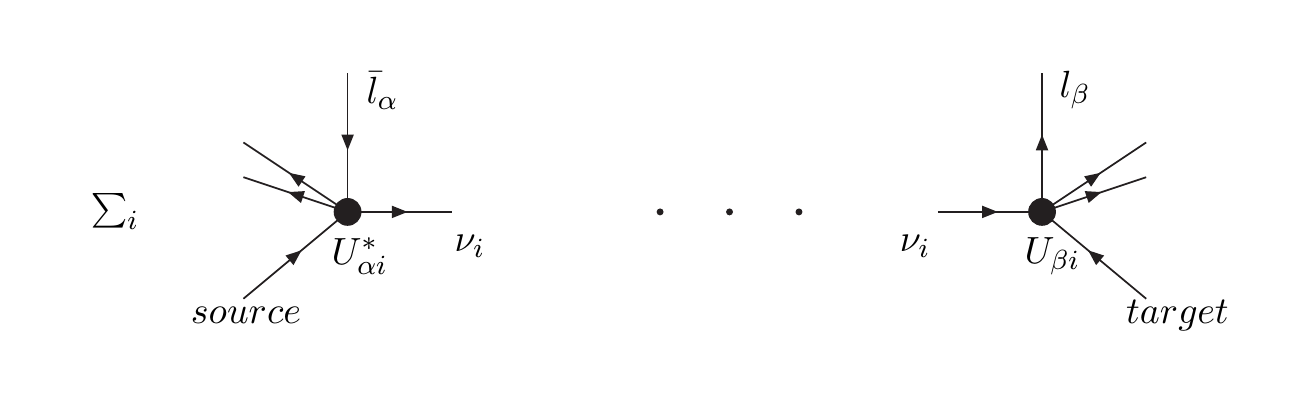}
\caption[]{Neutrino oscillations.}
\label{fig:nuoscmio}
\end{figure}

We have seen that the neutrino flavour fields $(\nu_e, \nu_\mu,\nu_\tau)$ that couple via CC to the leptons $(e,\mu, \tau)$  are unitary combinations of the mass eigenstates fields $(\nu_1, \nu_2, \nu_3)$:
\begin{eqnarray}
\left(\begin{array}{l}\nu_e \\
\nu_\mu \\
\nu_\tau \end{array}\right) = U_{\rm PMNS}(\theta_{12},�\theta_{13}, \theta_{23}, {\rm phases}) \left(\begin{array}{l}\nu_1 \\
\nu_2 \\
\nu_3 \end{array}\right).
\end{eqnarray}
In a neutrino oscillation experiment, neutrinos are produced  by a source  (e.g. pion or $\mu$ decays, nuclear reactions, etc) and are detected 
some macroscopic distance, $L$,  away from the production point. They  are produced and detected via  weak processes in combination with a given lepton flavour,that is in flavour states. As these states propagate undisturbed in space-time from the production to the detection regions, the different mass eigenstates, having slighly different phase velocities, pick up different phases, resulting in a non-zero 
probability that the state that arrives at the detector is in a different flavour combination to the one originally produced, see  \Fref{fig:nuoscmio}. The probability for this flavour transition oscillates with the distance travelled.

Two ingredients are mandatory for this phenomenom to take place:
\begin{itemize}
\item neutrinos must keep quantum coherence in propagation over macroscopic distances, which is only possible because they are so weakly interacting
\item there is sufficient uncertainty in momentum at production and detection so that a coherent flavour state can be produced\footnote{If the  momentum uncertainty is sufficiently small one could kinematically distinguish the mass eigenstate being produced/detected.}
\end{itemize}

The master formula for the oscillation probability of $\nu_\alpha$ turning into a $\nu_\beta$ is
\begin{eqnarray}
P(\nu_\alpha\rightarrow\nu_\beta) = \sum_{i,j} U^*_{\alpha i}
U_{\beta i} U_{\alpha j} U^*_{\beta j}  e^{-i \frac{\Delta m^2_{ji} L}{2 |\mathbf{p}|}},
\label{eq:master}
\end{eqnarray} 
where $\Delta m^2_{ji} \equiv m_i^2-m_j^2$,  $U_{\alpha i}$ are the elements of the 
PMNS matrix, $L$ is the baseline and $\mathbf{p}$ is the neutrino momentum. 

There are many ways to derive this formula. The simplest way that appears in most textbooks 
uses simple quantum mechanics, where neutrinos are treated as plane waves. A slightly more rigorous 
method treats neutrinos as wave packets. Finally, it is also possible to derive it from QFT, where
neutrinos are treated as intermediate virtual states. The different methods make more or less explicit the basic necessary conditions
of neutrino oscillations mentioned above, and therefore are more or less prone to quantum paradoxes.

\subsection{Plane wave derivation} 

Let us suppose that a neutrino of flavor $\alpha$ is produced at $t_0$. It is therefore a superposition of the mass eigenstates that we assume to be plane waves with spatial  momentum $\mathbf{p}$:
\begin{eqnarray}
|\nu_\alpha(t_0)\rangle = \sum_i U^*_{\alpha i} |\nu_i(\mathbf{p})\rangle.
\end{eqnarray}
The mass eigenstates are eigenstates of the free Hamiltonian:
\begin{eqnarray}
\hat{H}�|\nu_i(\mathbf{p})\rangle. = E_i(\mathbf{p}) |\nu_i(\mathbf{p})\rangle, ~~~E_i(\mathbf{p})^2 = {\mathbf p}^2+ m_i^2.
\end{eqnarray}
The time evolution operator from $t_0 \rightarrow t$ is given by $e^{-i \hat{H} (t-t_0)}$ and therefore 
the state at time $t$ is  given by
\begin{eqnarray}
|\nu_\alpha(t)\rangle = e^{-i \hat{H} (t-t_0)}|\nu_\alpha(t_0)\rangle= \sum_i U^*_{\alpha i}  e^{-i E_i({\mathbf p}) (t-t_0)} |\nu_i(\mathbf{p})\rangle.
\end{eqnarray}
The probability that at time $t$ the state is in flavour $\beta$ is
\begin{eqnarray}
P(\nu_\alpha \rightarrow \nu_\beta)(t) = \left|\langle\nu_\beta| \nu_\alpha(t)\rangle \right|^2 = \left|\sum_i U_{\beta i} U^*_{\alpha i}  e^{-i E_i({\mathbf p}) (t-t_0)}  \right|^2,
\label{eq:planeAcal}
\end{eqnarray}
where we have used the orthogonality relation $\langle \nu_i({\mathbf p})| \nu_j({\mathbf p})\rangle = \delta_{ij}$.

Since the neutrinos are ultrarelativistic, we can approximate
\begin{eqnarray}
E_i({\mathbf p})-E_j({\mathbf p}) \simeq {1 \over 2} \frac{m_i^2-m_j^2}{|{\mathbf p}|} + {\mathcal O}(m^4),
\end{eqnarray}
and $L \simeq (t-t_0)$, so that the master formula in eq.~(\ref{eq:master}) is recovered. 

The well-founded criticism to this derivation can be summarized in the following questions: 1) why  are all mass eigenstates of equal spatial momentum, ${\mathbf p}$ ? 2) is the plane wave treatment justified when the production and detection regions are localized ?
3) why is it necessary to do the $t-t_0 \rightarrow L$ conversion ?

A number of quantum paradoxes can be formulated  from these questions, that can be resolved only when the two basic conditions for neutrino oscillations above are made explicit. This can be achieved in a wave packet treatment.  

\subsection{Wave packet derivation}

Many authors have derived the master formula treating neutrinos involved as wave packets. For some recent references see \cite{Akhmedov:2009rb,Akhmedov:2010ms}. 

A neutrino of flavour $\alpha$ is produced  at time and position $(t_0, {\mathbf x}_0)=(0,{\mathbf 0})$ as a superposition of {\it source} wave packets, $f_i^S(\mathbf p)$, one for each mass eigenstate. The state at time and position $(t, \mathbf x)$ is therefore
\begin{eqnarray}
|\nu_\alpha(t)\rangle = \sum_i U^*_{\alpha i} \int_{\mathbf p} f^S_i({\mathbf p})  e^{-i E_i({\mathbf p}) t } %e^{i {\mathbf p}{\mathbf x}}
  |\nu_i({\mathbf p})\rangle.
\end{eqnarray}
For simplicity we will assume gaussian wave packets, with an average momentum $\mathbf Q_i$ and width $\sigma_S$:
\begin{eqnarray}
f_i^S(\mathbf p) \propto e^{-(\mathbf p-\mathbf Q_i)^2/2 \sigma_S^2}.
\end{eqnarray}
Note that we have lifted the assumption that all mass eigenstates have the same spatial momentum. 

A neutrino of flavour $\beta$ is detected at time and position $(T, \mathbf L)$ as a superposition of {\it detector} wave packets, $f_i^D({\mathbf p})$, created at this space-time position.  The state detected is therefore 
\begin{eqnarray}
|\nu_\beta(t)\rangle = \sum_j U^*_{\beta j} \int_{\mathbf p} f^D_j({\mathbf p})  e^{-i E_j({\mathbf p}) (t-T) } e^{-i {\mathbf p}
 {\mathbf L}} |\nu_j({\mathbf p})\rangle,
\end{eqnarray}
where we also assume gasussian wave packets at detection, with average momentum ${\mathbf Q}'_j$ and width $\sigma_D$:
\begin{eqnarray}
f_j^D(\mathbf p) \propto e^{-(\mathbf p-\mathbf Q'_j)^2/2 \sigma_D^2}.
\end{eqnarray}
The probability amplitude for the first state to turn into the second is therefore
\begin{eqnarray}
\mathcal A (\nu_\alpha \rightarrow \nu_\beta) = \langle \nu_\beta(t)|\nu_\alpha(t) \rangle = \sum_i U^*_{\alpha i} U_{\beta i} \int_{\mathbf p}  e^{-i E_i(\mathbf p) T } e^{i {\mathbf p} {\mathbf L}} f^S_i({\mathbf p})  f^{D*}_i({\mathbf p}) 
\label{eq:Acal}
\end{eqnarray}
For gaussian wave packets we can rewrite the product of the $S$ and $D$ wave packets as a gaussian wave packet:
\begin{eqnarray}
f_i^{D*}({\mathbf p}) f_i^S({\mathbf p}) \propto f_i^{ov}({\mathbf p}) e^{-(\mathbf Q_i-\mathbf Q'_i)^2/4 (\sigma_S^2+ \sigma_D^2)},
\end{eqnarray}
where the overlap wave packet
\begin{eqnarray}
f_i^{ov}({\mathbf p}) \equiv e^{-(\mathbf p-\bar{{\mathbf Q}}_i)^2/2 \sigma_{ov}^2}, ~~\bar{{\mathbf Q}}_i \equiv \left({\mathbf Q_i\over \sigma_S^2}+{\mathbf Q'_i\over \sigma_D^2}\right) \sigma_{ov}^2,~~\sigma_{ov}^2 \equiv {1\over 1/\sigma_S^2+ 1/\sigma_D^2}.
\end{eqnarray}
The momentum integral in eq.~(\ref{eq:Acal}) can be done analytically if we approximate
\begin{eqnarray}
E_i({\mathbf p}) \simeq E_i(\bar{\mathbf Q}_i) + \sum_k \left.{\partial E_i\over \partial p_k}\right|_{ \bar{\mathbf Q}_i} (p_k - (\bar{Q}_i)_k) +... = E_i(\bar{\mathbf Q}_i) + {\mathbf v}_i ({\mathbf p} - {\bar{\mathbf Q}_i}) +...,
\end{eqnarray}
where $\mathbf v_i$ is the overlap wave packet group velocity. 

The amplitude obtained is
\begin{eqnarray}
\mathcal A (\nu_\alpha \rightarrow \nu_\beta) \propto  \sum_i U^*_{\alpha i} U_{\beta i}   e^{-i E_i(\bar{{\mathbf Q}}_i)T } e^{i \bar{{\mathbf Q}}_i {\mathbf L}} e^{-(\mathbf Q_i-\mathbf Q'_i)^2/4 (\sigma_S^2+ \sigma_D^2)} e^{-(\mathbf L-\mathbf v_i T)^2 \sigma_{ov}^2/2}.
\end{eqnarray}
Note that the two last exponential factors impose momentum conservation (the average momentum of the source and detector wave packets 
should be equal up to the momentum uncertainty) and the classical relation $\mathbf L = \mathbf v_i T$ within the spatial  uncertainty, $\sigma_{ov}^{-1}$. 

Since we usually do not measure the detection time $T$ in a neutrino oscillation experiment, we should integrate the probability over this variable. 
For simplicity we assume $\mathbf Q_i \simeq\mathbf Q'_i$ and  parallel to $\mathbf L$. In this case,  the integral gives:
\begin{eqnarray}
P(\nu_\alpha \rightarrow \nu_\beta) &\propto& \int_{-\infty}^{\infty} d T |\mathcal A(\nu_\alpha \rightarrow \nu_\beta)|^2 \nonumber\\
&\propto& \sum_{i,j} U^*_{\alpha i}
U_{\beta i} U_{\alpha j} U^*_{\beta j}  e^{-i \frac{\Delta m^2_{ji} L}{2 |\mathbf{p}|}} \underbrace{e^{-\left({L\over L_{coh}(i,j)}\right)^2}}_{\rm coherence} \underbrace{e^{-\left({E_i(\bar{\mathbf Q}_i)-E_j(\bar{\mathbf Q}_j) \over 2 \sigma_{ov}}\right)^2}}_{\rm momentum~ uncertainty}
\label{eq:pwp}
\end{eqnarray}
where the coherence length
\begin{eqnarray}
L_{\rm coh}(i,j) \simeq \sigma_{ov} {|\mathbf v_i - \mathbf v_j|\over \sqrt{|\mathbf v_i ^2+ \mathbf v_j^2}},
\end{eqnarray}
represents the distance travelled by the two wave packets, moving at slightly different group velocities $v_i$ and $v_j$, such that the center of the two wave packets  have separated  spacially a distance of the order of the spatial uncertainty $\sigma_{ov}^{-1}$. For $L \geq L_{coh}(i,j)$ the coherence between the wave packets $i,j$ is lost and the corresponding terms in the oscillation probability exponentially suppressed. The last exponential factor in eq.~(\ref{eq:pwp}) leads to a suppression of the oscillation probability when the difference in average energies of the two wave packets $i,j$ is larger than the momentum uncertainty of the overlap wave packet, $\sigma_{ov}$. Note that $\sigma_{ov}$ is dominated by the smallest of the production and detection uncertainties, and therefore both should be large enough to ensure that the wave packets of the different mass eigenstates remain
coherent. To the extent that $L \ll L_{\rm coh}$ and $|E_i-E_j| \ll {\rm Min}(\sigma_{S}, \sigma_D)$, the probability reduces to the master formula, with one caveat: we have lost the normalization along the way. This is usually unavoidable in the wave packet derivation. The right normalization can be imposed only a posteriori, for example, from unitarity, $\sum_\beta P(\nu_\alpha \rightarrow \nu_\beta) = 1$. 

In summary, the wave packet derivation is clearly more physical, as it makes explicit the two necessary conditions  for neutrino oscillations to take place: coherence and sufficient momentum uncertainty.  

\subsection{QFT derivation}

Since we are dealing with relativistic quantum mechanics, QFT should be the appropriate framework to derive the oscillation probability. 

In QFT we consider  scattering processes where some asymptotic {\it in-states} that we can prepare in the infinite past  come close together at some finite time in an interaction region and scatter off  into other asymptotic {\it out-states} at time $t\rightarrow \infty$. The probability amplitude for this process is just the scalar product  
of the in and out states. In computing this amplitude we usually idealise the asymptotic states as plane waves, which is a good approximation 
provided the interaction region is small compared to the Compton wavelength of the scattering states. In reality however the proper normalization of the scattering probability as a probability per unit time and volume requires that the initial states are normalized wave packets.  

\begin{figure}
\centering
\includegraphics[
width=.5\linewidth]{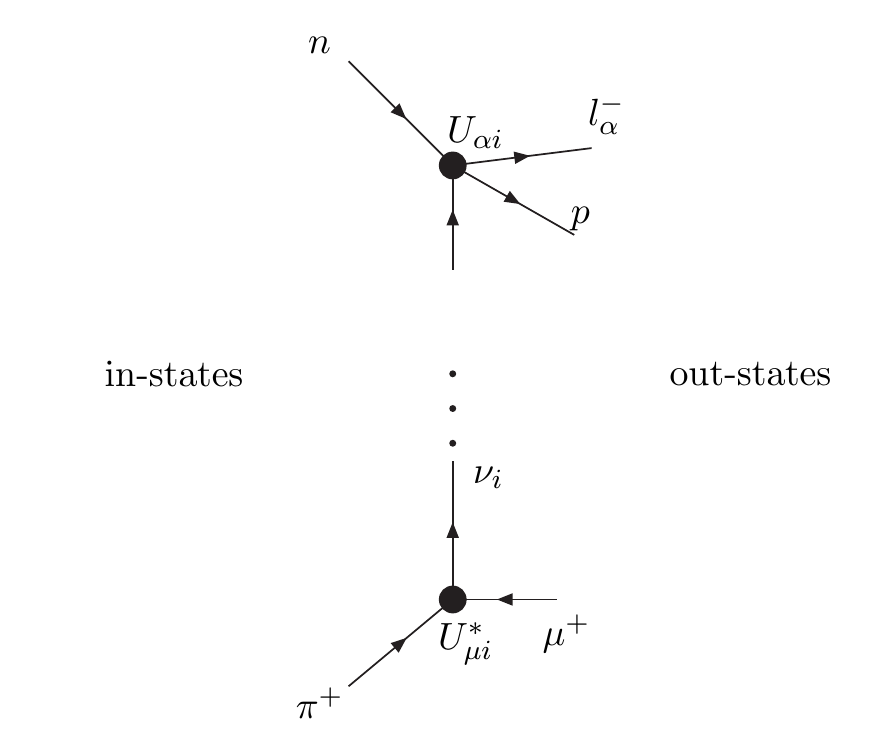}
\caption[]{Neutrino oscillations in QFT.}
\label{fig:qft}
\end{figure}

In a neutrino oscillation experiment, the asymptotic states are not the neutrinos, we cannot really prepare the neutrino states, but the particles that produce the neutrino at the source 
and those that interact with the neutrino in the detector. The neutrino is just a virtual particle being exchanged between the source and detector, see Fig.~\ref{fig:qft}, and in this perspective the interaction region is as large as the baseline and therefore macroscopic, in particular much larger 
than the Compton wavelength of the asymptotic states involved. It is mandatory therefore to consider the in-states as wave packets to ensure the localization of the source and detector. 

Consider for example  a neutrino beam produced from pions at rest and a detector some distance apart, where neutrinos interact with nucleons that are also at rest, via a quasi-elastic event:
\begin{eqnarray}
\pi n \rightarrow p \mu l_\beta. 
\end{eqnarray}
The in-states therefore will be the two wave packets representing a static pion that decays and is localized at time and position $(0,\mathbf{0})$ within the uncertainty better defined than the decay tunnel, and a nucleon  that is static and localized within the  detector, at time and position $(T,\mathbf L)$, when the interaction takes place. The out-states are the muon produced in pion decay and the lepton and hadron produced in the quasi-elastic event. The probability amplitude for the whole process includes the pion decay amplitude, the neutrino propagation  and the scattering amplitude at the detector. Therefore in order to 
extract from the full amplitude an oscillation probability, it must be the case that there is factorization of the whole probability into three factors that can be identified with the flux of neutrino from pion decay, an oscillation probability and a neutrino cross section. 

By explicit calculation \cite{Cerda}, it is possible to show that such factorization does indeed take place as long as kinematical effects of neutrino masses can be neglected. The oscillation probability defined as the ratio of the probability for the whole process and the  product of the neutrino flux from pion decay and the neutrino scattering cross-section is properly normalized.

\subsection{Neutrino oscillations in vacuum}

Let us analyse more closely the master formula eq.~(\ref{eq:master}). The probability is  
 a superposition of oscillatory functions of the baseline with wavelengths that depend on the
neutrino mass differences $\Delta m^2_{ij} = m_j^2-m_i^2$, and amplitudes that depend on different
combinations of the mixing matrix elements. 
Defining $W_{\alpha\beta}^{ij}\equiv \,[U_{\alpha i}U_{\beta i}^*
U_{\alpha j}^*U_{\beta j}]$ and using the unitarity of the mixing
matrix, we can rewrite the probability in the more familiar form:
\begin{eqnarray}
P(\nu_\alpha& \rightarrow & \nu_\beta) \, = \delta_{\alpha\beta}\, 
-4\; \sum_{j>i}\,\rm{Re}[W_{\alpha\beta}^{ij}]\, 
\sin^2\!\left(\frac{\Delta m^2_{ij}\,L}{4 E_\nu}\right)
\,\nonumber
\\ 
&\mp &\, 2  \,
\sum_{j>i}\, \rm{Im}[W_{\alpha\beta}^{ij}]\, 
\sin\!\left(\frac{\Delta m^2_{ij}\,L}{2 E_\nu}\right),
\label{eq:prob}
\end{eqnarray}
where the $\mp$ refers to neutrinos/antineutrinos and $|\vec{q}|\simeq E_\nu$. 

We refer to an \emph{appearance} or \emph{disappearance} oscillation
probability when the initial and final flavours are different
($\alpha\neq\beta$) or the same ($\alpha = \beta$), respectively. Note
that oscillation probabilities show the expected GIM suppression of
any flavour changing process: they vanish if the neutrinos are
degenerate.

\begin{figure}
\centering
\includegraphics[width=.47\linewidth]{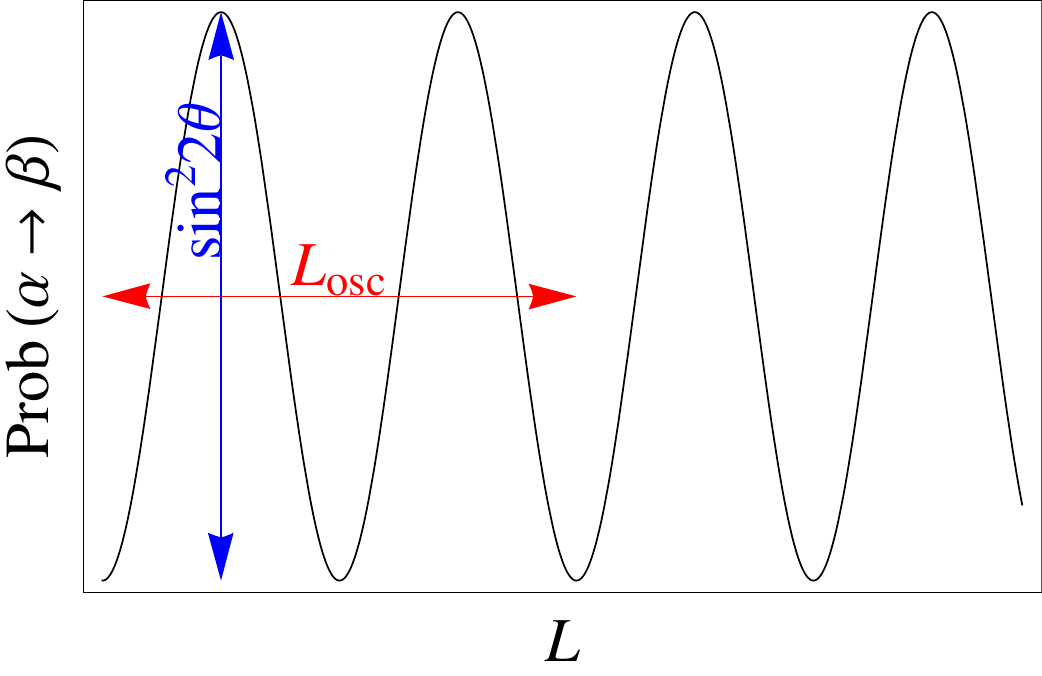}\quad
\includegraphics[width=.47\linewidth]{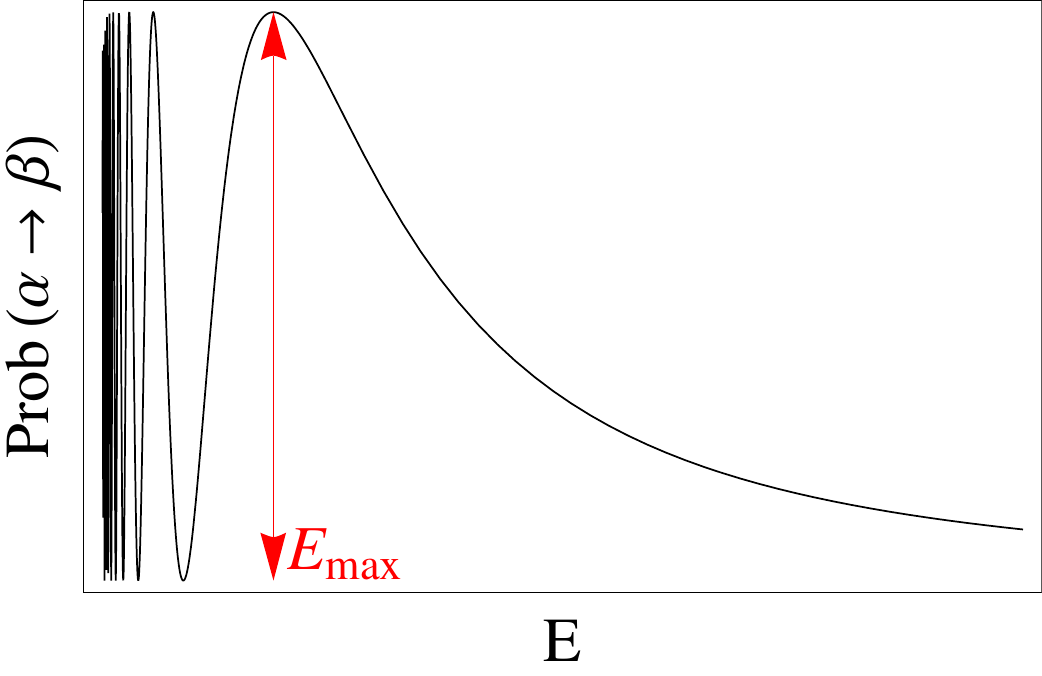}
\caption[]{Left: two-family appearance oscillation probability as a function of the
           baseline of $L$ at fixed neutrino energy. Right: same probability shown 
           as a function of the neutrino energy for fixed baseline.}
\label{fig:osc}
\end{figure}

In the simplest case of two-family mixing, the mixing matrix depends on just one
mixing angle:
\begin{eqnarray}
U_\text{PMNS} = \begin{pmatrix}
\cos \theta & \sin \theta \cr
-\sin \theta & \cos \theta 
          \end{pmatrix}\SPp,
\end{eqnarray}
and there is only one mass square difference $\Delta m^2$.  The
oscillation probability of \Eref{eq:prob} simplifies to the well-known
expression where we have introduced convenient physical units:
\begin{eqnarray}
P(\nu_\alpha\rightarrow \nu_\beta) 
&=& \sin^2 2\theta \; \sin^2\!\left(1.27 \frac{\Delta m^2(\UeVZ^2) \,L (\UkmZ)}{ E_\nu(\UGeVZ)}\right), \quad
\alpha\neq\beta\SPp.\nonumber\\
P(\nu_\alpha \rightarrow \nu_\alpha)&=& 1- P(\nu_\alpha\rightarrow \nu_\beta).
\label{eq:wk}
\end{eqnarray} 
The probability is the same for neutrinos and antineutrinos, because
there cannot be CP violation when there are only two families. Indeed CPT implies
that the disappearance probabilities are the same for neutrinos an antineutrinos, 
and therefore according to eq.~(\ref{eq:wk}) the same must hold for the appearance probability.  
The latter is a
sinusoidal function of the distance between source and detector, with
a period determined by the oscillation length:
\begin{eqnarray} 
L_\text{osc}~(\UkmZ) = 
\pi \frac{E_\nu(\UGeVZ)}{1.27 \Delta m^2 (\UeVZ^2)}\SPp,
\end{eqnarray}
which is proportional to the neutrino energy and inversely
proportional to the neutrino mass square difference.  The amplitude of
the oscillation is determined by the mixing angle. It is maximal for
$\sin^2 2 \theta =1$ or $\theta=\pi/4$. The oscillation probability 
as a function of the baseline  is shown on the left plot of \Fref{fig:osc}. 

In many neutrino oscillation experiments the baseline
is not varied but the oscillation probability can be measured as a function of 
the neutrino energy. This is shown on the right plot of \Fref{fig:osc}. In this case, the
position of the first maximum contains information on the mass splitting:
\begin{eqnarray}
E_{\rm max}(\UGeVZ) = 1.27 { \Delta m^2 (\UeVZ^2) L(\UkmZ)\over \pi/2}. 
\end{eqnarray}
An optimal neutrino oscillation experiment in vacuum is such that the ratio of the neutrino energy and baseline are
tunned to be of the same order as the mass splitting, $E/L \sim \Delta m^2$. If $E/L \gg \Delta m^2$, the oscillation phase is small and 
the oscillation probability depends on the combination $P(\nu_\alpha\rightarrow\nu_\beta) \propto \sin^2 2 \theta (\Delta m^2)^2$, and the 
mixing angle and mass splitting cannot be disentagled. The opposite limit $E/L \ll \Delta m^2$  is the fast oscillation regime, where one can only measure an energy or baseline-smeared oscillation probability 
\begin{eqnarray}
\langle P(\nu_\alpha\rightarrow\nu_\beta) \rangle \simeq  {1 \over 2} \sin^2 2 \theta.
\end{eqnarray}
It is interesting, and reassuring, to note that this averaged oscillation regime gives the same result as the flavour transition probability in the case of incoherent propagation ($L \gg L_{\rm coh}$):
\begin{eqnarray}
P\left(\nu_\alpha \rightarrow \nu_\beta\right) = \sum_i |U_{\alpha i} U_{\beta i}|^2 = 2 \cos^2\theta \sin^2\theta  = \frac{1}{2} \sin^2 2\theta .
\end{eqnarray}
Flavour transitions via incoherent propagation are sensitive to mixing but not to the neutrino mass splitting. 
The 'smoking gun' for neutrino oscillations is not the flavour
transition, which can occur in the presence of neutrino mixing without
oscillations, but the peculiar $L/E_\nu$ dependence.  An optimal
experiment that intends to measure both the mixing and the mass splitting requires
running $E/L\sim \Delta m^2$.

\subsection{Neutrino propagation in matter}

When neutrinos propagate in matter (Earth, sun, etc.),  their
propagation is modified owing to coherent forward scattering on electrons
and nucleons \cite{wolf}:

\begin{center}
\includegraphics[width=0.8\linewidth]{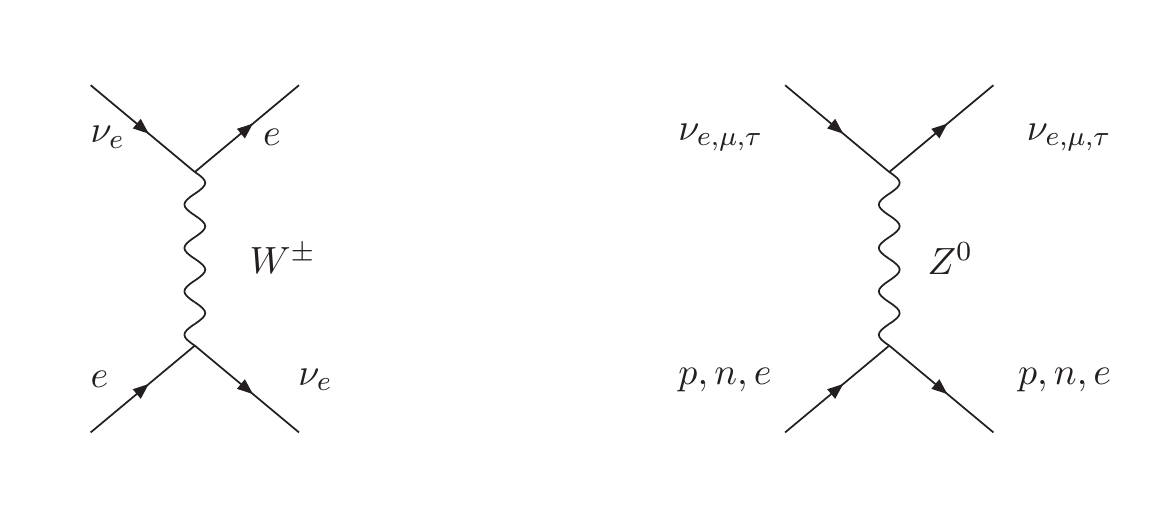}\quad
\end{center}

The effective Hamiltonian density 
resulting from the charged current interaction is
\begin{eqnarray}
\mathcal{H}_{\text{CC}}=2  \sqrt{2} G_F \,[\bar{e}\gamma_\mu P_L \nu_e]
[\bar{\nu}_e\gamma^\mu P_L e]
= 2 \sqrt{2} G_F\,[\bar{e}\gamma_\mu P_L e]
[\bar{\nu}_e\gamma^\mu P_L \nu_e]. 
\end{eqnarray}
Since the medium is not polarized, the expectation value of the
electron current is simply the number density of electrons:
\begin{eqnarray}
\langle \bar{e}\gamma_\mu P_L e \rangle_{\rm unpol. medium} & = & \delta_{\mu 0} {N_e\over 2}
.
\end{eqnarray}
Including also the neutral current interactions in the same way, the 
effective Hamiltonian for neutrinos in the presence of matter is
\begin{eqnarray}
\langle {\mathcal H}_{CC}+ {\mathcal H}_{NC} \rangle_{\rm medium}=  \bar{\nu} V_m \gamma^0 (1-\gamma_5)\nu \\\nonumber\\
V_m = \begin{pmatrix}
       \frac{G_F}{\sqrt{2}} \left(N_e  -\frac{N_n}{2}\right) 
         & 0 
           & 0\\
       0 
         &   \frac{G_F}{\sqrt{2}} \left(-\frac{N_n}{2}\right)  
           & 0 \\
       0 
         & 0 
           & \frac{G_F}{\sqrt{2}} \left(-\frac{N_n}{2}\right)
\end{pmatrix}\SPp,
\end{eqnarray}
where $N_n$ is the number density of neutrons.  Due to the neutrality of matter, the proton and electron contributions to the neutral current potential cancel. 

The plane wave solutions to the modified Dirac equation satisfy a 
different dispersion relation and as a result,  
the phases of neutrino oscillation phenomena change. The new dispersion relation becomes
\begin{equation}
E - V_m - M_\nu = 
  \left(\pm |\vec{p}| - V_m \right) 
  \frac{1}{E + M_\nu - V_m} \left( \pm |\vec{p}| - V_m \right) \;\;\; h=\pm, 
\end{equation}
where $h=\pm$ indicate the two helicity states and we have neglected
effects of $\mathcal{O}(V M_\nu)$. This is a reasonable approximation
since $m_\nu \gg V_m$. For the positive energy states we then have
\begin{equation}
E > 0 \;\;\;\; E^2 = |\vec{p}|^2 + M_\nu^2 + 4 E V_m \;\;\ h=-\;\;\;  E^2 = |\vec{p}|^2 + M_\nu^2,\;  h=+,   
\end{equation}
while for the negative energy ones $V_m \rightarrow - V_m$ and $h\rightarrow -h$. 

The effect of matter can be simply accommodated in an effective mass matrix:
\begin{equation}
{\tilde M}_\nu^2 = M_\nu^2 \pm 4 E V_m.
\end{equation}
The effective mixing matrix $\tilde{V}_\text{MNS}$ is the one that takes us from the original flavour basis
to that which diagonalizes this effective mass matrix:
\begin{eqnarray}
\begin{pmatrix} {\tilde m}^2_1 & 0 & 0 \\
                0 & {\tilde m}^2_2 & 0 \\
                0 & 0 & {\tilde m}^2_3 
\end{pmatrix} = {\tilde V}^\dagger_\text{MNS} 
\left( M_\nu^2 \pm  4 E 
     \begin{pmatrix} V_e & 0   & 0 \\
                     0 & V_\mu & 0 \\
                     0 & 0 & V_\tau 
     \end{pmatrix} 
\right) {\tilde V}_\text{MNS}.
\end{eqnarray}
The effective mixing angles and masses depend on the energy. 

The matter potential
in the center of the sun is $V_m \sim 10^{-12}\UeV$ and in the Earth
$V_m \sim 10^{-13}\UeV$. In spite of these tiny values, these effects
are non-negligible in neutrino oscillations.

\subsection{Neutrino oscillations in constant matter}

In the case of two flavours, the effective mass and mixing angle have relatively 
simple expressions:
\begin{eqnarray}
\Delta {\tilde m}^2 &=& \sqrt{\left(\Delta m^2\cos 2\theta\mp 2 \sqrt{2} E \,G_F\,N_e 
\right)^2+\left(\Delta m^2 \sin 2\theta\right)^2},\\
\sin^2 2\tilde\theta &=&\frac{\left(\Delta m^2 \sin 2\theta\right)^2}
{(\Delta {\tilde m}^2)^2}
%\left(\Delta m^2\cos 2\theta \mp 2 \sqrt{2}\,G_F E \,N_e\right)^2
%+\left(\Delta m^2 \sin 2\theta\right)^2}\,
%\tan 2 {\tilde\theta} = {\Delta m^2 \sin 2\theta \over \Delta m^2 \cos\theta - 2 \sqrt{2}\,G_F E \,N_e}.
\end{eqnarray} 
where the sign $\mp$ corresponds to neutrinos/antineutrinos.  The
corresponding oscillation amplitude has a resonance \cite{Mikheev:1986gs},
when the neutrino energy satisfies
\begin{eqnarray}
\sqrt{2}\,G_F\,N_e \mp \frac{\Delta m^2}{2 E}\cos 2\theta = 0 \;\;\; \Rightarrow \;\;\;\sin^2 2 \tilde \theta = 1,\;\;\; \Delta {\tilde m}^2 = \Delta m^2 \sin 2 \theta .
\label{res1}
\end{eqnarray} 
The oscillation amplitude is therefore maximal, independently of the value of the vacuum 
mixing angle. 

We also note that
\begin{itemize}
\item oscillations vanish at $\theta=0$, because the oscillation length 
becomes infinite for $\theta=0$;
\item the resonance is only there for $\nu$ or $\bar{\nu}$  but not 
both; 
\item the resonance condition depends on the sign$(\Delta m^2 \cos 2 \theta)$:
\begin{center}
resonance observed in $\nu$ $\rightarrow$ sign($\Delta m^2 \cos 2 \theta$) $> 0$,\\
resonance observed in $\bar{\nu}$ $\rightarrow$ sign($\Delta m^2 \cos 2 \theta$) $< 0$.
\end{center} 
\end{itemize}

\begin{figure}
\begin{center}
\includegraphics[width=0.5\linewidth]{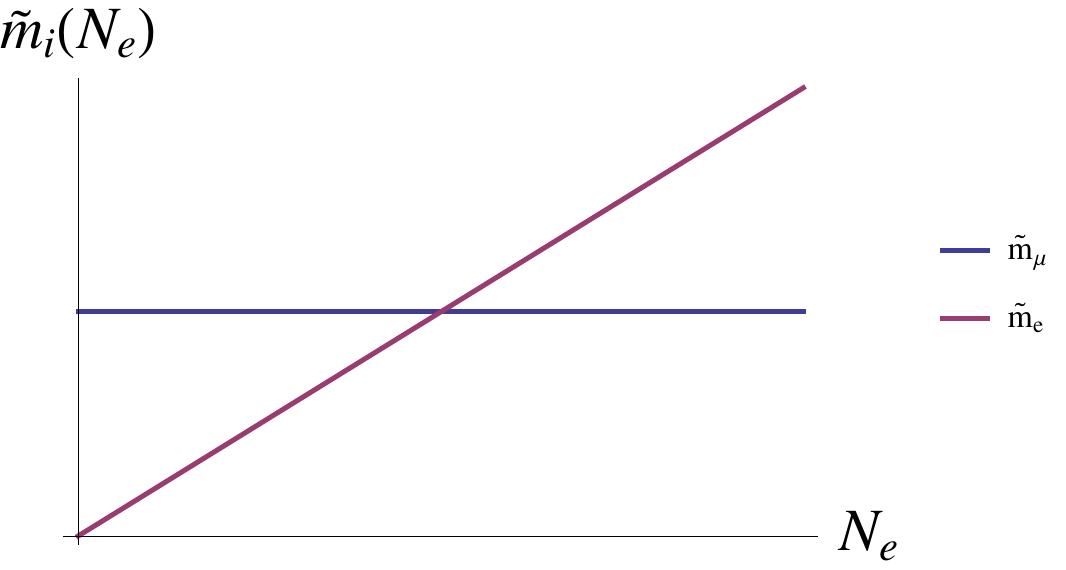}\includegraphics[width=0.5\linewidth]{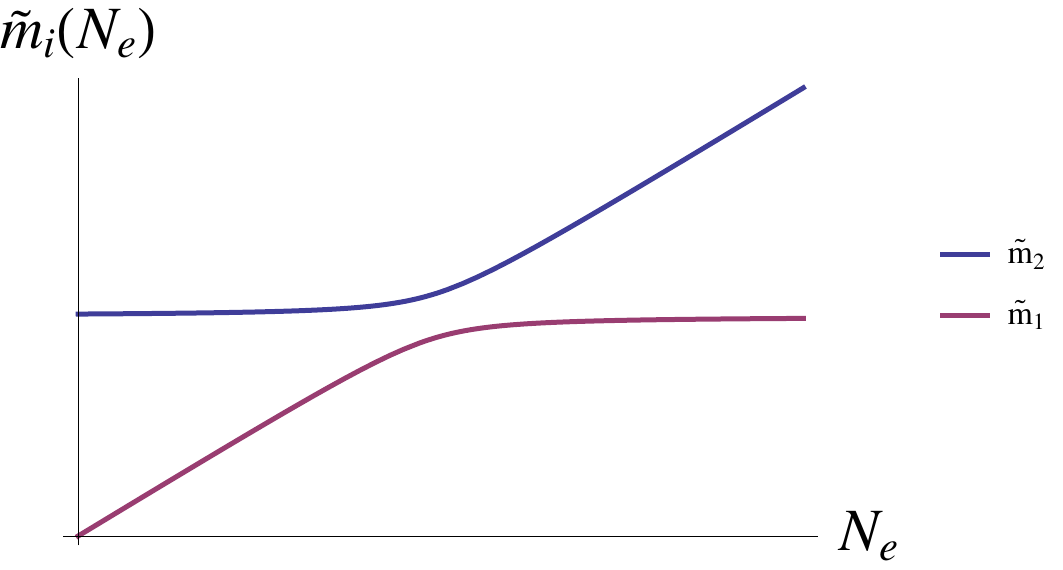}\quad
\caption{Mass eigenstates as a function of the electron number density at fixed neutrino energy for $	\theta=0$ (left) and $\theta \neq 0$ (right).  }
\label{fig:levelx}
\end{center}
\end{figure}

The origin of this resonance is a would-be level crossing in the case of vanishing mixing. In the case
of two families, for $\theta=0$, the mass eigenstates as a function of the electron number density, at fixed 
neutrino energy, are
depicted in Fig.~\ref{fig:levelx} for $\Delta m^2 >0$. As soon as the mixing is lifted from zero, no matter how small, 
the crossing cannot take place. The resonance condition corresponds to the minimum level-splitting point.
\subsection{Neutrino oscillations in variable matter}

In the sun the density of electrons is not constant. However, if the 
variation is sufficiently slow, the eigenstates  will 
 change slowly with the density, and we can 
assume that the neutrino produced in an eigenstate in the center of the sun, remains in the same 
eigenstate along the trajectory. This is the so-called \emph{adiabatic approximation}.

We consider here 
two-family mixing for simplicity. At any point in 
the trajectory, it is possible to diagonalize the Hamiltonian 
fixing the matter density to that at the given point. The
resulting eigenstates can be written as
\begin{eqnarray}
& |{\tilde\nu}_1\rangle = |\nu_e\rangle \,\cos\tilde\theta - |\nu_\mu\rangle \, \sin\tilde \theta , 
\\
& |{\tilde\nu}_2\rangle = |\nu_e\rangle \,\sin\tilde\theta  + |\nu_\mu\rangle \,\cos\tilde\theta .
\label{eq:eigenmat}  
\end{eqnarray}
Neutrinos are produced close to the centre $x=0$ where the electron density is
$N_e(0)$. Let us suppose that it satisfies
\begin{eqnarray}
2 \sqrt{2} G_F N_e(0) \gg \Delta m^2 \cos 2 \theta.
\end{eqnarray}
Then the diagonalization of the mass matrix at this point gives
\begin{eqnarray}
\tilde \theta \simeq \frac{\pi}{2} \Rightarrow  |\nu_e\rangle \simeq |{\tilde\nu}_2\rangle,
\end{eqnarray}
in such a way that an electron neutrino is mostly the second mass
eigenstate.  When neutrinos exit the sun, at $x=R_\odot$, the matter
density falls to zero, $N_e(R_\odot) = 0$, and the local effective
mixing angle is the one in vacuum, ${\tilde \theta} = \theta$.  If
$\theta$ is small, the eigenstate ${\tilde\nu}_2$ is mostly $\nu_\mu$
according to \Eref{eq:eigenmat}.

%%%%%%%%%%%%%%%%%%%%%%%%%%%% fig. mswtrian  %%%%%%%%%%%%%%%%%%%%%%%%%%%%%%
\begin{figure}
\centering 
\includegraphics[width=.5\linewidth]{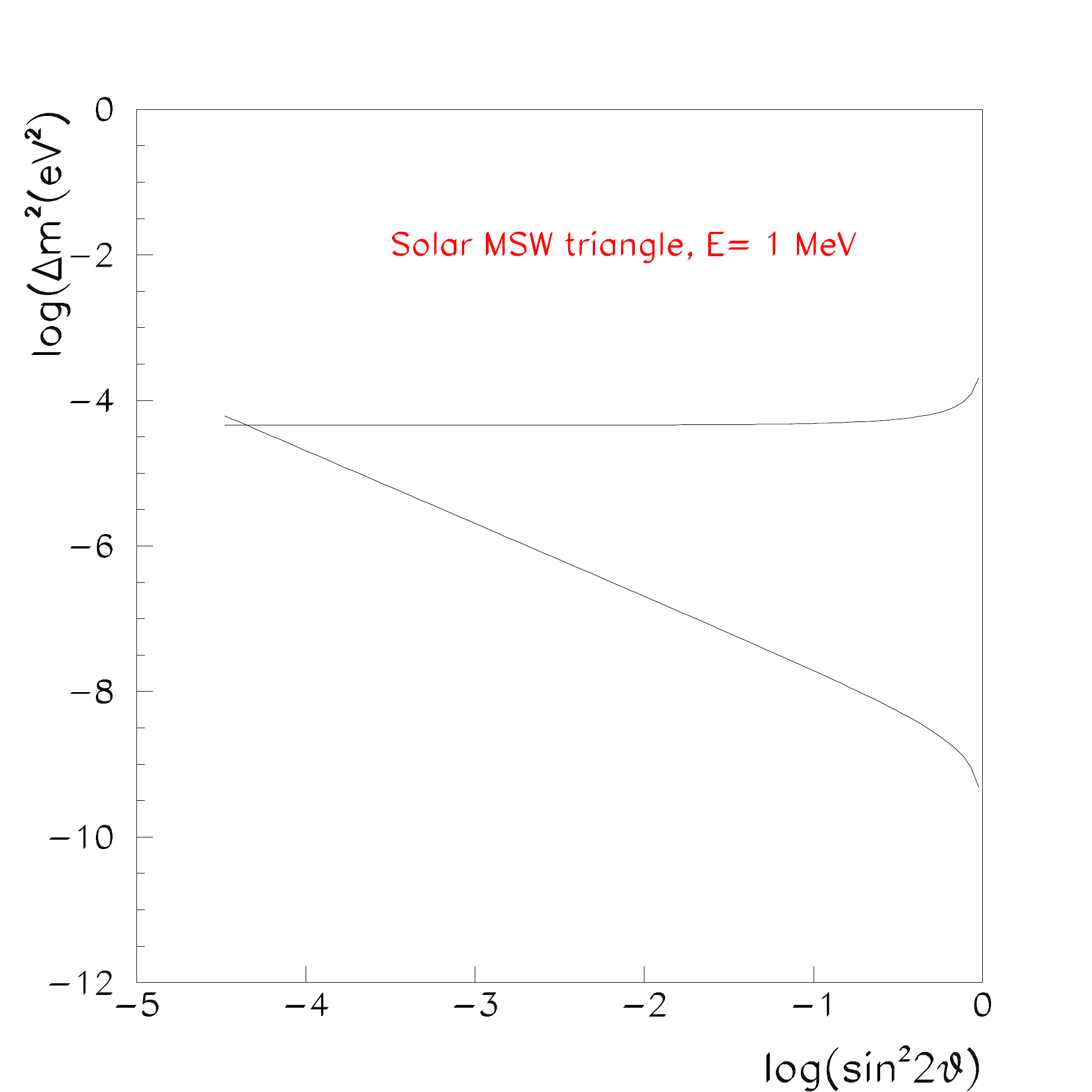}
\caption[]{MSW triangle: in the region between the two lines the
           resonance and adiabaticity conditions are both satisfied
           for neutrinos of energy 1\UMeV.}
           \label{fig:mswtrian} 
\end{figure}

Therefore an electron neutrino produced at $x=0$ is mostly the eigenstate
$\tilde \nu_2$, but 
this eigenstate outside the sun is mostly $\nu_\mu$. There is maximal $\nu_e \rightarrow \nu_\mu$ conversion if the adiabatic approximation is a good one. 
This is the famous MSW effect \cite{wolf,Mikheev:1986gs}. 
 The conditions for this to happen are:
\begin{itemize}
\item \emph{Resonant condition}: the density at the production is above the critical one 
\begin{equation}
N_e(0) > \frac{\Delta m^2 \cos 2\theta}{2 \sqrt{2} E G_F}. 
\label{eq:rcond}
\end{equation}
\item \emph{Adiabaticity}: the splitting of the levels is large compared 
to energy injected in the system by the variation of $N_e(r)$. A measurement of this is given by $\gamma$ which should be much larger than one:
\begin{equation}
\gamma = \frac{\sin^2 2 \theta}{\cos 2\theta}
         \frac{\Delta m^2}{2 E} \frac{1}{|\nabla \log N_e(r)|} > \gamma_\text{min} > 1, 
\end{equation}
where $\nabla = \partial/\partial r$.
\end{itemize}
At fixed energy both conditions give the famous MSW triangles, if
plotted on the plane $(\log(\sin^2 2 \theta), \log(\Delta m^2))$:
\begin{eqnarray}
\log\!\left(\Delta m^2\right) < 
\log\!\left( \frac{2 \sqrt{2} G_F N_e(0) E}{\cos 2\theta}\right)  \\
\log\!\left(\Delta m^2\right) > 
\log\!\left( \gamma_{\rm min} 2 E 
  \nabla \log N_e \frac{\cos 2\theta}{\sin^2 2\theta}\right). 
\end{eqnarray}
For example, taking $N_e(r) = N_c \exp(-r/R_0),  R_0=
R_\odot/10.54, N_c = 1.6 \times 10^{26}\Ucm^{-3}, E =1\UMeV$, these
curves are shown in \Fref{fig:mswtrian}.

\begin{figure}
\centering 
\includegraphics[width=.7\linewidth]{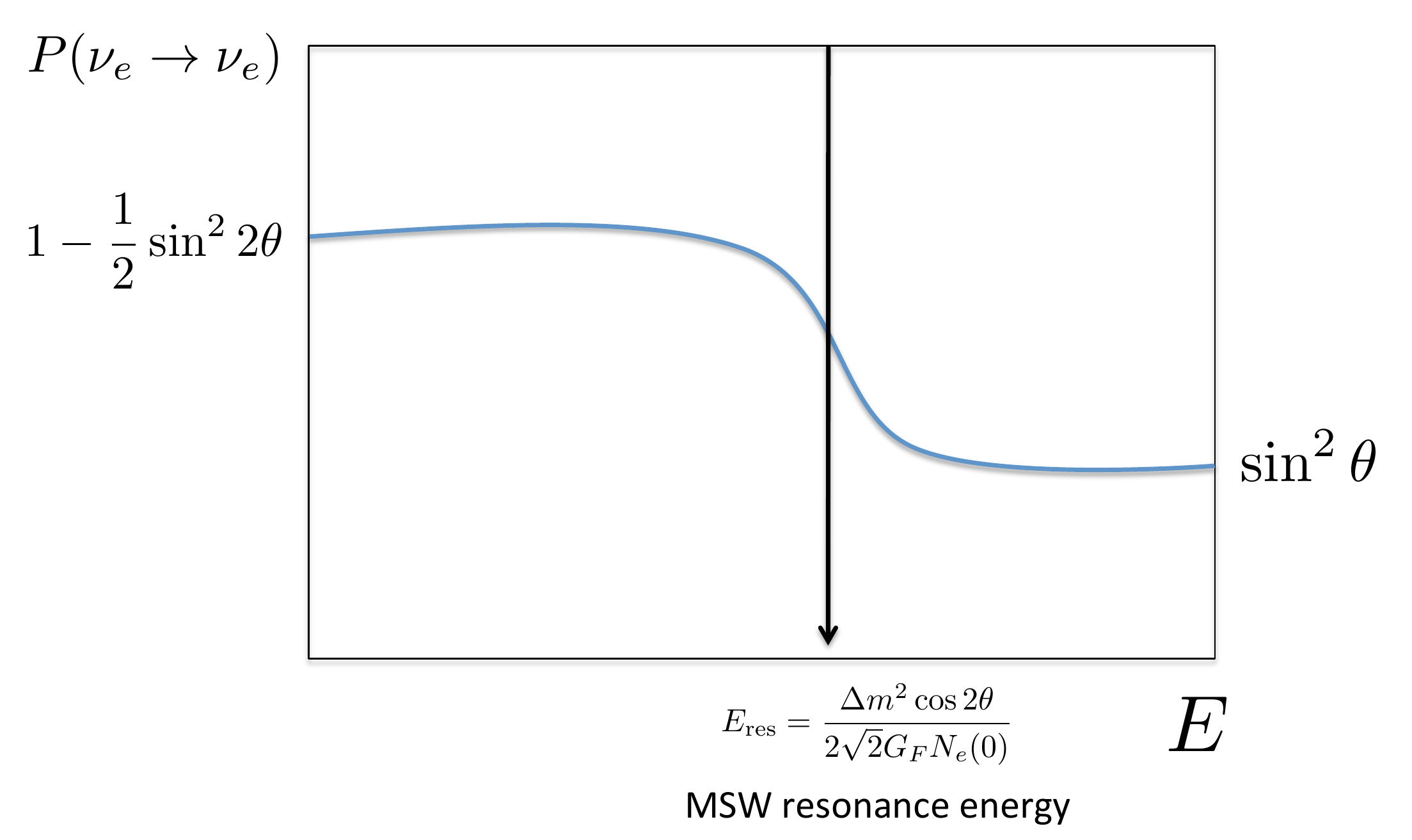}
\caption[]{Schematic survival probability of solar neutrinos as a function of the energy.}
           \label{fig:mswmio} 
\end{figure}

It should be stressed that neutrino oscillations are not responsible for the flavour transition of solar neutrinos. The survival probability of the solar $\nu_e$ in the adiabatic approximation is the incoherent sum of the contribution of each of the mass eigenstates:
\begin{eqnarray}
P(\nu_e \rightarrow \nu_e) = \sum_i |\langle \nu_e | {\tilde\nu}_i(R_\odot) \rangle|^2 |\langle  {\tilde\nu}_i(0)|\nu_e \rangle|^2,
\end{eqnarray}  
where $\tilde\nu_i(r)$ is the $i$-th mass eigenstate for the electron number density, $N_e(r)$, at a distance $r$ from the center of the sun. 
If the mass eigenstates contribute incoherently, how can we measure the neutrino mass splitting ?
The answer is that the resonance condition of eq.~(\ref{eq:rcond}) depends on the neutrino energy. If we define
\begin{eqnarray}
E_{\rm res} \equiv {\Delta m^2 \cos 2 \theta \over 2 \sqrt{2} G_F N_e(0)},
\end{eqnarray}
the MSW effect will affect neutrinos with $E > E_{\rm res}$,  while for $E< E_{\rm res}$, the oscillation probability is close to that in 
vacuum for averaged oscillations. The spectrum of the solar neutrino flux includes energies both above and below $E_{\rm res}$:
\begin{eqnarray}
P(\nu_e\rightarrow \nu_e)  &\simeq 1-{1\over 2} \sin^2 2 \theta, &  E \ll E_{\rm res}
 \nonumber\\
P(\nu_e\rightarrow \nu_e)  &
\simeq \sin^2 \theta, &  E \gg E_{\rm res}
\end{eqnarray}
 The sensitivity to $\Delta m^2$ relies on the ability to 
locate the resonant energy.  This behaviour is esquematically depicted in Fig.~\ref{fig:mswmio}.

\section{Evidence for neutrino oscillations} 

Nature has been kind enough to provide us with two natural sources of
neutrinos (the sun and the atmosphere) where neutrino flavour
transitions have been observed in a series of ingenious experiments,
that started back in the 1960s with the pioneering experiment of
R.~Davies.  This effort was rewarded with the Nobel prize of 2002 to R.~Davies
and M.~Koshiba {\it for the detection of cosmic neutrinos}. 

\subsection{ Solar neutrinos}
\label{sec:solar}

\begin{figure}
\centering
\includegraphics[width=.6\linewidth]{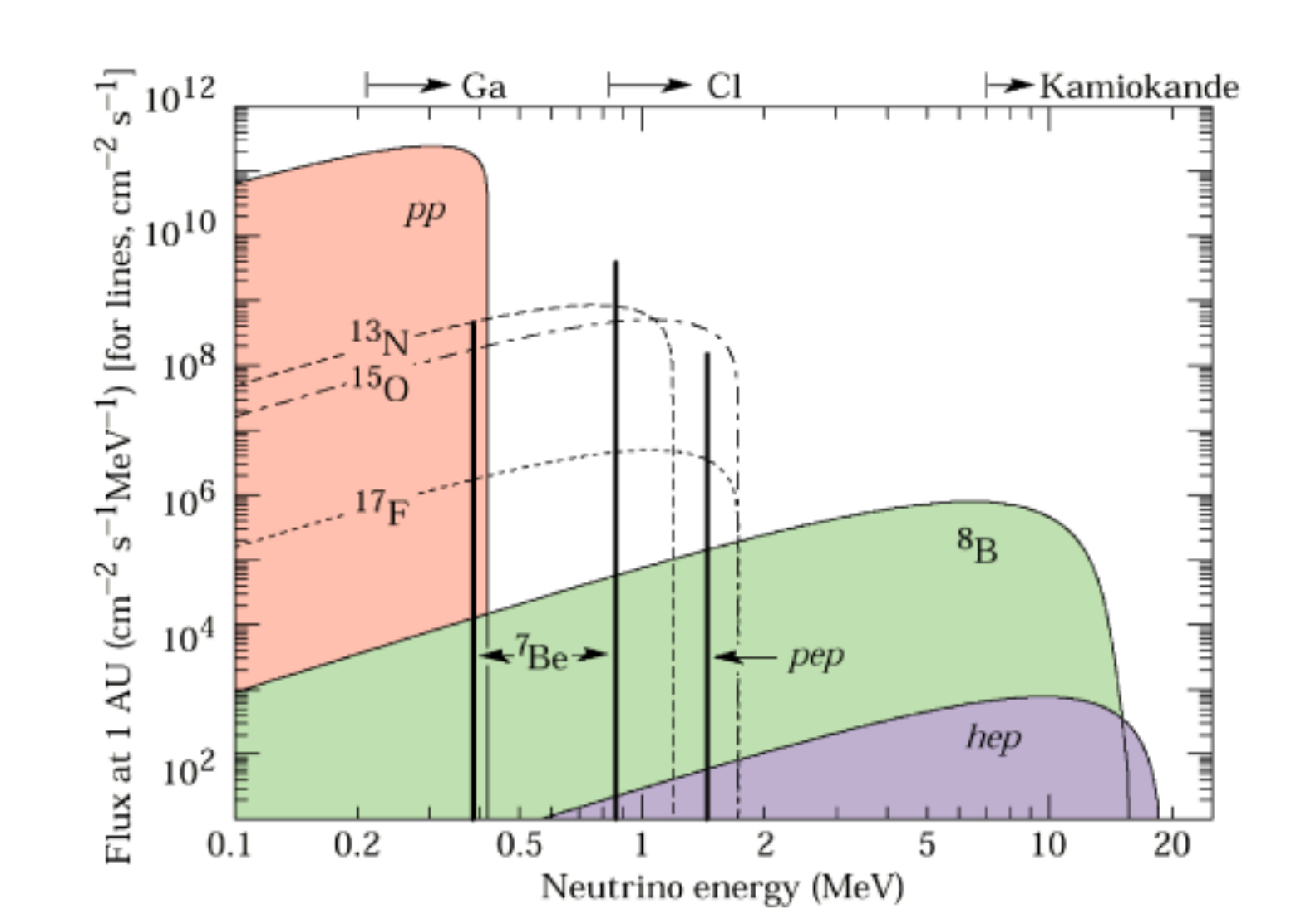}
\caption[]{Spectrum of solar neutrinos\cite{Bahcall:2000nu}. The arrows  indicate
           the threshold of the different detection
           techniques.}
\label{fig:bp2000}
\end{figure}

The sun, like all stars, is an intense source of neutrinos produced in the chain of 
 nuclear reactions that burn hydrogen into helium:
\begin{eqnarray}
4 p \longrightarrow \Isotope[4]{He} + 2 e^+ + 2 \nu_e. 
\end{eqnarray}
The theory of stelar nucleosynthesis was stablished at the end of the 30's 
by H.~Bethe \cite{Bethe:1939bt}.  The  spectrum of the solar  $\nu_e$, for massless neutrinos,  
is shown in \Fref{fig:bp2000}. The prediction of this flux, obtained 
by J.~Bahcall and collaborators \cite{Bahcall:2000nu}, is the result of a detailed 
simulation of the solar interior and has been improved over many years. 
It is the so-called standard solar model (SSM).

Neutrinos coming from the sun have been detected with several
experimental techniques that have a different neutrino energy
threshold as indicated in \Fref{fig:bp2000}. On the one hand, the
radiochemical techniques, used in the experiments Homestake
(chlorine, $^{37}$Cl)\cite{homestake}, Gallex/GNO 
\cite{gallex-gno} and Sage \cite{sage} (using gallium, $^{71}$Ga, and
germanium, $^{71}$Ge, respectively), can count the total number of
neutrinos with a rather low threshold ($E_\nu > 0.81\UMeV$ in Homestake
and $E_\nu > 0.23\UMeV$ in Gallex and Sage), but they cannot get any
information on the directionality, the energy of the neutrinos, nor the
time of the event.  

On the other hand, Kamiokande \cite{kamio}
pioneered a new technique to observe solar neutrinos using water
Cherenkov detectors that can measure the recoil electron in 
  elastic neutrino scattering on electrons:  $\nu_e \; + \; e^- \;\rightarrow \; \nu_e \; +
e^-$. This  is a real-time experiment that provides
information on the directionality and the energy of the neutrinos. The threshold
on the other hand is much higher, $\sim 5\UMeV$. All these
experiments have consistently observed a number of solar neutrinos
between 1/3 and 1/2 of the number expected in the SSM and for a long time
this was referred to as the \emph{solar neutrino problem or deficit}.

The progress in this field over the last decade has been enormous
culminating in a solution to this puzzle that no longer relies on
the predictions of the SSM. There have  been three milestones. 

\begin{figure}
\centering
\includegraphics[width=.47\linewidth]{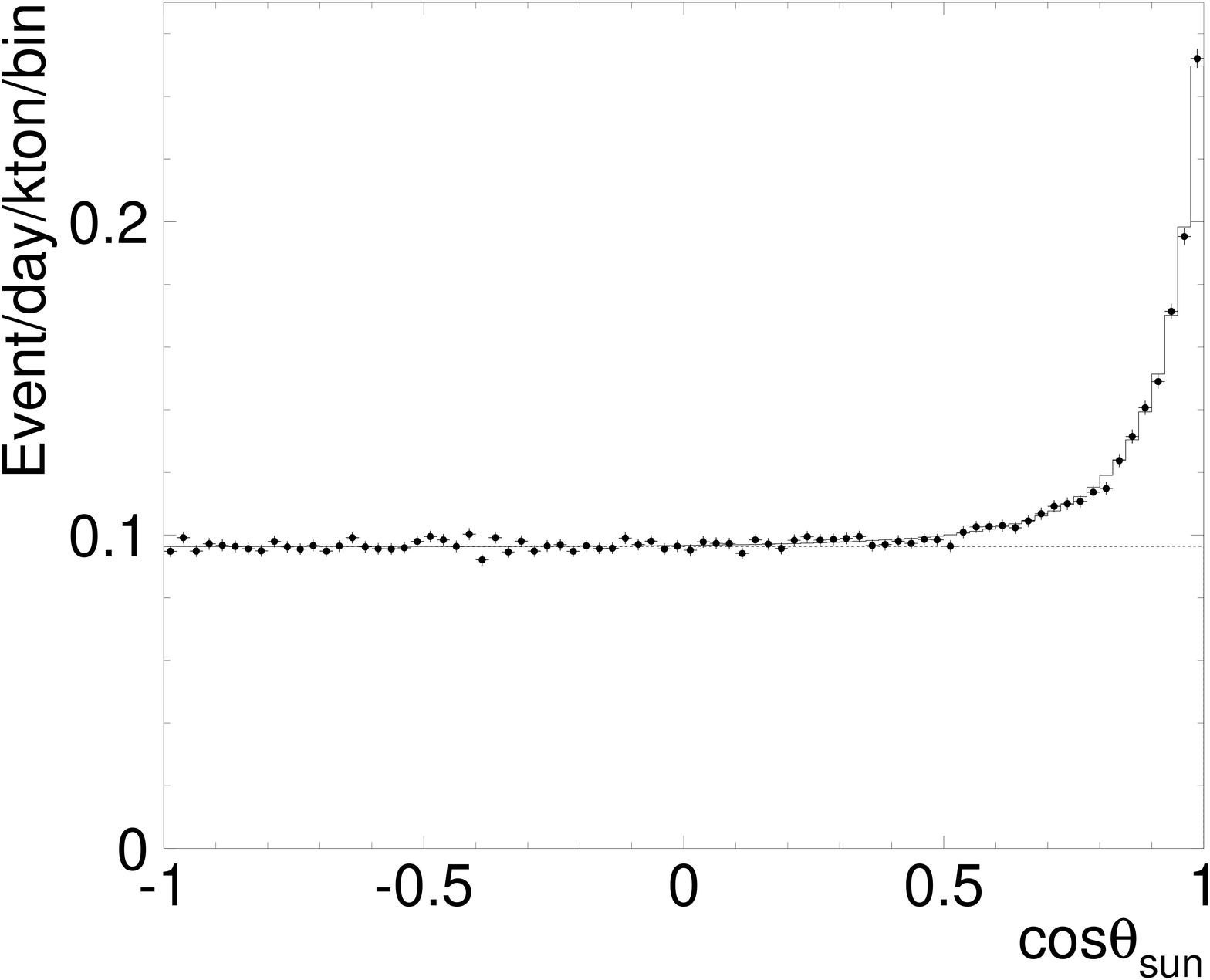}\hfill
\includegraphics[width=.47\linewidth]{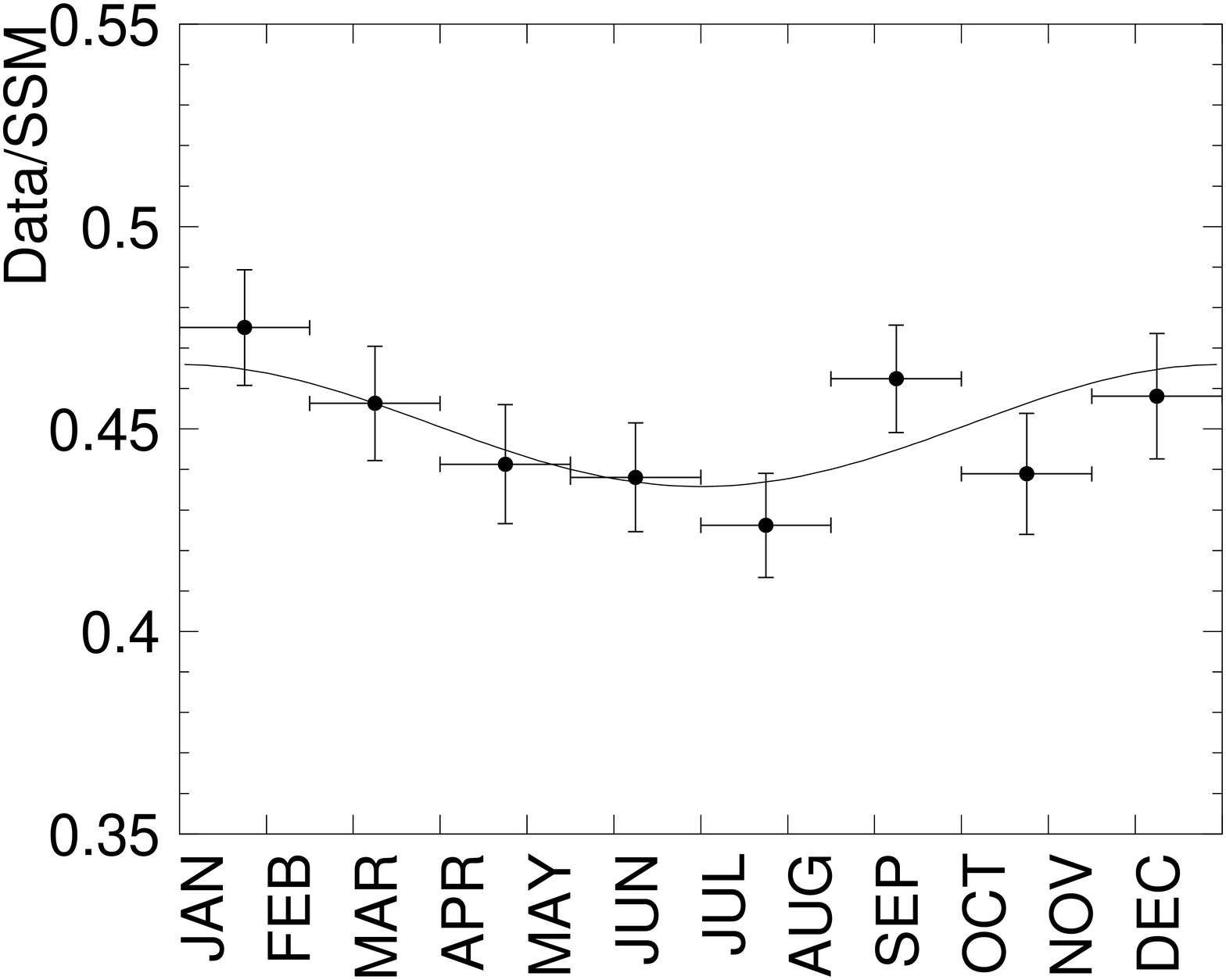}
\caption[]{Left: distribution of solar neutrino events as a function
           of the zenith angle of the sun. Right: seasonal variation of the solar
           neutrino flux in SuperKamiokande (from Ref.~(\cite{Hosaka:2005um})).}
\label{fig:mile1}
\end{figure}

{\bf 1998}: The experiment SuperKamiokande \cite{Fukuda:1998ua} measured the solar neutrino
 deficit with unprecedented precision, using the elastic reaction (ES):
 \begin{eqnarray}
(\text{ES})&
 \quad\nu_e \; + \; e^- \;\rightarrow \; \nu_e \; + e^- 
  & \quad E_\text{thres} > 5\UMeV. 
\end{eqnarray}
 The measurement of
 the direction of the events demonstrated that the neutrinos measured
 definitely come from the sun: the left plot of \Fref{fig:mile1} shows
 the distribution of the events as a function of the zenith angle of
 the sun.  A seasonal variation of the flux is expected since the
 distance between the Earth and the sun varies seasonally. The right
 plot of \Fref{fig:mile1} shows that the measured variation is in
 perfect agreement with that expectation.
%If the deficit of $\nu_e$ in the Sun is interpreted in terms of
%neutrino oscillations, two very important observables to discriminate
%between different solutions are the spectral distribution of the
%events shown in the left plot of \Fref{fig:mile2}, which shows a
%rather flat spectrum, and the day/night asymmetry. The latter is
%important because neutrinos arriving from the Sun at night have to
%cross the Earth and some of the possible solutions are such that
%matter effects in neutrino propagation in the Earth are relevant.
%The analysis of solar data in year 2000 in terms of neutrino mixing
%of the $\nu_e$ into some other type indicated a number of possible solutions
%as shown in \Fref{fig:sol2000}.

\begin{figure}
\centering
\includegraphics[width=.6\linewidth]{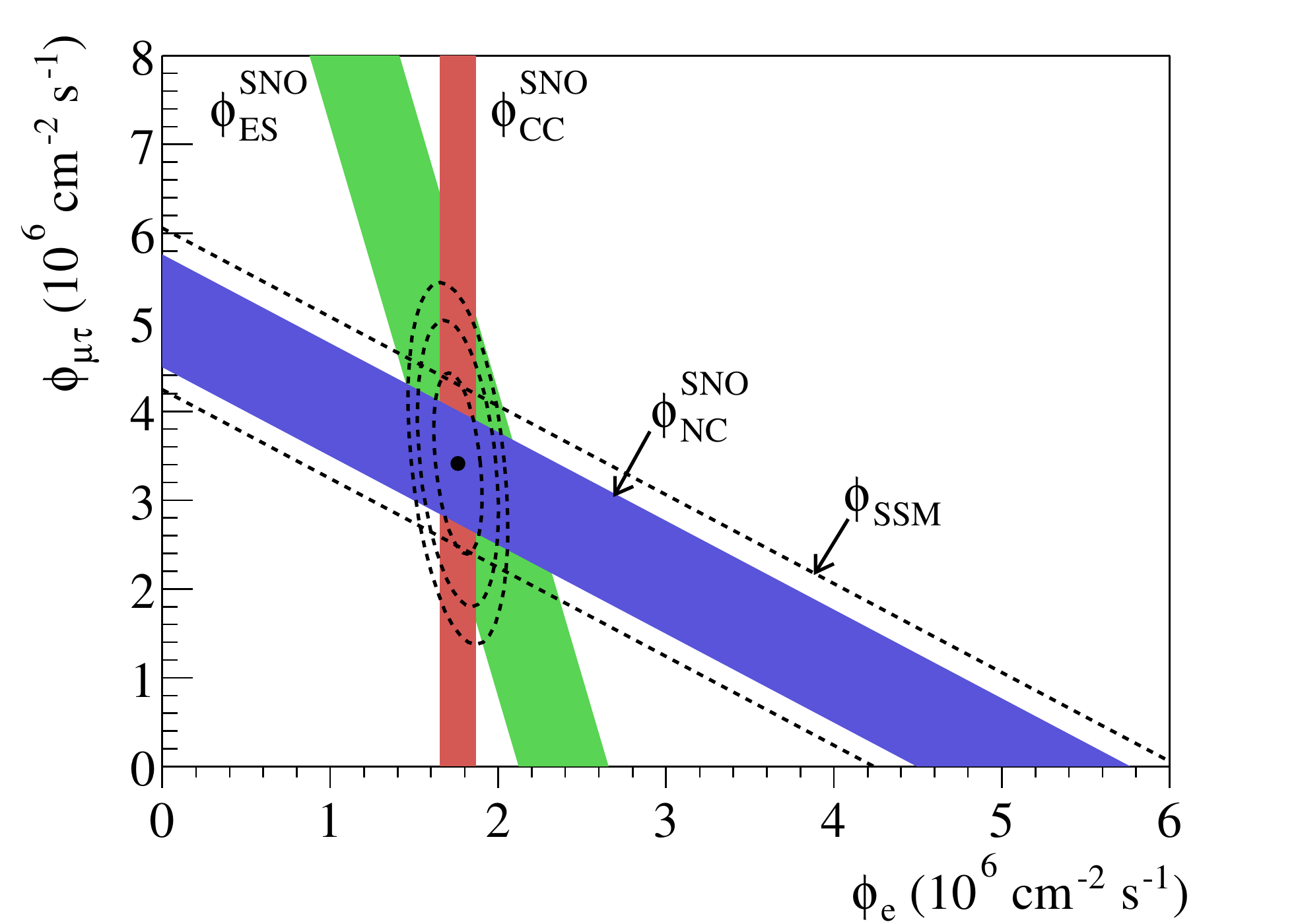}
\caption[]{Flux of $\nu_\mu$ and $\nu_\tau$ versus the flux of $\nu_e$
           in the solar neutrino flux as measured from the three
           reactions observable in the SNO experiment. The dashed band
           shows the prediction of the SSM, which agrees perfectly
           with the flux measured with the NC reaction (from
           Ref.\cite{Ahmad:2002jz}).}
\label{fig:compa}
\end{figure}

{\bf  2001}: The SNO experiment \cite{sno,Ahmad:2002jz}  measured the flux
of solar neutrinos using also the two reactions:   
\begin{eqnarray}
(\text{CC})&
 \quad\nu_{e}  + d \rightarrow p + p +  e^{-}  \;\;\;\;\;
  & \quad E_\text{thres} > 5\UMeV \\
(\text{NC})&
 \quad\nu_{x}  + d \rightarrow p + n +  \nu_x \;\;\;x=e,\mu,\tau 
  & \quad E_\text{thres} > 2.2 \UMeV  
\end{eqnarray}
Since the CC reaction is only sensitive to electron neutrinos, while
the NC one is sensitive to all the types that couple to the $Z^0$
boson, the comparison of the fluxes measured with both reactions can
establish if there are $\nu_\mu$ and $\nu_\tau$ in the solar flux
independently of the normalization given by the SSM. 
%The neutrino
%fluxes measured by the three reactions by SNO are:
%\begin{equation}
%\phi^{\text{CC}} = 1.67(9)   \times 10^6\Ucm^{-2}\UsZ^{-1}, \;\;\;
%\phi^{\text{NC}} = 5.54(48)  \times 10^6\Ucm^{-2}\UsZ^{-1}, \;\;\; 
%\phi^{\text{ES}} = 1.77(26)  \times 10^6\Ucm^{-2}\UsZ^{-1} .
%\end{equation}
The result is shown on the Nobel-prize-winning plot  \Fref{fig:compa}.
These measurements demonstrate that the sun shines $(\nu_\mu,
\nu_\tau)$ about twice more than it shines $\nu_e$, which
constitutes the first direct demonstration of flavour transitions in
the solar flux! Furthermore the NC flux that measures all active
species in the solar flux, is compatible with the total $\nu_e$ flux
expected according to the SSM.

%The post-SNO global fits of all solar data shown in
%\Fref{fig:sno} (left) in terms of neutrino oscillations are quite
%different from those in \Fref{fig:sol2000}. Of all the possible
%solutions, only the one at the largest mixing angle and mass square
%difference survives, the famous LMA solution.

\begin{figure}
\centering
\includegraphics[width=.4\linewidth]{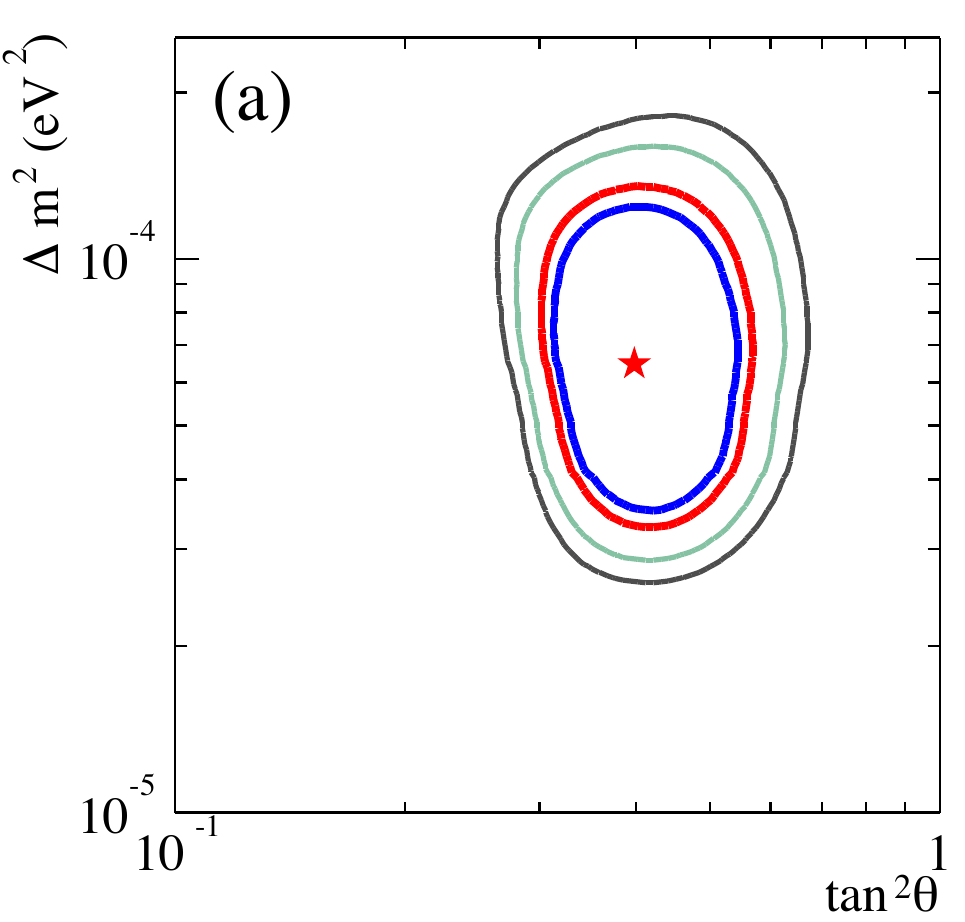}
\caption[]{Analysis of all solar data at SNO  in terms of neutrino
           oscillations (from Ref.~\cite{sno}).}
\label{fig:sno}
\end{figure}
All solar neutrino data can be interpreted in terms of neutrino masses and mixings. The analysis in terms of 
two neutrino families is shown in the left plot of Fig.~\ref{fig:sno}. The solar $\nu_e$ deficit can be explained 
for a $\Delta m^2_{\rm solar} \simeq 7$--$8 \times 10^{-5} \UeVZ$ and a relatively large mixing angle.
The fortunate circumstance that  
\begin{eqnarray}
\Delta m^2_{\rm solar} \sim \langle E_\nu(1\UMeV)\rangle/L(100\Ukm) 
\end{eqnarray}
implies that one could look for this oscillation measuring  reactor neutrinos  at baselines of $\sim100$ km. This was the third milestone.

{\bf 2002}: The solar oscillation is confirmed with reactor neutrinos
in the KamLAND experiment \cite{kamland}. This is 1kton of liquid
scintillator which measures the flux of reactor neutrinos produced in
a cluster of nuclear plants around the Kamioka mine in Japan. The average distance is
$\langle L \rangle = 175\Ukm$.  Neutrinos are detected via inverse
$\beta$-decay which has a threshold energy of about $2.6\UMeV$:
\begin{eqnarray}
\bar{\nu}_e + p \rightarrow e^+ + n \qquad E_\text{th} > 2.6\UMeV \SPp.
\end{eqnarray}

\Fref[b]{fig:kamland2008} shows the KamLAND results
\cite{Abe:2008aa} on the antineutrino spectrum,   as well as the survival probability 
as a function of the ratio $E_\nu/L$.
\begin{figure}
\centering
\includegraphics[width=.45
\linewidth]{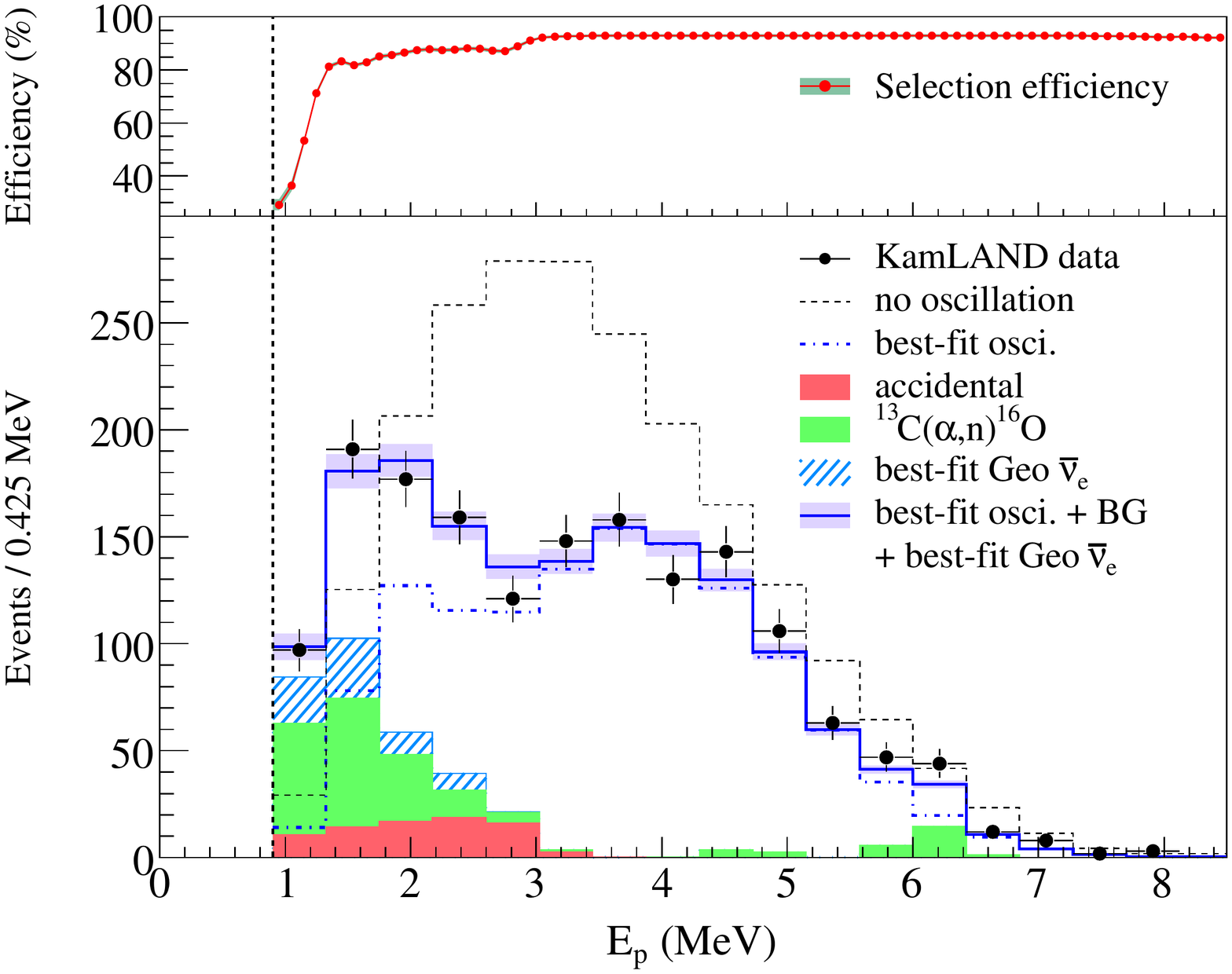}
\includegraphics[width=.45\linewidth]{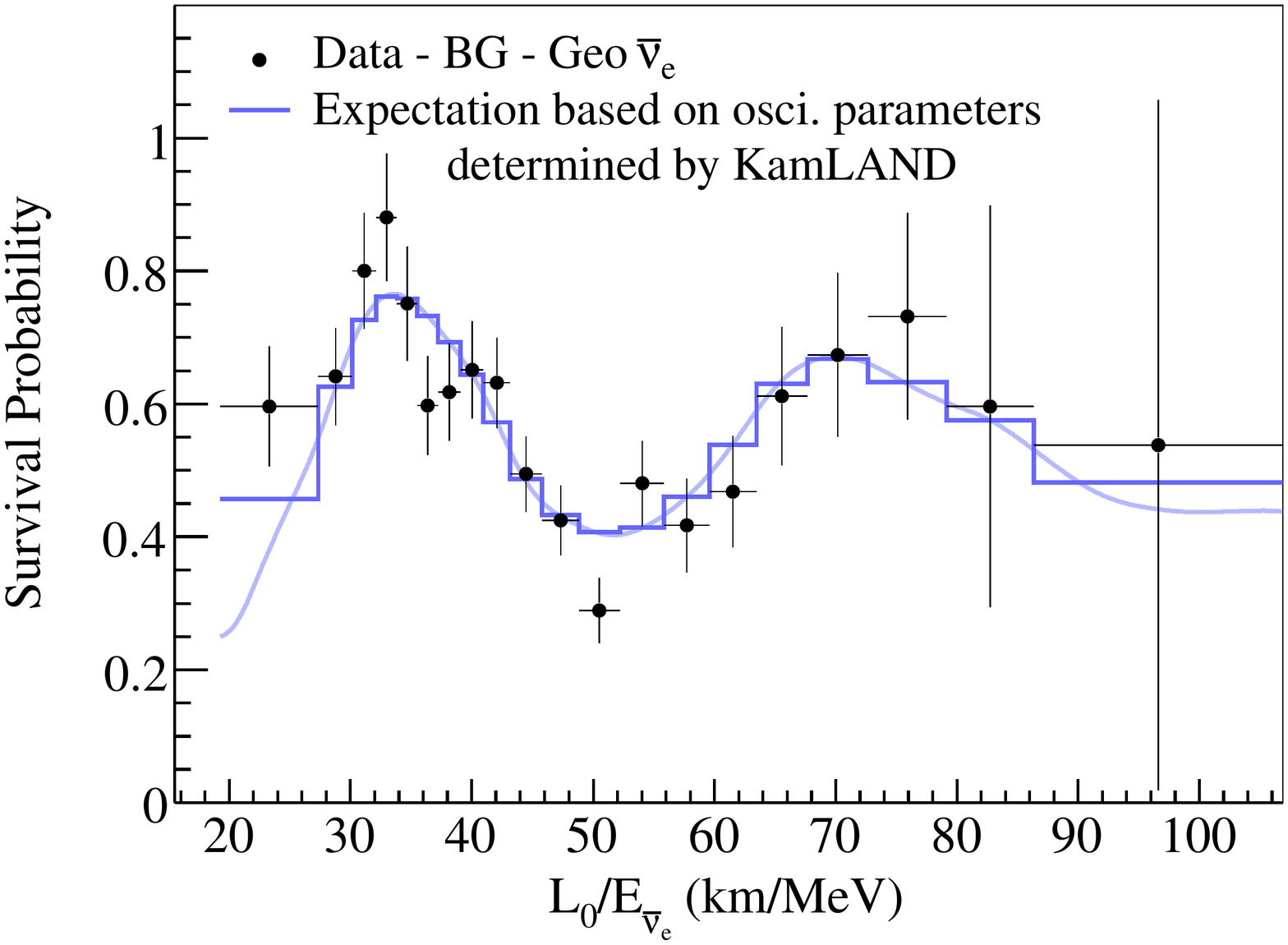}
\caption[]{Spectral distribution of the $\bar{\nu}_e$ events in
           KamLAND (left) and $E_\nu/L$ dependence (right). The data
           are compared to the expectation in the absence of
           oscillations and to the best fit oscillation
           hypothesis (from Ref.~\cite{Abe:2008aa}).}
\label{fig:kamland2008}
\end{figure}
 The low-energy contribution of geoneutrinos is clearly visible. This measuremet  could have important
implications in geophysics.  

\begin{figure}
\centering
\includegraphics[width=.5\linewidth]{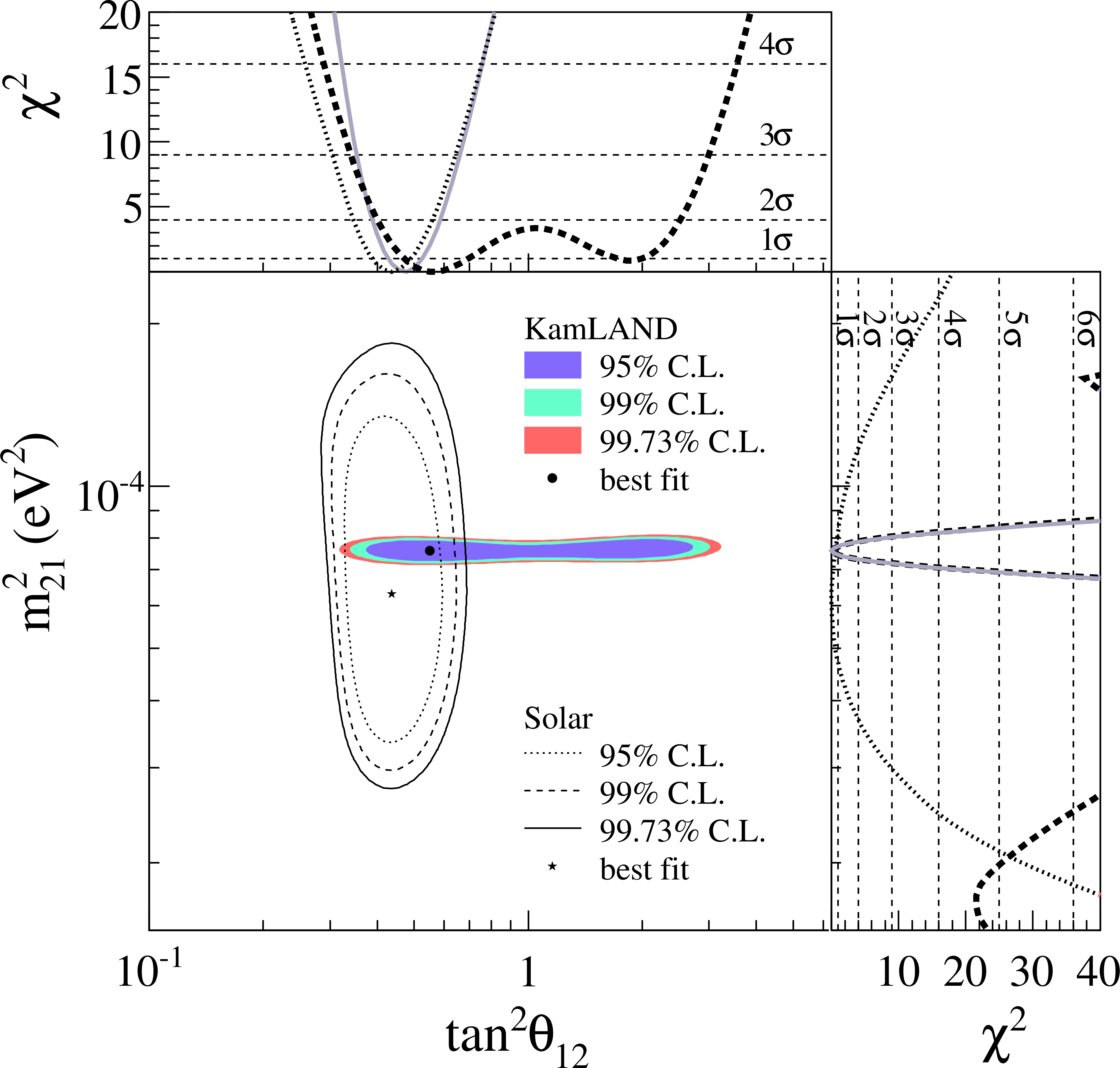}
\caption[]{Analysis of all solar and KamLAND data in terms of
           oscillations (from Ref.~\cite{Abe:2008aa}).}
\label{fig:kamosc}
\end{figure}

Concerning the sensitivity to the
oscillation parameters, \Fref{fig:kamosc} shows the present
determination of the solar oscillation parameters from KamLAND and
other solar experiments. The precision in the determination of $\Delta
m^2_\text{solar}$ is spectacular and shows that solar neutrino experiments are
entering the era of precision physics.

\begin{figure}
\centering
\includegraphics[width=.6\linewidth]{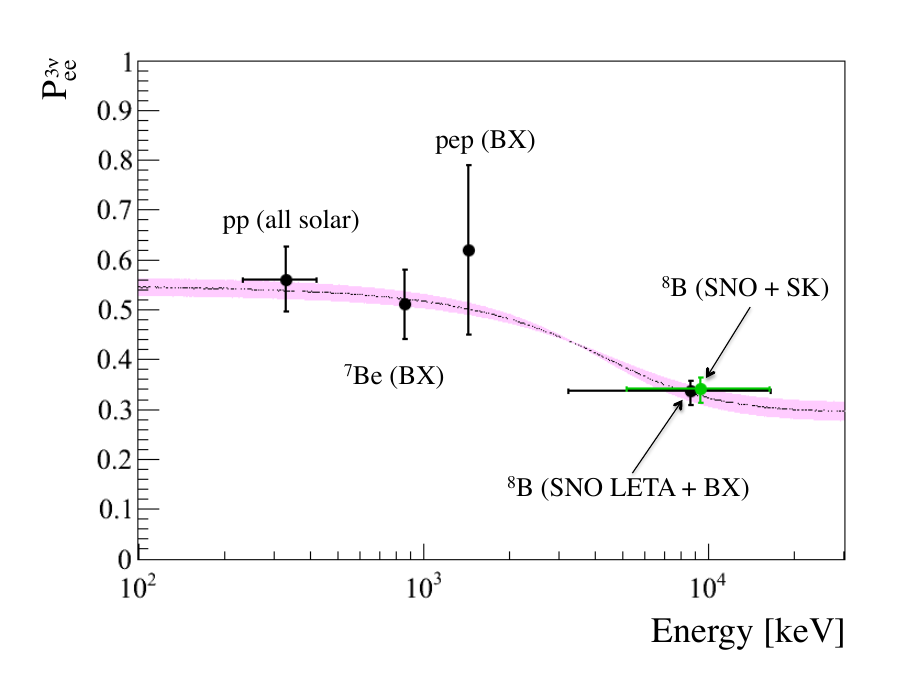}\hfill
\caption[]{Comparison of solar neutrino fluxes measured by the
           different solar neutrino experiments ( from Ref.~\cite{Bellini:2013lnn}).}
\label{fig:borexino}
\end{figure}

The last addition to this success story is the Borexino \cite{Bellini:2013lnn} experiment. This is the lowest-threshold real-time solar
neutrino experiment and the only one capable of measuring the flux of
the monocromatic $^7$Be neutrinos and pep neutrinos. Their recent results are shown in Fig.~\ref{fig:borexino}. 
The result is in agreement with the oscillation interpretation of other
solar and reactor experiments and it adds further information to
disfavour alternative exotic interpretations of the data.
In summary, solar neutrinos experiments have made fundamental
discoveries in particle physics and are now becoming useful for other
applications, such as a precise understanding of the sun and the
Earth.

\subsection{Atmospheric neutrinos }
 
\begin{figure}
\begin{center}
\includegraphics[width=0.8\linewidth]{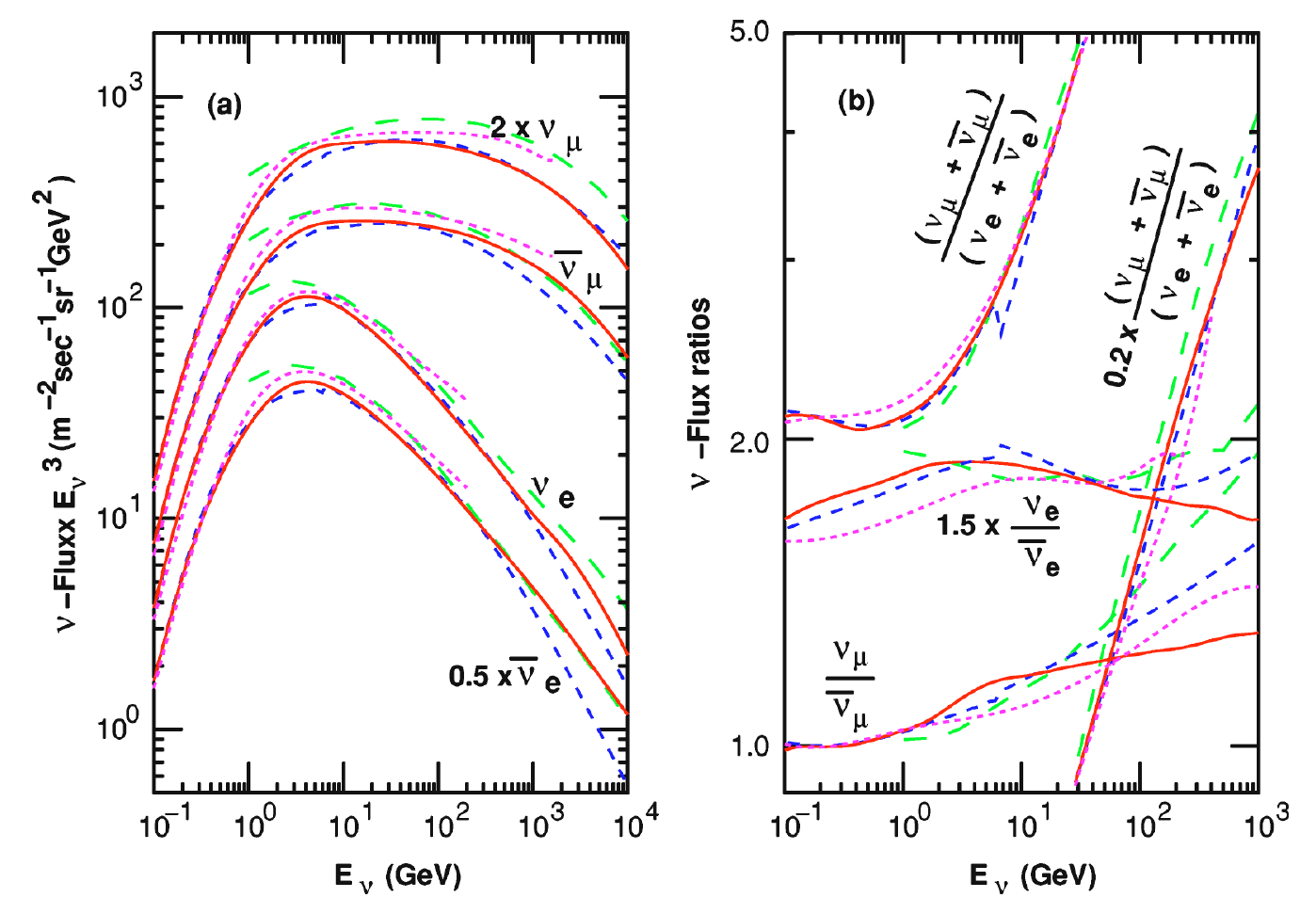}
\caption[]{Comparison of the predictions of different Monte Carlo simulations of the
           atmospheric neutrino fluxes averaged over all directions
           (left) and of the flux ratios $(\nu_\mu +
           \bar\nu_\mu)/(\nu_e + \bar\nu_e)$, $\nu_\mu / \bar\nu_\mu$, 
           and $\nu_e/\bar\nu_e$ (right). The solid line corresponds
           to a recent full 3D simulation.  Taken from the last
           reference in Ref.~\cite{Honda:2004yz}.}
\label{fig:atmflux}
\end{center}
\end{figure}

Neutrinos are also produced in the atmosphere when primary cosmic rays
impinge on it producing $K,\pi$ that subsequently decay. The fluxes of
such neutrinos can be predicted within a 10--20$\%$ accuracy to be
those in the left plot of \Fref{fig:atmflux}.

\begin{figure}
\centering
\includegraphics[width=.5\linewidth]{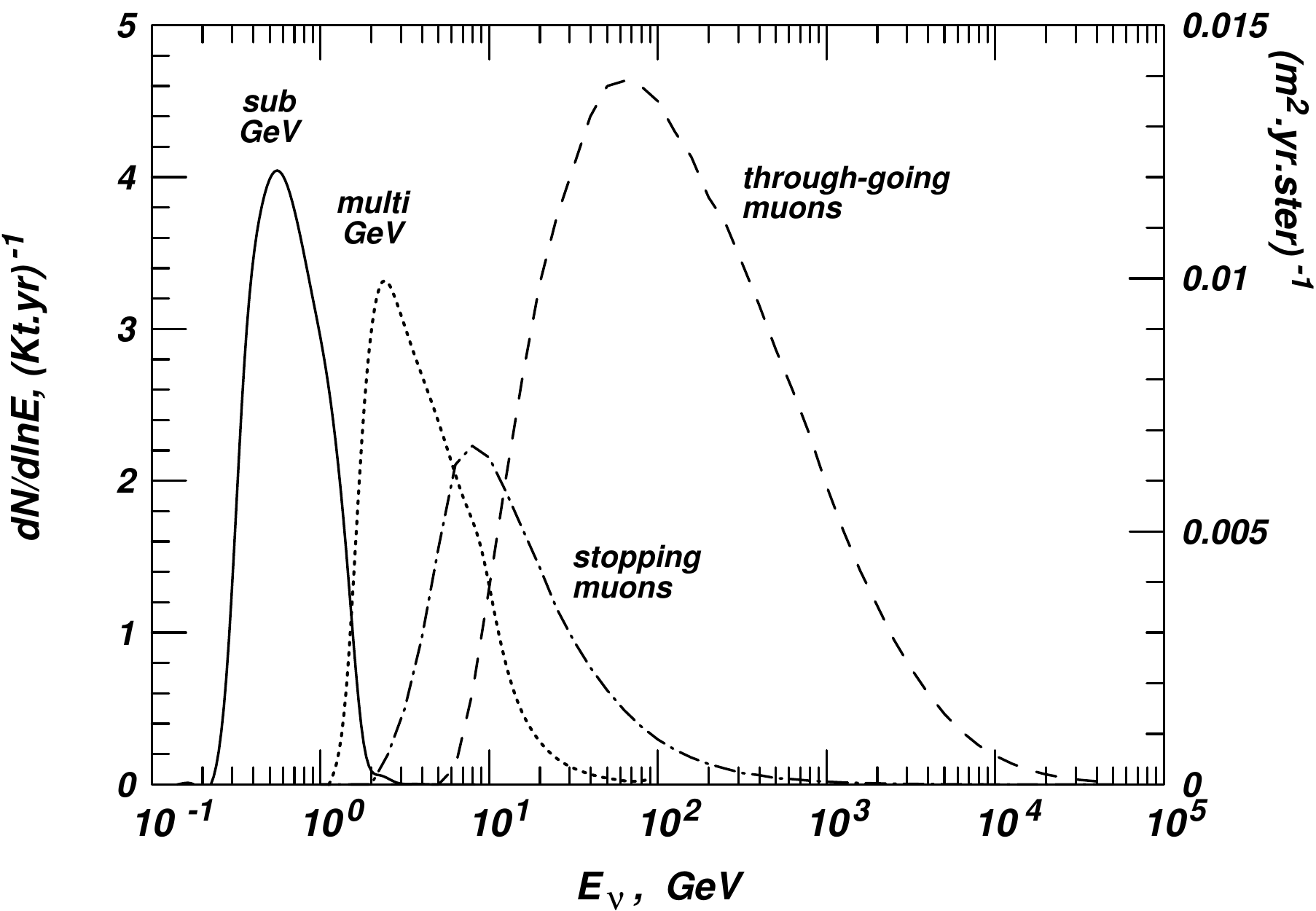}\quad
\includegraphics[width=.35\linewidth]{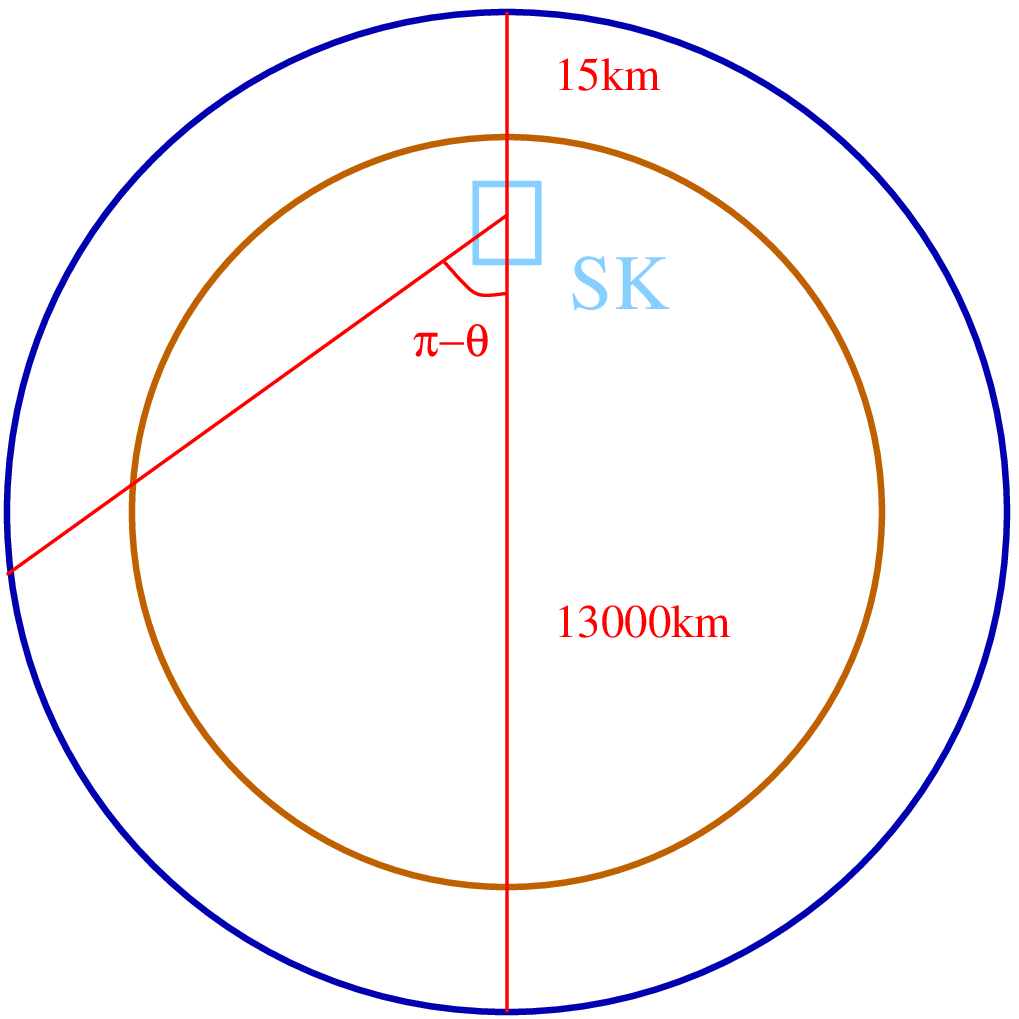}
\caption[]{Left: Parent neutrino energies of the different samples
           considered in Superkamiokande: sub-GeV, multi-GeV, stopping
           and through-going muons. Right: Distances travelled by
           atmospheric neutrinos as a function of the zenith
           angle.}
\label{fig:sksample}
\end{figure}

Clearly, atmospheric neutrinos are an ideal place to look for neutrino
oscillation since the $E_\nu/L$ span several orders of magnitude, with
neutrino energies ranging from a few hundred MeV to $10^3\UGeV$ and
distances between production and detection varying from
$10$--$10^4\Ukm$, as shown in \Fref{fig:sksample} (right).

Many of the uncertainties in the predicted fluxes cancel when the
ratio of muon to electron events is considered. The first indication
of a problem was found when a deficit was observed precisely in this
ratio by several experiments: Kamiokande, %\cite{kamio-atmos}, 
IMB,
%\cite{imb}, 
Soudan2 %\cite{soudan2} 
and Macro.
 %\cite{macro}.

\begin{figure}
\begin{center}
\includegraphics[width=.85\linewidth]{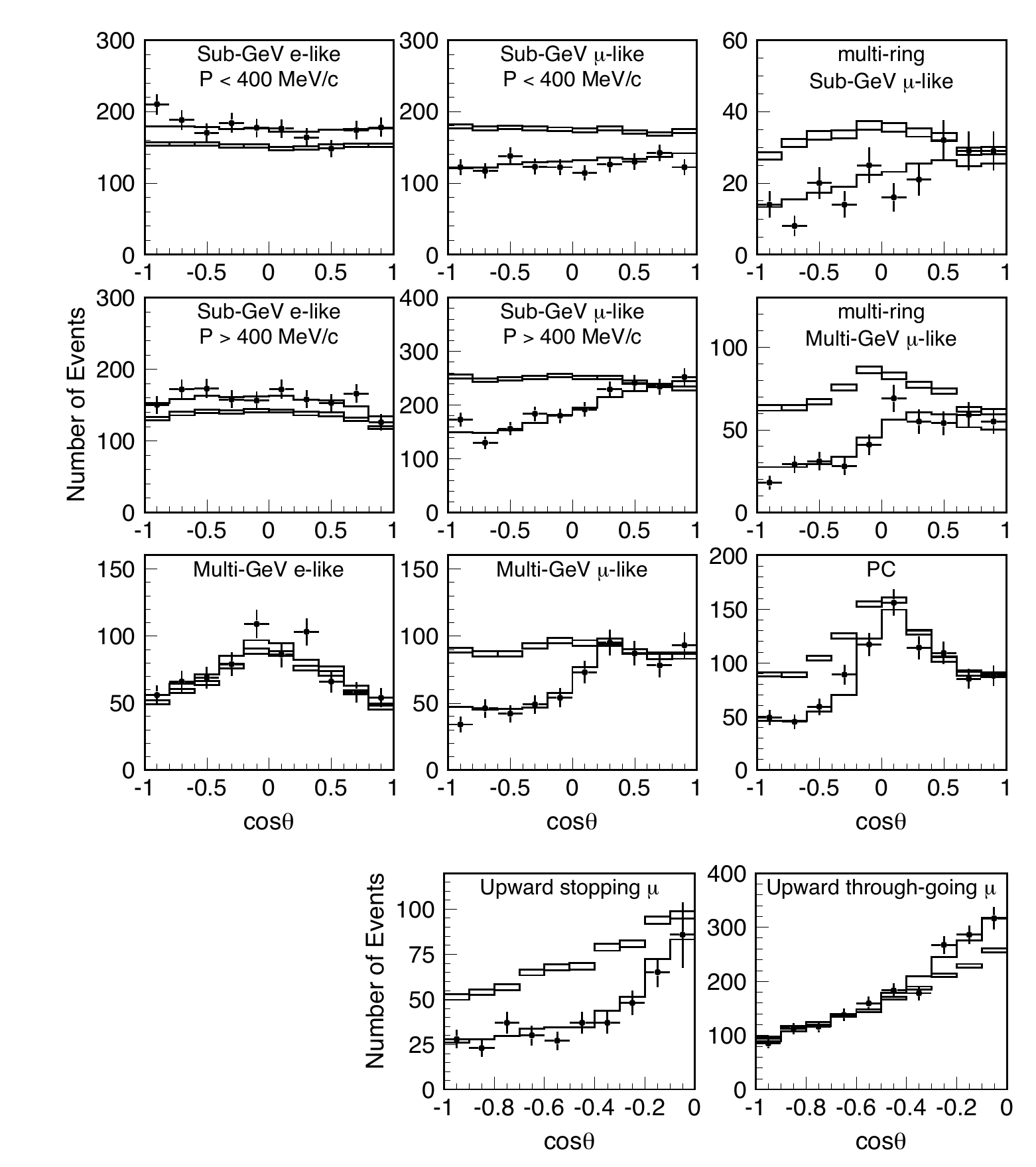}
\caption[]{Zenith angle distribution for fully-contained single-ring
           $e$-like and $\mu$-like events, multi-ring $\mu$-like
           events, partially contained events, and upward-going
           muons. The points show the data and the boxes show
           the Monte Carlo events without neutrino oscillations.  The
           solid lines show the best-fit expectations for $\nu_\mu
           \leftrightarrow \nu_\tau$ oscillations (from Ref.~
           \cite{sk-zen}).}
\label{fig:skzen}
 \label{fig:zenith_angle}
\end{center}
\end{figure}

In 1998, SuperKamiokande clarified  the origin of
this anomaly \cite{sk-atmos}.  This experiment can distinguish muon
and electron events, measure the direction of the outgoing lepton (the
zenith angle with respect to the Earth's axis) which is correlated to
that of the neutrino ( the higher the energy the higher the
correlation), in such a way that they could measure the variation of
the flux as a function of the distance travelled by the
neutrinos. Furthermore, they considered different samples of events:
sub-GeV (lepton with energy below $1\UGeV$) ), multi-GeV (lepton with
energy above $1\UGeV$), together with stopping and through-going muons
that are produced on the rock surrounding Superkamiokande. The
different samples correspond to different parent neutrino energies as
can be seen in \Fref{fig:sksample} (left).
The number of events for the different samples as a function of the zenith 
angle of the lepton are shown in the Nobel-prize-winning plot \Fref{fig:skzen}. 

\begin{figure}
\centering
\includegraphics[width=.55\linewidth]{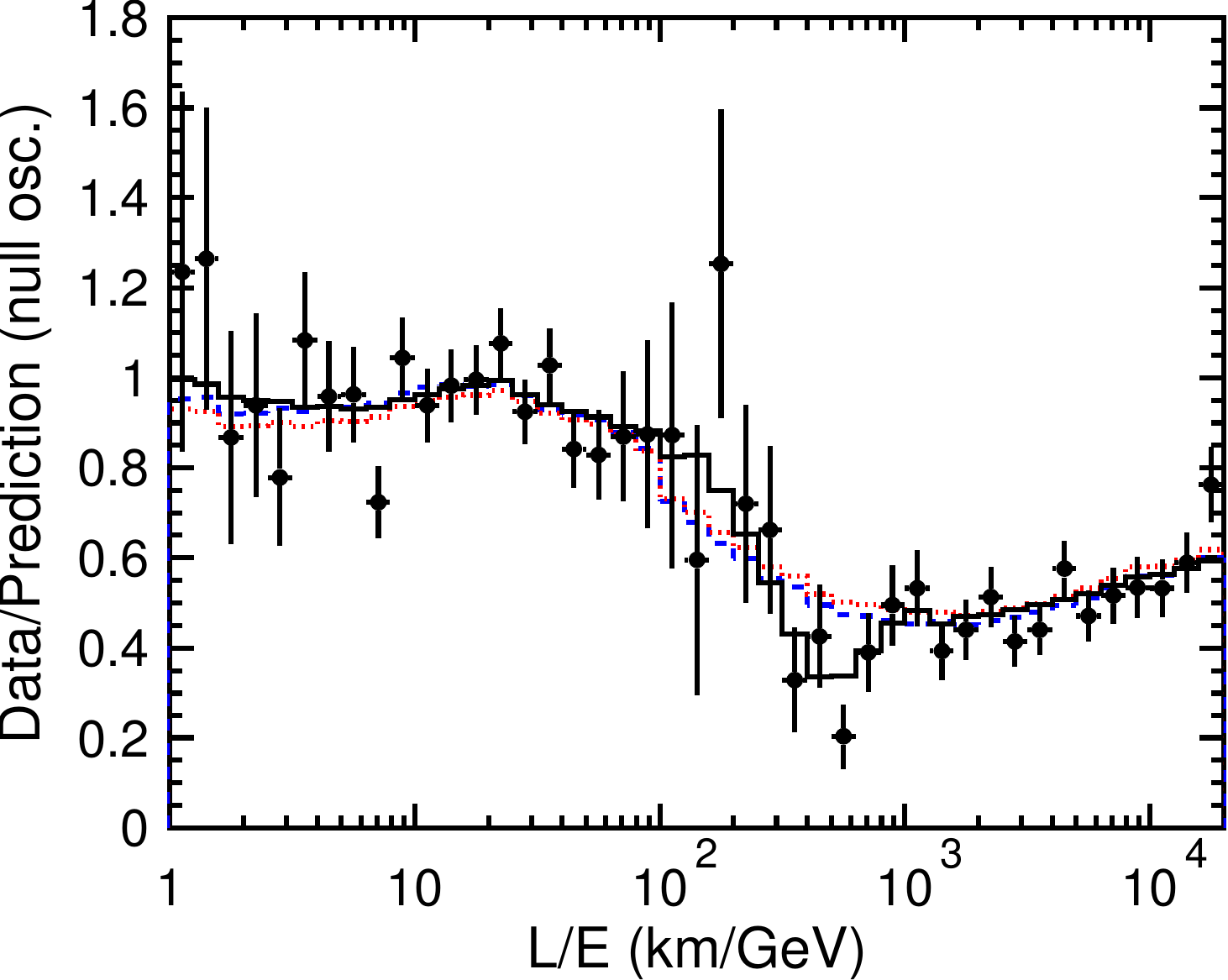}
\caption[]{Ratio of the data to the non-oscillated Monte Carlo events
           (points) with the best-fit expectation for 2-flavour
           $\nu_\mu \leftrightarrow \nu_\tau$ oscillations (solid
           line) as a function of $E_\nu/L$ (from Ref.~\cite{sk_atmos_el}).}
\label{fig:sk_el}
\end{figure}

While the electron events observed are in rough agreement with predictions,
a large deficit of muon events was found with a strong dependence on
the zenith angle: the deficit was almost 50$\%$ for those events
corresponding to neutrinos coming from below $\cos \theta = -1$, while
there is no deficit for those coming from above.  The perfect fit to the oscillation
hypothesis is rather non-trivial given the sensitivity of this
measurement to the $E_\nu$ (different samples) and $L$ (zenith angle)
dependence. The significance of the $E_\nu/L$ dependence has also been
 measured by the SuperKamiokande Collaboration
\cite{sk_atmos_el}, as shown in \Fref{fig:sk_el}. The best fit value of the oscillation parameters
indicate $\Delta m^2 \simeq 3\times 10^{-3}$ eV$^2$ and maximal mixing. 

Appropriate neutrino beams to search for the atmospheric oscillation
can easily be produced at accelerators if the detector is located at a
long baseline of a few hundred kilometres, and also with reactor neutrinos in a baseline
of  ${\mathcal O}(1)$km, since
\begin{equation}
|\Delta m^2_\text{atmos}| \sim \frac{E_\nu(1-10\UGeV)}{L(10^2-10^3\Ukm)} \sim \frac{E_\nu(1-10\UMeV)}{L(0.1-1\Ukm)}.
\end{equation}
A \emph{conventional} accelerator neutrino beam, as the one used in the FSS experiment, is produced from protons hitting a
target and producing $\pi$ and $K$:
\begin{eqnarray}
p \;\;\rightarrow\;\; \text{Target} \rightarrow \pi^+, K^+ \rightarrow 
 & \nu_\mu  (\% \nu_e, \bar{\nu}_{\mu}, \bar{\nu}_e) \\
   &  \nu_\mu   \rightarrow \nu_x .
\end{eqnarray}
Those of a selected charge are focused and are left to decay in a long
decay tunnel producing a neutrino beam of mostly muon neutrinos (or
antineutrinos) with a contamination of electron neutrinos of a few per
cent. The atmospheric oscillation can be established by studying, as a
function of the energy, either the disappearance of muon neutrinos, the appearance of electron neutrinos or, 
if the energy of the beam is large enough, the appearance of $\tau$
neutrinos.

\begin{figure}
\centering 
\includegraphics[width=.66\linewidth]{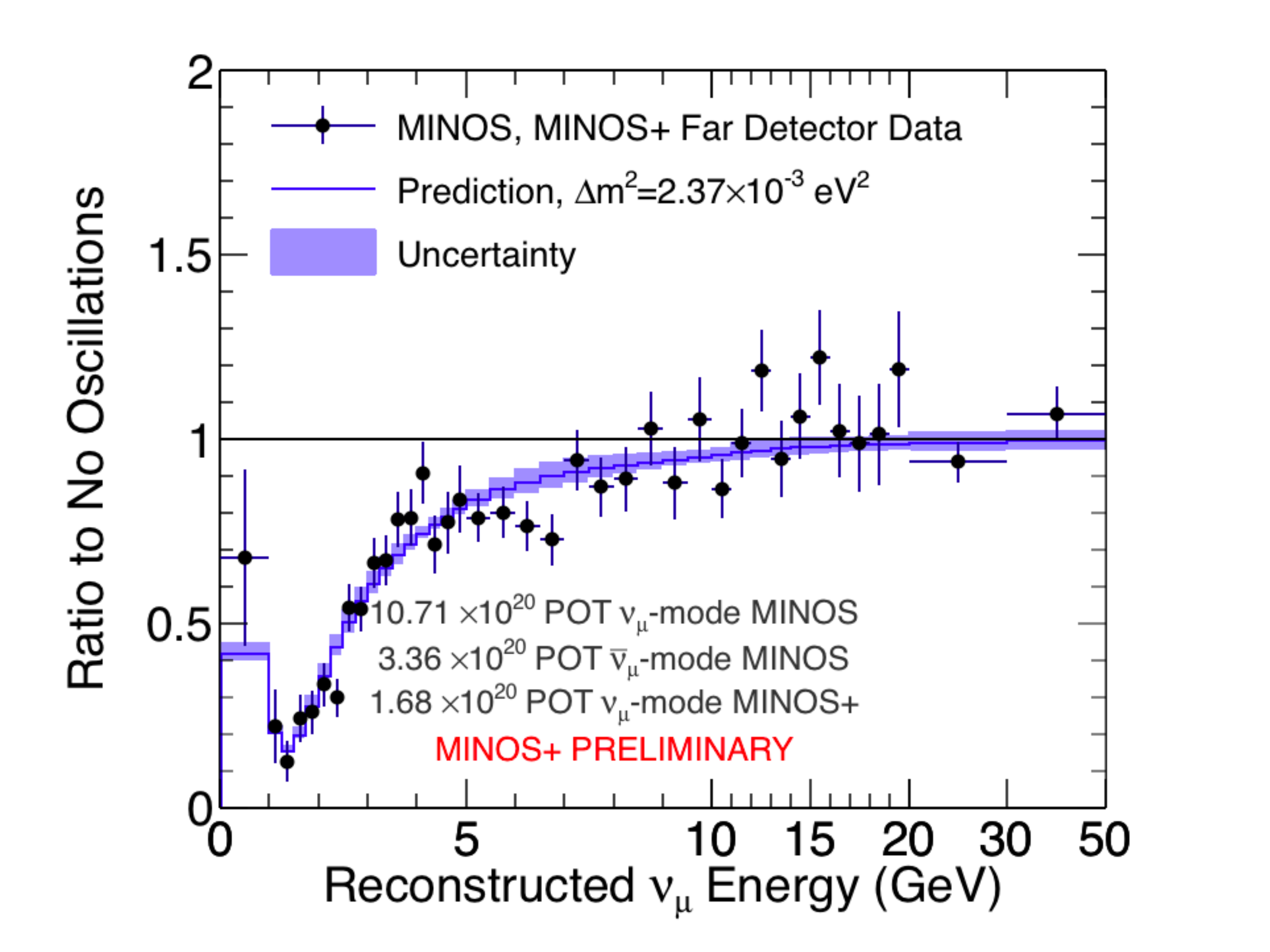}
\caption[]{ Ratio of measured to expected (in absence of
           oscillations) neutrino events in MINOS as a functions of
           neutrino energy compared to the best fit oscillation solution (from Ref.~\cite{minos}). }
\label{fig:minos}
\end{figure}

Three conventional beams confirmed the atmospheric oscillation from the measurement of the disappearance of $\nu_\mu$ neutrinos: K2K (L = 235\Ukm),
MINOS (L = 730\Ukm) and from the appearance of $\nu_\tau$ OPERA (L = 730\Ukm). 
 \Fref{fig:minos} shows the measurement of the $\nu_\mu$ survival probability  as a function of the reconstructed
 neutrino energy in the MINOS experiment.

Three reactor neutrino experiments, Daya Bay \cite{An:2012eh}, RENO \cite{Ahn:2012nd} and Double Chooz \cite{Abe:2011fz}, have discovered that the electron neutrino flavour also oscillates with  the atmospheric wavelength: 
 electron antineutrinos from reactors disappear at distances of ${\mathcal O}(1\Ukm)$, but with a small amplitude. 
 
Finally the T2K experiment has measured for the first time the appearance of $\nu_e$ in an accelerator $\nu_\mu$ beam \cite{t2k} in the atmospheric range.

The agreement of all these measurements with the original atmospheric oscillation signal was excellent. 

\section{The three-neutrino mixing scenario}

As we have seen the evidence summarized in the previous section points to 
 two distinct neutrino mass square differences related to the solar and atmospheric oscillation frecuencies:
\begin{equation}
\underbrace{ |\Delta m^2_\text{solar}| \;\;\;}_{\sim 8 \cdot 10^{-5}\UeV^2}     \ll\;\;\;\;
\underbrace{|\Delta m^2_\text{atmos}|}_{\sim 2.5\cdot 10^{-3}\UeV^2} \;\;\;\; %\ll\;\;\;
%\underbrace{|\Delta m^2_\text{LSND}|}_{> 0.1\UeV^2} 
\end{equation}
The mixing of the three standard neutrinos $\nu_e, \nu_\mu,
\nu_\tau$ can accommodate  both. 
  The two independent neutrino mass square
differences are conventionally assigned to the solar and atmospheric ones in the following way:
\begin{equation}
\Delta m^2_{13} = m_3^2 - m_1^2 = \Delta m^2_\text{atmos}, \qquad  
\Delta m^2_{12} = m_2^2 - m_1^2 = \Delta m^2_\text{solar}\SPp.
\end{equation}
The PMNS mixing matrix  depends on three angles and one or more CP phases (see eq.~(\ref{mns}) for the standard parametrization). 
Only one CP phase, the so-called Dirac phase $\delta$, appears in neutrino oscillation probabilities.
 
With this convention, the mixing angles $\theta_{23}$ and
$\theta_{12}$ in the parametrization of \Eref{mns} correspond
approximately to the ones measured in atmospheric and solar
oscillations, respectively. This is because solar and atmospheric
anomalies approximately decouple as independent 2-by-2 mixing
phenomena thanks to the hierarchy between the two mass splittings,
$|\Delta m^2_\text{atmos}| \gg |\Delta m^2_\text{solar}|$ , on the one
hand, and the fact that the angle $\theta_{13}$, which measures the
electron component of the third mass eigenstate element $\sin
\theta_{13} = \left(U_\text{PMNS}\right)_{e 3}$, is small.

To see this, let us first consider the situation in which $E_\nu/L
\sim| \Delta m^2_{\rm atmos}|$. We can thus neglect the solar mass square
difference in front of the atmospheric one and $E_\nu/L$. The
oscillation probabilities obtained in this limit are given by
\begin{eqnarray}
P(\nu_e\to\nu_\mu)&\simeq&s_{23}^2\,\sin^2 2\theta_{13}\,
  \sin^2\!\left(\frac{\Delta m_{13}^2 L}{4 E_\nu}\right), \\
P(\nu_e\to\nu_\tau)&\simeq&c_{23}^2\,\sin^2 2\theta_{13}\,
  \sin^2\!\left(\frac{\Delta m_{13}^2 L}{4 E_\nu}\right),\,\\
P(\nu_\mu\to\nu_\tau)&\simeq&c_{13}^4\,\sin^2 2\theta_{23}\,
  \sin^2\!\left(\frac{\Delta m_{13}^2 L}{4 E_\nu}\right). 
\label{approx_atmos}
\end{eqnarray}
The results for antineutrinos are the same. 
All flavours oscillate therefore with the atmospheric frecuency, but only two angles enter these formulae: $\theta_{23}$ and
$\theta_{13}$. The latter is the only one that enters the
disappearance probability for $\nu_e$ or $\bar{\nu}_e$ in this regime:
\begin{equation}
P(\nu_e\to\nu_e)= P(\bar{\nu}_e\to\bar{\nu}_e)=1 - P(\nu_e\to\nu_\mu) - P(\nu_e\to\nu_\tau) 
\simeq \sin^2 2\theta_{13}\,\sin^2\left(\frac{\Delta m_{13}^2 L}{4 E_\nu}\right)\SPp. 
\end{equation}
This is precisely the measurement of reactor neutrino experiments like Chooz, Daya Bay, RENO and Double Chooz. 
Therefore the oscillation amplitude of these experiments is a direct measurement of the angle $\theta_{13}$, which has 
been measured to be small.

Note that in the limit $\theta_{13}\rightarrow 0$, the only
probability that survives in  \Eref{approx_atmos} is the $\nu_\mu \rightarrow \nu_\tau$ one,
which has the same form as a 2-family mixing formula \Eref{eq:wk} if
we identify
\begin{equation}
(\Delta m^2_\text{atmos},\theta_\text{atmos}) \rightarrow 
(\Delta m^2_{13}, \theta_{23})\SPp.
\end{equation}   
Therefore the close-to-maximal mixing angle observed in atmospheric neutrinos and the accelerator neutrino experiments like MINOS is identified with $\theta_{23}$. 

Instead if we consider experiments in the solar range, $E_\nu/L \sim \Delta m^2_{solar}$, the atmospheric oscillation
its too rapid and gets averaged out. The survival probability for
electrons in this limit is given by:
\begin{equation}
P(\nu_e\to \nu_e) = P(\bar{\nu}_e\to \bar{\nu}_e) 
  \simeq c_{13}^4 
    \left(1-\sin^2 2\theta_{12}\,\sin^2\!\left(\frac{\Delta m_{12}^2 L}{4 E_\nu}\right)\, 
    \right) 
  + s_{13}^4.  
\end{equation}
Again it depends only on two angles, $\theta_{12}$ and $\theta_{13}$, 
and in the limit in which the latter is zero, the survival probability
measured in solar experiments has the form of two-family mixing if we
identify
\begin{eqnarray}
(\Delta m^2_\text{solar},\theta_\text{solar}) \rightarrow (\Delta m^2_{12}, \theta_{12})\SPp.
\end{eqnarray}
The results that we have shown in the previous section of solar and atmospheric experiments
have been analysed in terms of 2-family mixing. The previous argument
indicates that when fits are done in the context of 3-family mixing
nothing changes too much.

\begin{figure}
\centering
\includegraphics[width=.7\linewidth]{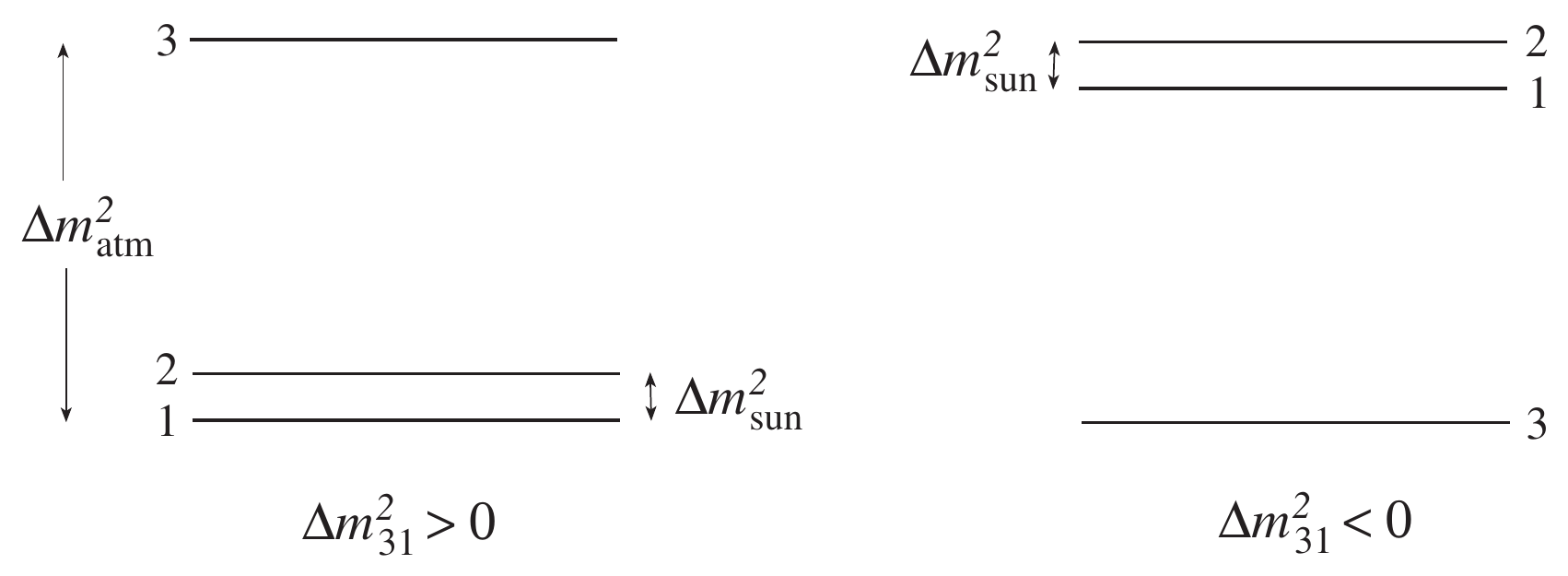}
\caption[]{Possible neutrino spectra consistent with solar and
           atmospheric data.}
\label{fig:spectrum}
\end{figure}

On the other hand, the fact that reactor experiments have already measured the disappearance 
of reactor $\bar{\nu}_e$ in the atmospheric range implies that the effects of $\theta_{13}\simeq 9^\circ$ are
not negligible, and therefore a proper analysis of all the oscillation data requires performing global fits in the 3-family 
scenario. 
\Fref[b]{fig:ggm} shows the $\Delta \chi^2$ as a function of each of the six parameters  from a recent global analysis \cite{Gonzalez-Garcia:2014bfa}. 
\begin{figure}
\centering
\includegraphics[width=.6\linewidth]{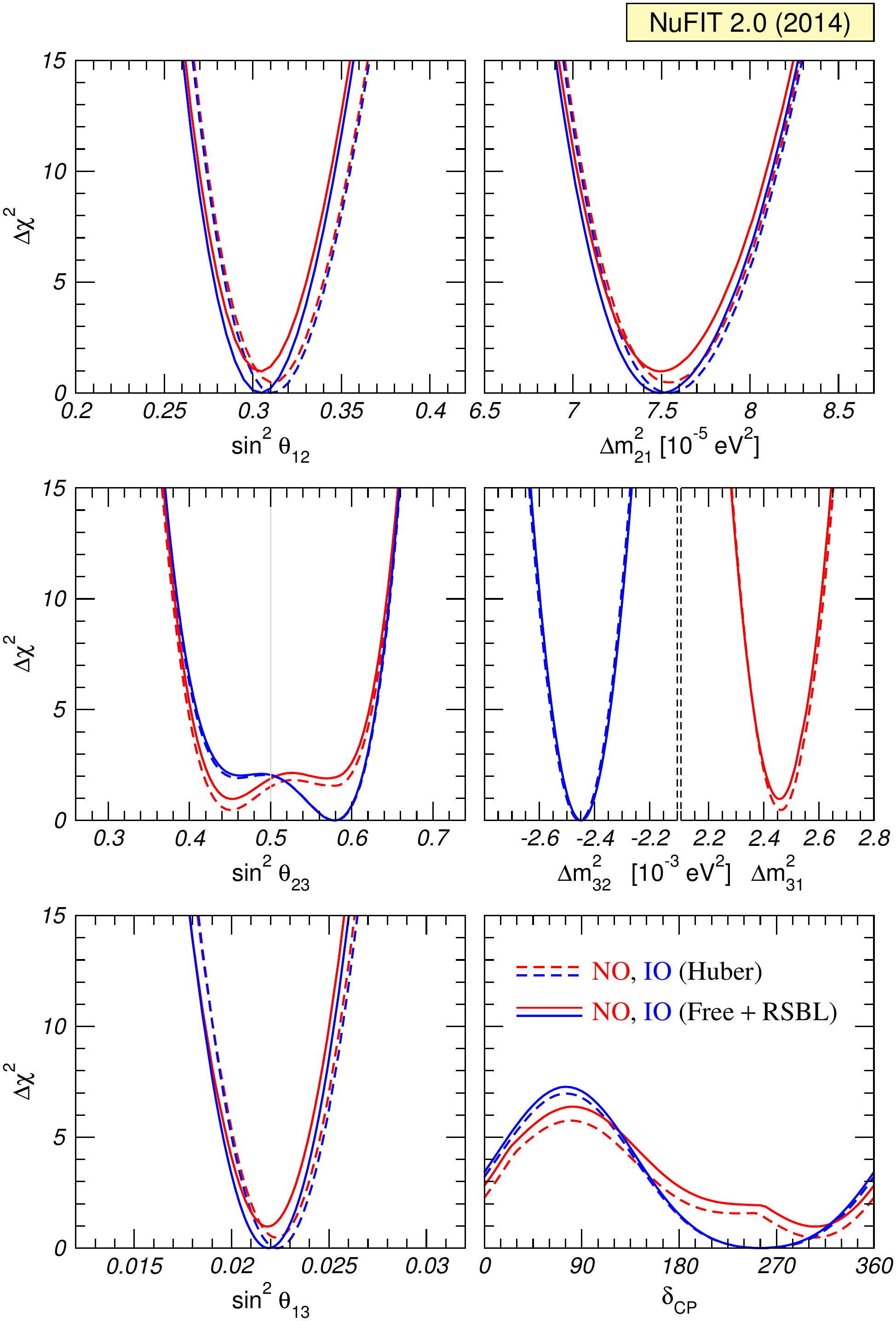}
\caption[]{Fits to the standard $3\nu$-mixing scenario including all
           available neutrino oscillation data (from Ref.~\cite{Gonzalez-Garcia:2014bfa}).}
\label{fig:ggm}
\end{figure}
There are two parameters in which we observe to distinct minima, these corresponds to 
degeneracies that cannot be resolved with present data. The first corresponds to the neutrino mass
ordering or hierarchy:  present data cannot distinguish between the normal (NH or NO) and inverted ordering (IH or IO) represented
in Fig.~\ref{fig:spectrum}. 
Note that we denote by $\Delta m^2_{13}=\Delta m^2_{\rm atmos}$ the atmospheric splitting for NO and $\Delta m^2_{23} = -\Delta m^2_{\rm atmos}$. The second degeneracy corresponds to the octant choice of $\theta_{23}$. Present data
are mostly sensitive to $\sin^2 2\theta_{23}$. If this angle is not maximal, there are  two possible choices that are roughly equivalent
$\theta_{23} \leftrightarrow \pi/4 -\theta_{23}$. Due to this degeneracy, the largest angle is also the one less accurate. 
The $1\sigma$ limits for NO are:
\begin{eqnarray}
\theta_{23}/^\circ = 42^{+ 3}_{-1.6},  \qquad
\theta_{12}/^\circ = 33.5^{+0.78}_{-0.75}, \qquad
\theta_{13}/^\circ = 8.5(2),\nonumber\\
\Delta m^2_{12} = 7.5(2) \times 10^{-5}\UeV^2, \qquad
\Delta m^2_{13} = 2.46(5) \times 10^{-3}\UeV^2. \SPp.
\end{eqnarray}
The CP phase $\delta$ remains completely unconstrained at 3$\sigma$.  As we will see, the dependence on the phase requires sensitivity to both frecuencies simultaneously. There is however at 2$\sigma$  some hint of a preference for $\delta > 180^\circ$
.  For more details see \cite{Gonzalez-Garcia:2014bfa}.

\begin{figure}
\centering
\includegraphics[width=.5\linewidth]{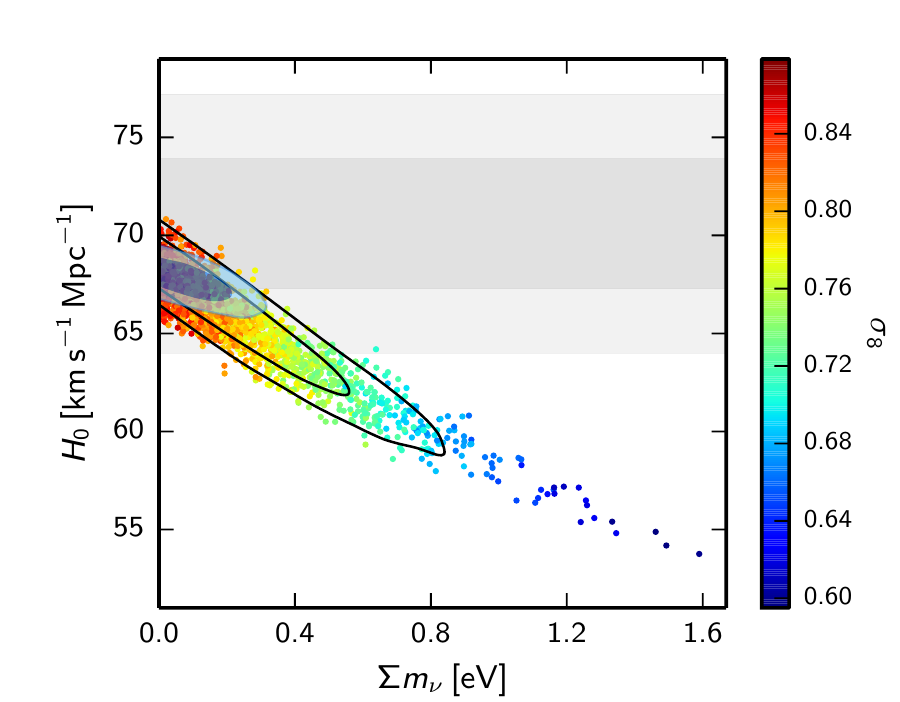}
\caption[]{Constraints on the sum of neutrino masses (in the standard 3$\nu$ scenario from cosmology\cite{Ade:2015xua}).}
\label{fig:planck}
\end{figure}

Neutrino oscillations cannot provide information on the absolute neutrino mass scale. The best sensitivity to this scale is at present
coming from cosmology. Indeed neutrinos properties are imprinted in the history of the universe.  In particular the features of the cosmic 
microwave background (CMB) and the large scale galaxy distribution depends sizeably on the sum of neutrino masses. The last results from Planck \cite{Ade:2015xua} are shown in Fig.~\ref{fig:planck}. Their conservative limit at 95$\%$CL is impressive:
\begin{eqnarray}
\sum_i m_i \leq 0.23 \UeVZ.
\end{eqnarray}

\section{Prospects in neutrino oscillation experiments}

An ambitious experimental program is underway to pin down the remaining unknowns and reach a 1$\%$ precision in the lepton flavour parameters. The neutrino ordering,  the octant of $\theta_{23}$ and the CP violating phase, $\delta$, can be search for in neutrino oscillation experiments with improved capabilities.

\begin{figure}
\centering
\includegraphics[width=.475\linewidth]{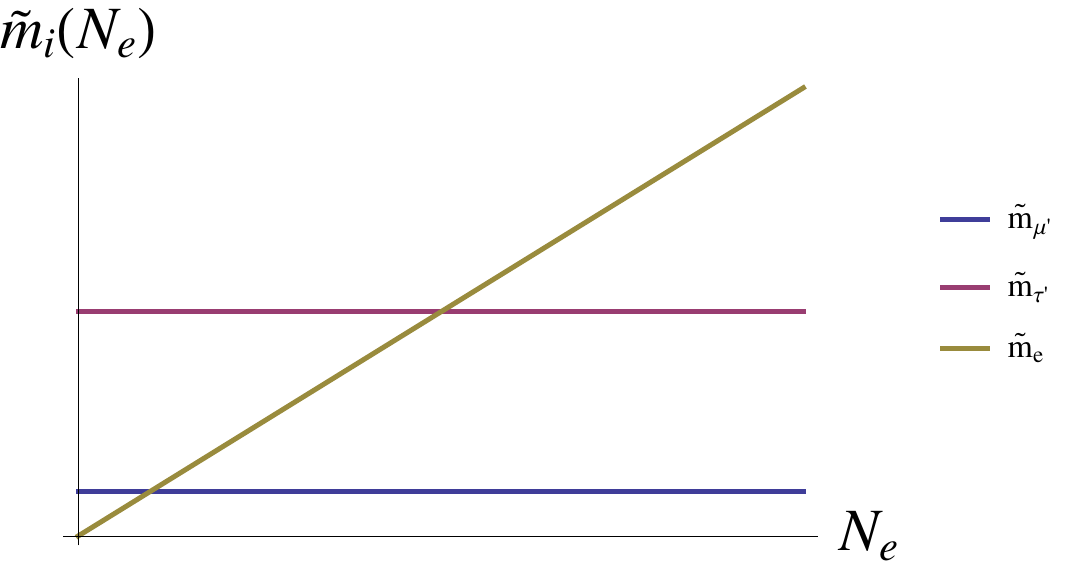}\includegraphics[width=.475\linewidth]{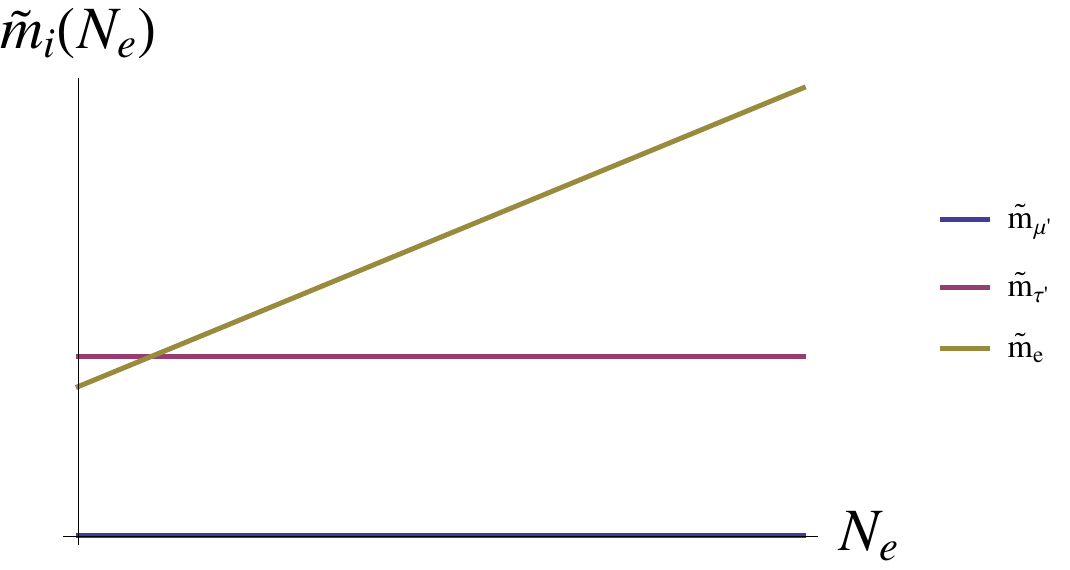}
\caption[]{Level crossings for $\nu$ in the three neutrino scenario for NO (left) and IO (right) at vanishing
$\theta_{12}$ and $\theta_{13}$.}
\label{fig:3nux}
\end{figure}

Concerning the neutrino ordering, the best hope to identify the spectrum exploits the MSW effect in the propagation of $\UGeV$ neutrinos 
through the Earth matter. In the case of three neutrinos propagating in matter, the $\nu$ mass eigenstates as a function of the electron density
for vanishing $\theta_{12}, \theta_{13}$ are depicted in Fig.~\ref{fig:3nux} for NO and IO. For NO we see that there are two level crossings
giving rise to two MSW resonances. The first one is essentially the one relevant for solar neutrinos, as it affects the smallest mass splitting, with the resonance condition:
\begin{eqnarray}
E^{(1)}_{\rm res} = {\Delta m^2_{12} \cos 2 \theta_{12}\over 2 \sqrt{2} G_F N_e }.
\end{eqnarray}
The second one affects the largest mass splitting
\begin{eqnarray}
E^{(2)}_{\rm res} = {\Delta m^2_{13}\cos 2 \theta_{13}\over 2 \sqrt{2} G_F N_e }.
\end{eqnarray}
For IO, only the first resonance appears in the $\nu$ channel. 

For $\bar{\nu}$  the dependence on $N_e$ of the first eigenstate has a negative slope and therefore only the second resonance appears for IO. 

\begin{figure}
\centering
\includegraphics[width=.5\linewidth]{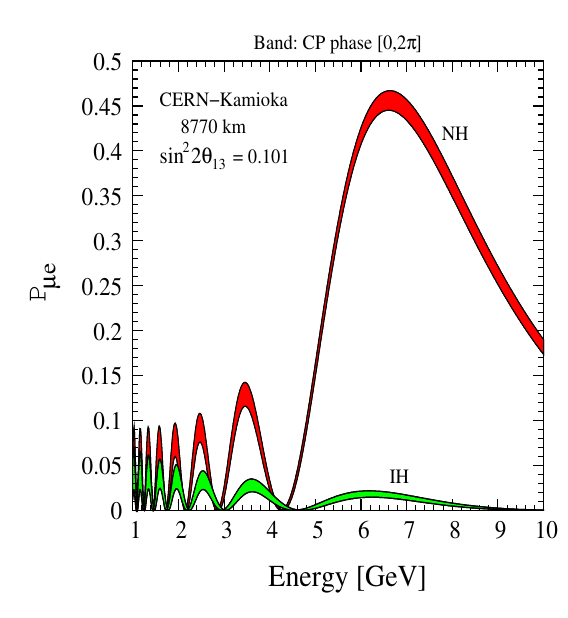}
\caption[]{$P_{\mu e}$ as a function of neutrino energy for $L$ corresponding to the distance  CERN-Kamioka for NH/IH. The bands corresponds to the uncertainty in $\delta$ (from Ref.\cite{sanjib}).}
\label{fig:sanjib}
\end{figure}

The existence of the atmospheric resonance implies a large enhancement of the oscillation probability $P(\nu_e \leftrightarrow \nu_\mu)$ for NO for energies near the resonant energy and at sufficiently long baseline. For IO the enhancement occurs in  $P(\bar{\nu}_e \leftrightarrow {\bar \nu}_\mu)$ instead. For the typical matter densities of the Earth crust and mantle and the value of the atmospheric mass splitting, the 
resonant energy for neutrinos travelling through Earth is $\simeq 6~\UGeVZ$, an energy that can be reached in  accelerator neutrino beams. The measurement of the neutrino ordering becomes almost a digital measurement sending a conventional $\nu$ beam sufficiently far as shown in Fig.~\ref{fig:sanjib}, which shows the oscillation probability $P(\nu_\mu\rightarrow \nu_e)$ as a function of the neutrino energy
at a distance corresponding to the baseline from CERN-Kamioka (8770 km).

The first experiment that will be sensitive to this effect is the NOvA experiment, optimized like T2K to see the $\nu_e$ appearance signal, with a baseline of $810$km, which is however a bit short to see a large enhancement. Nevertheless if lucky NOvA could discriminate 
the ordering at $3\sigma$. 

The atmospheric resonance must also affect  atmospheric neutrinos at the appropriate energy and baseline. Unfortunately the atmospheric flux contains both neutrinos and antineutrinos in similar numbers, and the corresponding events cannot be tell apart, because present atmospheric neutrino detectors cannot measure the lepton charge. If we superimpose the neutrino and antineutrino signals, both orderings will give rise to an enhancement in the resonance region, since no matter what the ordering is, either the neutrino or antineutrino channel will have
a resonance.  Nevertheless with sufficient statistics, there is discrimination power and in fact the biggest neutrino telescopes, ICECUBE and 
KM3NeT have proposed to instrument more finely some part of their detectors (PINGU and ORCA projects) to perform this measurement. Also the next generation of atmospheric neutrino detectors, such as HyperKamiokande, with a factor ${\mathcal O}(20)$ more mass than the present SuperKamiokande, or the INO detector that is designed to measure the muon charge in atmospheric events, could discriminate between the two orderings. 

\begin{figure}
\centering
\includegraphics[width=.5\linewidth]{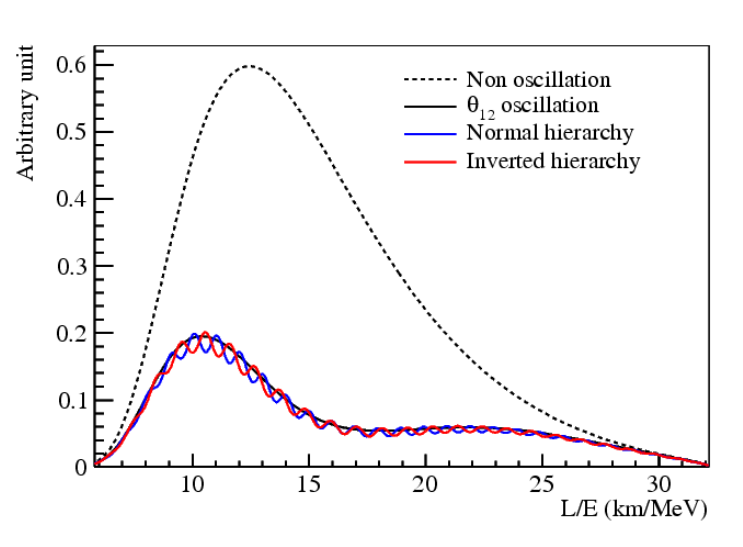}
\caption[]{Reactor neutrino spectrum in JUNO for NO/IO ( from Ref.~\cite{juno}).}
\label{fig:juno}
\end{figure}

A very different strategy has been proposed for reactor neutrino experiments (e.g JUNO project). The idea is to measure very precisely the reactor neutrinos at a
baseline of roughly $50$~km, where the depletion of the flux due to the solar oscillation is maximal. At this optimal distance, one can get 
a superb measurement of the solar oscillation parameters, $(\theta_{12}, \Delta m^2_{12})$, and, with sufficient energy resolution, one could 
detect the modulation of the signal due to the atmospheric oscillation \cite{Petcov:2001sy,Choubey:2003qx}. Fig.~\ref{fig:juno} shows how this modulation is sensitive 
to the neutrino ordering. A  leap ahead is  needed to reach the required energy resolution.  

\subsection{Leptonic CP violation}

As we have seen,  the CP phase, $\delta$, in the mixing matrix  induces CP
violation in neutrino oscillations, that is a difference between
$P(\nu_\alpha \rightarrow \nu_\beta)$ and $P(\bar{\nu}_\alpha
\rightarrow \bar{\nu}_\beta)$, for $\alpha \neq\beta$. As we saw in
the general expression of \Eref{eq:prob}, CP violation is possible
if there are imaginary entries in the mixing matrix that make
Im$[W_{\alpha\beta}^{jk}] \neq 0$. By CPT, disappearance probabilities cannot
violate CP however, because under CPT
\begin{equation}
P(\nu_\alpha \rightarrow \nu_\beta) = 
P(\bar{\nu}_\beta \rightarrow \bar{\nu}_\alpha)\SPp,
\end{equation}
so in order to observe a CP or T-odd asymmetry the initial and final
flavour must be different, $\alpha\neq\beta$:
\begin{equation}
A_{\alpha \beta}^{CP}\equiv\frac{P(\nu_\alpha\rightarrow \nu_\beta)-
P(\bar\nu_\alpha\rightarrow \bar\nu_\beta)}{P(\nu_\alpha\rightarrow
 \nu_\beta)+
P(\bar\nu_\alpha\rightarrow \bar\nu_\beta)} , \quad
 A_{\alpha \beta}^{T}\equiv\frac{P(\nu_\alpha\rightarrow \nu_\beta)-
P(\nu_\beta\rightarrow \nu_\alpha)}{P(\nu_\alpha\rightarrow \nu_\beta)+
P(\nu_\beta\rightarrow \nu_\alpha)}\SPp. 
\end{equation}
In the case of 3-family mixing it is easy to see that the CP(T)-odd
terms in the numerator are the same for all transitions $\alpha\neq
\beta$:
\begin{equation}
A^\text{CP(T)-odd}_{\nu_\alpha \nu_\beta} = 
\frac{\sin\delta c_{13}\sin
      2\theta_{13} \overbrace{\sin 2\theta_{12} \frac{\Delta m^2_{12} L}{4
      E_\nu}}^\text{solar} \overbrace{\;\sin 2\theta_{23}\sin^2\frac{\Delta
      m^2_{13} L}{4 E_\nu}}^\text{atmos}}
     {P^\text{CP-even}_{\nu_\alpha\nu_\beta}} .\;\;\;
\end{equation}
As expected, the numerator is GIM suppressed in all the $\Delta
m_{ij}^2$ and all the angles, because if any of them is zero, the
CP-odd phase becomes unphysical. Therefore an experiment
which is sensitive to CP violation must be sensitive to both 
mass splittings simultaneously. In this  situation, it is not clear a priori what the optimization of $E/L$ should be.

It can be shown that  including only statitistical errors, the signal-to-noise for this asymmetry
is maximized for  $\langle E_\nu\rangle/L \sim
|\Delta m^2_\text{atmos}|$. In
this case, only two small parameters remain in the CP-odd terms: the
solar splitting, $\Delta m^2_\text{solar}$ (\ie small compared to the other
scales, $\Delta m^2_\text{atmos}$ and $\langle E_\nu \rangle/L$), and
the angle $\theta_{13}$. The asymmetry is then larger in the
subleading transitions: $\nu_e\rightarrow \nu_\mu(\nu_\tau)$, because
the CP-even terms in the denominator are also suppressed by the same
small parameters.  A convenient approximation for the
$\nu_e\leftrightarrow\nu_\mu$ transitions is obtained expanding to
second order in both small parameters \cite{golden}:
\begin{eqnarray}
P_{\nu_ e\nu_\mu ( \bar \nu_e \bar \nu_\mu ) } & = & 
  s_{23}^2 \,\sin^2 2 \theta_{13} \, 
    \sin^2\!\left (\frac{\Delta m^2_{13} \, L}{4E_\nu} \right ) \quad 
\equiv P^\text{atmos} \nonumber\\ 
&+ &
  c_{23}^2 \, \sin^2 2 \theta_{12} \, 
    \sin^2\!\left(\frac{\Delta m^2_{12} \, L}{4 E_\nu} \right ) \quad
\equiv P^\text{solar} \nonumber \\ 
&+ & 
  \tilde J \cos \left ( \pm \delta - 
    \frac{\Delta m^2_{13} \, L}{4 E_\nu} \right ) \; 
  \frac{\Delta m^2_{12} \, L}{4 E_\nu} 
    \sin\left(\frac{\Delta m^2_{13} \, L}{4 E_\nu} \right )  \quad
\equiv P^\text{inter},
\end{eqnarray}
where ${\tilde J} \equiv c_{13}\; \sin 2 \theta_{13} \; \sin 2
\theta_{12}\; \sin 2 \theta_{23}$.  The first term corresponds
to the atmospheric oscillation, the second one is the solar one and
there is an interference term which has the information on the phase
$\delta$ and depends on both mass splittings.

%The difference between the neutrino and antineutrino
%oscillation probabilities for the leading and subleading channels are
%shown in \Fref{fig:leadsub}.

%\begin{figure}
%\centering
%\includegraphics[width=.5\linewidth]{nue_numu}\hfill
%\includegraphics[width=.5\linewidth]{numu_nutau}
%\caption[]{Comparison of the
%           $\nu_e\leftrightarrow\nu_\mu$/$\bar{\nu}_e\leftrightarrow\bar{\nu}_\mu$
 %          (left) and
 %          $\nu_\mu\leftrightarrow\nu_\tau$/$\bar{\nu}_\mu\leftrightarrow\bar{\nu}_\tau$
%           (right) oscillation probabilities for $E_\nu = 500\UMeV$,
%           $\theta_{13} = 8^\circ$ and $\delta=90^\circ$ as a function
%           of the distance}
%\label{fig:leadsub}
%\end{figure}

These results correspond to vacuum propagation, but usually these experiments require 
the propagation of neutrinos  in the Earth matter. 
The oscillation probabilities in matter can also be approximated by a similar
series expansion \cite{golden}. The result has the same structure
as in vacuum:
\begin{eqnarray}
P_{\nu_e\nu_\mu({\bar \nu}_e {\bar \nu}_\mu)}  = 
  s^2_{23}\, 
    \sin^2 2 \theta_{13}\left(\frac{\Delta_{13}}{B_\pm}\right)^2 \; \; 
    \sin^2\!\left( \frac{B_{\pm} L}{2} \right) \nonumber \\
  +  c_{23}^2 \, 
    \sin^2 2 \theta_{12} \, \left(\frac{\Delta_{12}}{A}\right)^2 
    \sin^2\!\left(\frac{A \, L}{2} \right ) \nonumber \\
  + {\tilde J} \;\frac{\Delta_{12}}{A}\;  
    \sin(\frac{A L}{2}) \;\frac{\Delta_{13}}{B_\pm}\; 
    \sin\!\left(\frac{B_\pm L}{2}\right) \; 
    \cos\!\left( \pm \delta - \frac{\Delta_{13}  \, L}{2} \right )\SPp, 
\end{eqnarray}
where 
\begin{equation}
B_{\pm} = | A \pm \Delta_{13} | \qquad
\Delta_{ij} = \frac{\Delta m^2_{ij}}{2 E_\nu} \qquad
A= \sqrt{2} G_F N_e\SPp. 
\end{equation}
The oscillation probability for neutrinos and antineutrinos now differ 
not just because of leptonic CP violation, but also due to the matter 
effects, that as we have seen can be resonant. In particular, the atmospheric
term which is the dominant one,  shows the expected  resonant enhancement in
the  neutrino or antineutrino oscillation probability (depending on
 the ordering). \\\\

The sensitivity to the 
interference term requires very good knowledge of the leading atmospheric term and the 
present degeneracies (the octant and the neutrino ordering) directly affect 
the leading term compromising therefore the $\delta$ sensitivity. Either 
both uncertainties are solved before this measurement, or there must be sufficient sensitivity
from the energy dependence of the signal to resolve all unkowns simultaneously. 

\begin{figure}
\centering
\includegraphics[width=.5\linewidth]{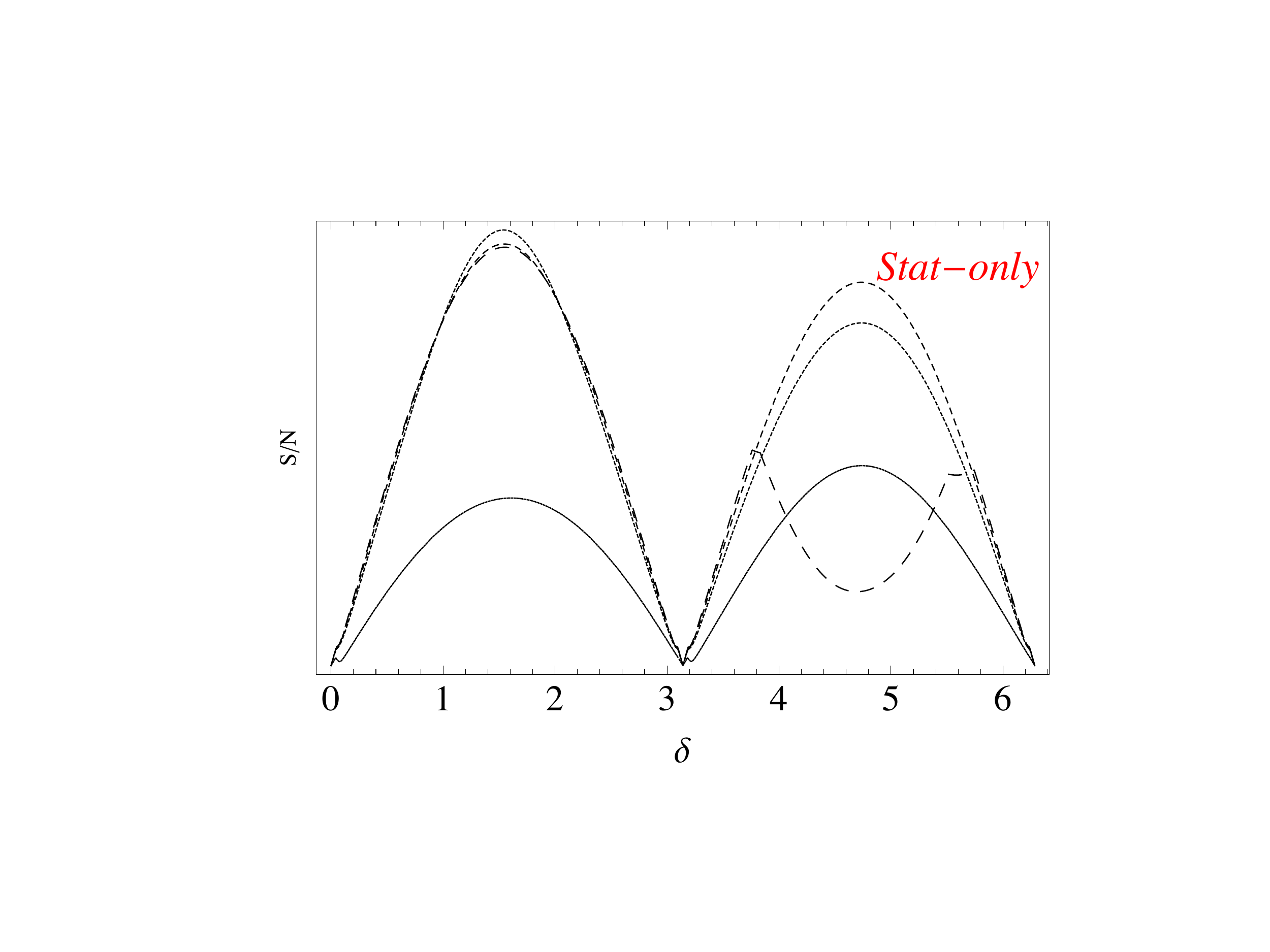}
\caption[]{Signal-to-noise for the discovery of CP violation at fixed $E/L \sim |\Delta m^2_{\rm atm}|$ as a function of the true value of $\delta$ for $L=295$km (long-dashed), $L=650$km  (short-dashed), $L=1300$km (dotted), $L=2300$km (solid). The ordering is assumed to be unknown.}
\label{fig:opt}
\end{figure}

A rough optimization of $L$ for fixed $E/L$ for discovering CP violation is shown in Fig.~\ref{fig:opt}. It shows the signal-to-noise 
as a function of the true value of $\delta$, assuming only statistical errors, but including the expected dependence of the cross sections and fluxes. At very short baselines, the sensitivity is compromised due to the lack of knowledge
of the neutrino ordering. In a wide intermediate region around ${\mathcal O}(1000)$km
the sensitivity is optimal, and at much larger baselines the sensitivity deteriorates 
because  the matter effects completely hide CP-violation. 

\begin{figure}
\centering
\includegraphics[width=.4\linewidth]{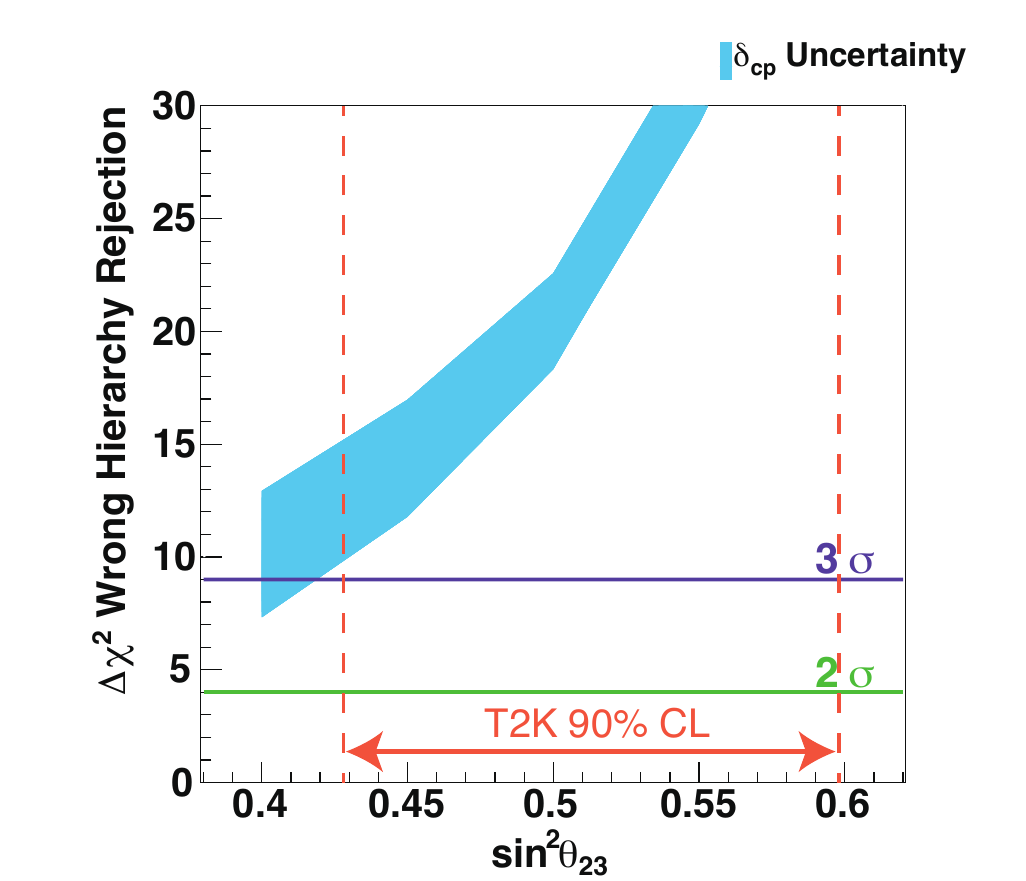}\includegraphics[width=.5\linewidth]{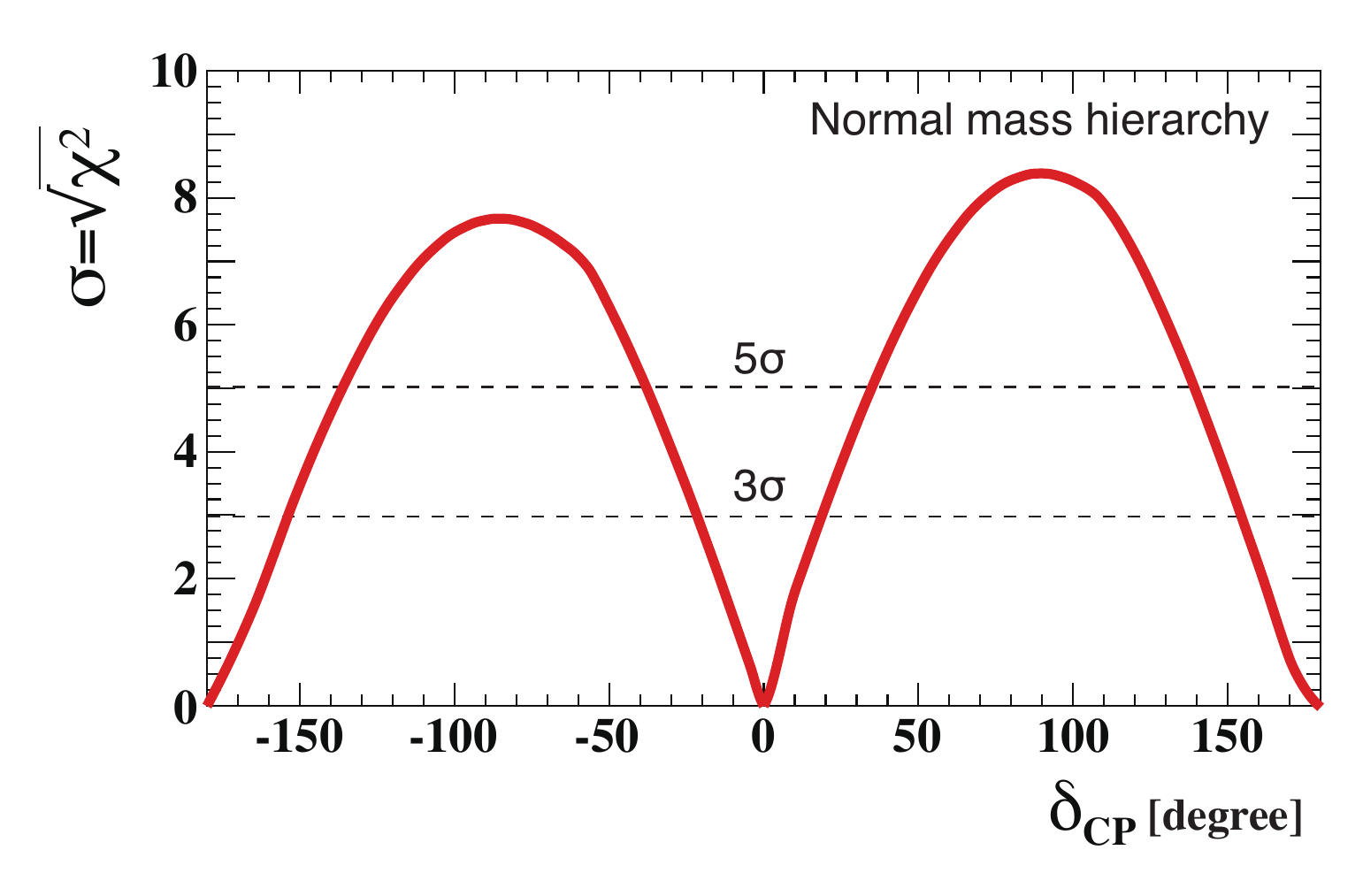}
\caption[]{Prospects for determining the ordering (left) and discovering CP violation (right)  in HyperKamiokande (from Ref.~\cite{hyperk}).}
\label{fig:future1}
\end{figure}

\begin{figure}
\centering
\includegraphics[width=.4\linewidth]{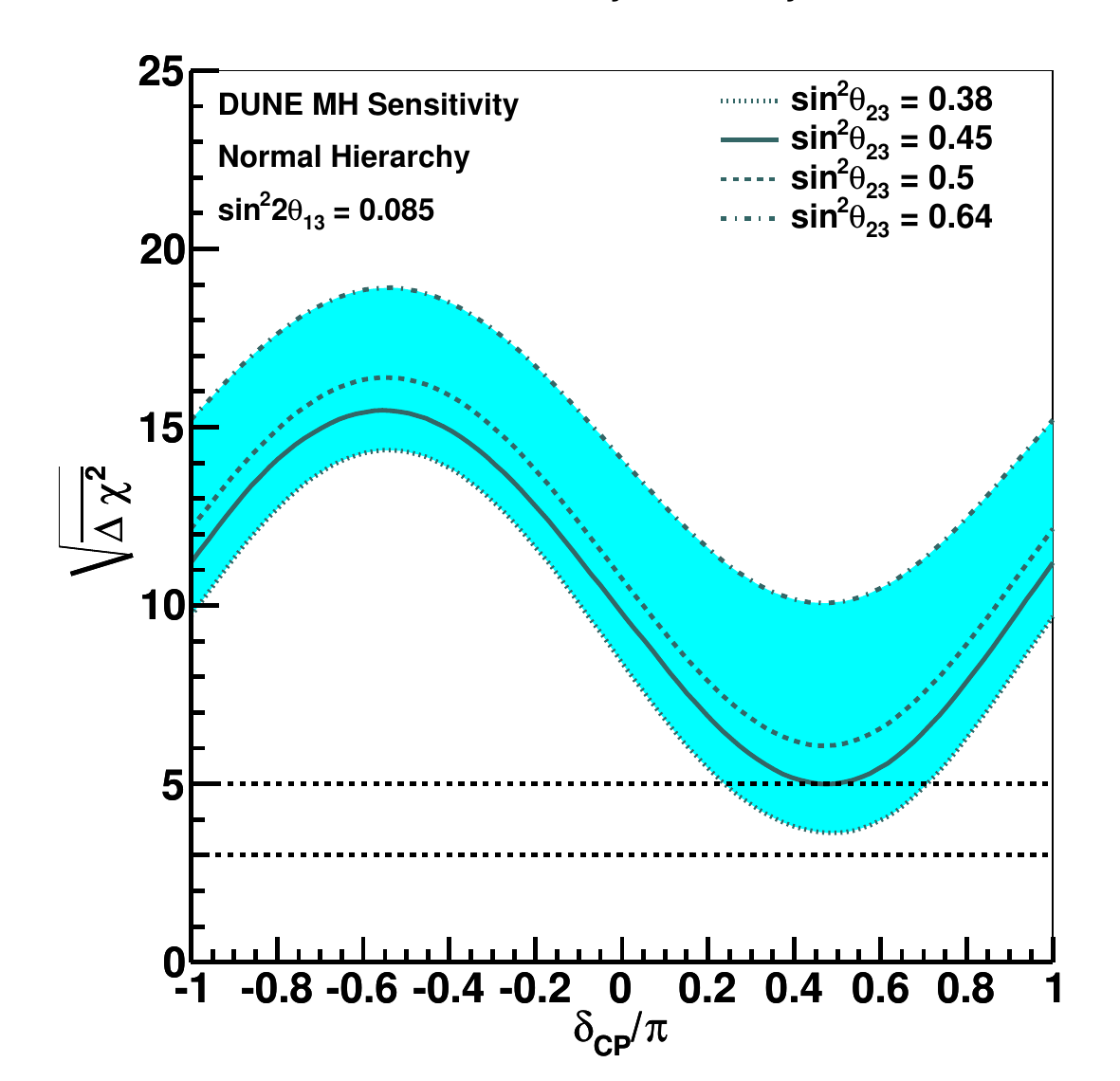}\includegraphics[width=.4\linewidth]{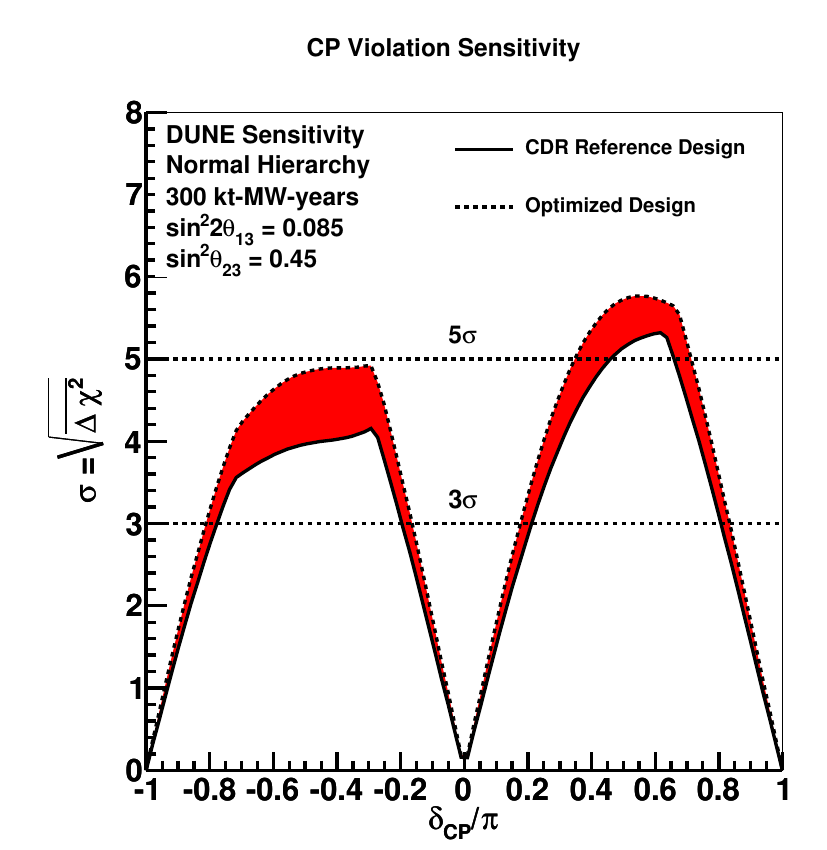}
\caption[]{Prospects for determining the ordering (left) and discovering CP violation (right) in DUNE ( from Ref.\cite{dune}).}
\label{fig:future2}
\end{figure}

Several projects have been proposed to search for leptonic CP violation, including conventional beams, but also 
novel neutrino beams from muon decays (neutrino factories), from radioactive ion decays ($\beta$-beams) or from 
spalation sources (ESS). The relatively large value of $\theta_{13}$ has refocussed the interest in using the less challenging  
conventional beams and two projects are presently being developed: the HyperKamiokande detector, an upscaled
version of SuperKamiokande that will measure atmopheric neutrinos with unprecedent precision, and also intercept
a neutrino beam from JPARC at a relatively short baseline $L=295$km, and the DUNE project that involves a $\sim$ 30 kton liquid
argon neutrino detector and a neutrino beam from Fermilab to the Soudan mine at a baseline of $L=1500$km. 
The expected sensitivities to  the neutrino ordering and to CP violation of both projects are shown in Figs.~\ref{fig:future1}., ~\ref{fig:future2}.

\section{Outliers: the LSND anomaly}

\begin{figure}
\centering
\includegraphics[width=.6\linewidth]{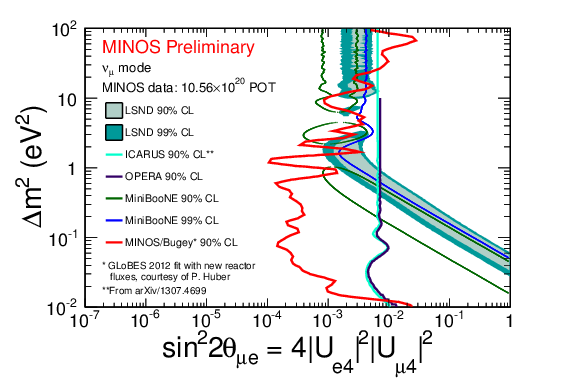}
\caption[]{Sterile neutrino search in the disappearance of $\nu_\mu$'s in MINOS  (from Ref.\cite{minosplus}).}
\label{fig:minosplus}
\end{figure}

The  long-standing puzzle brought by the LSND experiment  is still unresolved. This experiment \cite{lsnd} observed a surplus of electron events
in a muon neutrino beam from $\pi^+$ decaying in flight (DIF) and a
surplus of positron events in a neutrino beam from $\mu^+$ decaying at
rest (DAR). The interpretation of this data in terms of neutrino oscillations, that is a non-vanishing $P(\nu_\mu \rightarrow \nu_e)$,
gives the range shown by a coloured band in
\Fref{fig:minosplus}. 

\[
\begin{array}{lll} 
\pi^+ \rightarrow
 & \mu^+  
  & \nu_\mu \\ 
 &&\nu_\mu \rightarrow \nu_e \quad\text{DIF}~(28\pm 6/10\pm2)\\
 & \mu^+  
  & \rightarrow e^+ \nu_e  \bar{\nu}_\mu\\
 && \hspace{10mm} \bar{\nu}_\mu \rightarrow\bar{\nu}_e \quad\text{DAR}~(64\pm18/12\pm3)
\end{array}
\]
A significant fraction of this region was already excluded by the experiment KARMEN
\cite{karmen} that has unsuccessfully searched for $\bar{\nu}_\mu
\rightarrow \bar{\nu}_e$ in a similar range.

The experiment MiniBOONE  was designed to further investigate the LSND signal, with inconclusive results \cite{miniboone}. They did not confirm the anomaly, but found some anomaly at lower energies. 

\begin{figure}
\centering
\includegraphics[width=.7\linewidth]{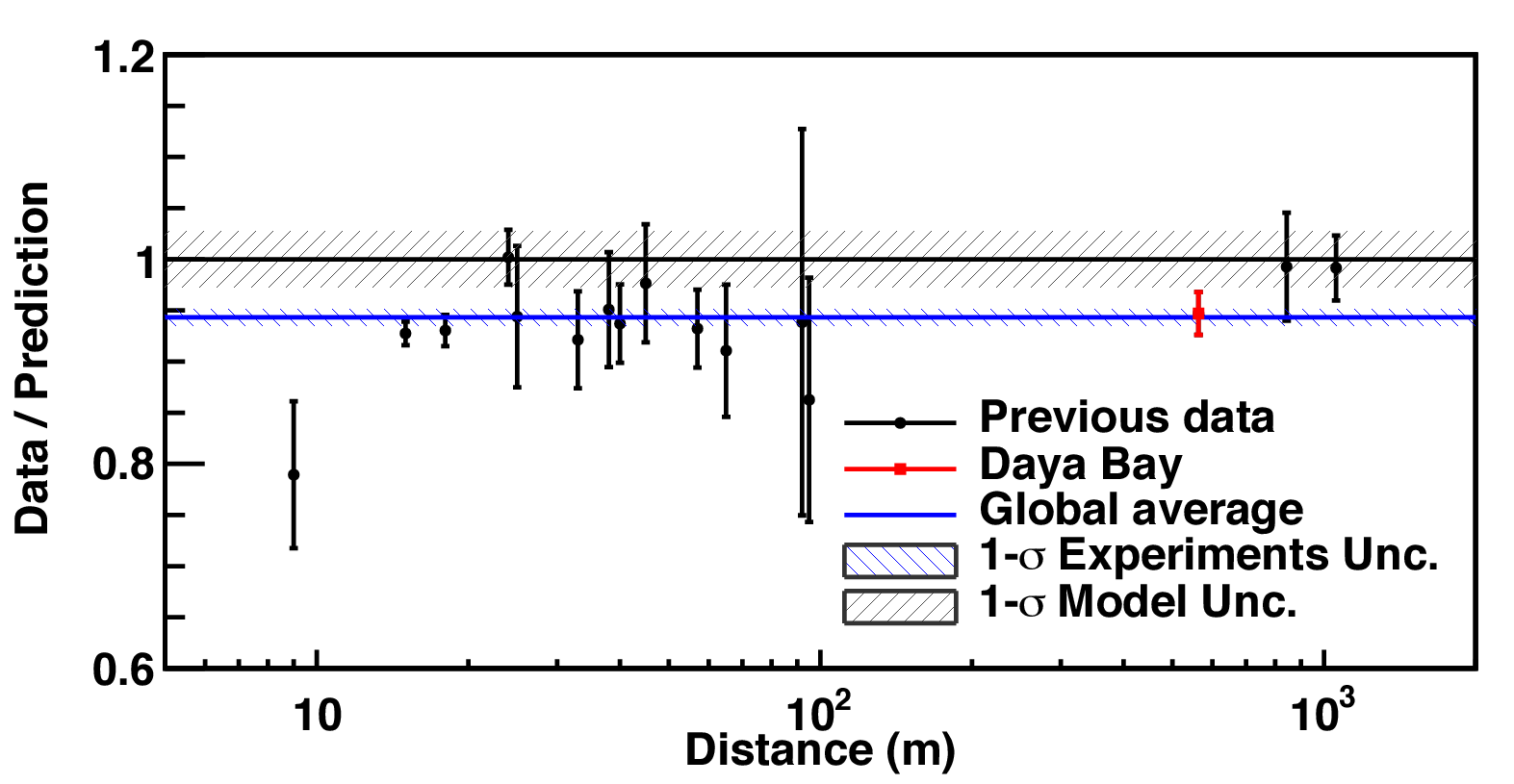}
\caption[]{Reactor neutrino flux measured by various near detectors compared with the recent flux predictions (from Ref.~\cite{dayabayflux}).}
\label{fig:dayabayflux}
\end{figure}

On the other hand, recently the results of various short baseline  (tens of meters) reactor neutrino experiments were revised, after an update on the reactor
neutrino flux predictions \cite{Mueller:2011nm,Mention:2011rk,Huber:2011wv}, which increased  theses fluxes by a few per cent. While the measured  neutrino flux was found to be in agreement with predictions before, after this revision some reactor neutrinos seem to disappear before reaching near detectors, $L={\mathcal O}(10)$m. This is the so-called reactor anomaly shown in Fig.~\ref{fig:dayabayflux}. This result brought some excitement because if this disappearance is due to oscillations, it  might reinforce the oscillation interpretation of the LSND anomaly. 

The required mass splitting to describe both anomalies is $\Delta m^2_{\rm LSND} \simeq 1$eV$^2$, which is much larger than the solar and atmospheric, and therefore requires the existence of at least a fourth neutrino mass eigenstate, $i$. If such a state can explain the LSND anomaly, it must couple to both electrons and muons. Unfortunately the smoking gun would require that also accelerator $\nu_\mu$  disappear with the same wavelength and this has not been observed: 
\begin{eqnarray}
P(\nu_\mu \rightarrow \nu_e) &\propto& |U_{ei} U_{\mu i}|^2 \hspace{2cm}   {\rm LSND}\nonumber\\
1-P(\nu_e \rightarrow \nu_e) &\propto& |U_{ei}|^4   \hspace{2.5cm}   {\rm reactor} \nonumber\\
1-P(\nu_\mu \rightarrow \nu_\mu) &\propto& |U_{\mu i}|^4 \hspace{2.5cm}    {\rm not~observed} \nonumber
\end{eqnarray}
 The strongest constraint on the disappearance of $\nu_\mu$ in the LSND range has been recently set by MINOS+\cite{minosplus} and
 is shown in Fig.~\ref{fig:minosplus} together with the region favoured by the LSND anomaly. An improvement of this sensitivity is expected also from the measurement of atmospheric neutrinos in ICECUBE. 

 A number of experiments are being constructed to clarify the reactor anomaly. Hopefully in the near future they will settle this long-standing puzzle. 
  
 \section{Neutrinos and BSM Physics}
 
 The new lepton flavour sector of the SM has opened new perspectives into the flavour puzzle. As we have seen neutrinos are massive but significantly lighter than the remaining charged fermions. Clearly the gap of Fig.~\ref{fig:hierar} calls for an explanation. The leptonic mixing matrix 
 is also very different to that in the quark sector. The neutrino mixing matrix is approximately given by \cite{Gonzalez-Garcia:2014bfa}
\begin{equation}
|U_\text{PMNS}| _{3\sigma}\simeq 
\begin{pmatrix}
        0.80-0.84 & 0.51-0.58  & 0.137-0.158    \\
        0.22-0.52 & 0.44-0.70 & 0.61-0.79 \\
        0.25-0.53 & 0.46-0.71 & 0.59-0.78 
\end{pmatrix}\SPp.
\end{equation}
The CKM matrix is presently constrained \cite{Agashe:2014kda} to be:
\begin{equation}
|V_\text{CKM}| \simeq 
\begin{pmatrix}
        0.97427(14) & 0.22536(61) & 0.00355(15)   \\
        0.22522(61) & 0.97343(15) & 0.0414(12) \\
        0.00886(33) & 0.0405(12) & 0.99914(5)
\end{pmatrix}\SPp.
\end{equation}
There is a striking difference between the two (and not only in the precision of the entries...). 
The CKM matrix  is close 
to the unit matrix:
\begin{equation}
 V_\text{CKM}  \simeq 
\begin{pmatrix}
  1 & O(\lambda) & O(\lambda^3)   \\
  O(\lambda) & 1 & O(\lambda^2)   \\
  O(\lambda^3) & O(\lambda^2) & 1 
\end{pmatrix}, 
\quad \lambda \sim 0.2,
\end{equation}
 while the leptonic one has large off-diagonal entries.  With a similar level of precision, it is close to the tri-bimaximal mixing pattern \cite{Harrison:2002er}
\[
U_{\rm PMNS}\simeq V_\text{tri-bi} \simeq 
\begin{pmatrix}
 \sqrt{\frac{2}{3}} &  \sqrt{\frac{1}{3}} & 0\\
-\sqrt{\frac{1}{6}} &  \sqrt{\frac{1}{3}} & \sqrt{\frac{1}{2}} \\
 \sqrt{\frac{1}{6}} & -\sqrt{\frac{1}{3}} & \sqrt{\frac{1}{2}} 
\end{pmatrix} .
\]
Discrete flavour symmetries have been extensively studied as the possible origin of this pattern. For recent review see \cite{King:2013eh}.

While we do not have yet a compelling explanation of the different mixing patterns, we do have one for 
  the gap between neutrino and other fermion masses. We saw that 
if the light neutrinos are Majorana particles and get their mass via the Weinberg interaction of Fig.~\ref{fig:majorana}, 
they are signalling  BSM physics. Neutrino masses
are suppressed because they arise from  a
new scale of physics that could be $\Lambda \gg v$. Generically such BSM would induce not only neutrino masses but also other effects represented at low-energies by the $d=6$ effective operators of eq.~(\ref{eq:eft}). Unfortunately the list of $d=6$ operators is too long to be of guidance: which one might be more relevant is to a large extent model dependent. 

\begin{figure}[h]
\centering
\includegraphics[width=.6\linewidth]{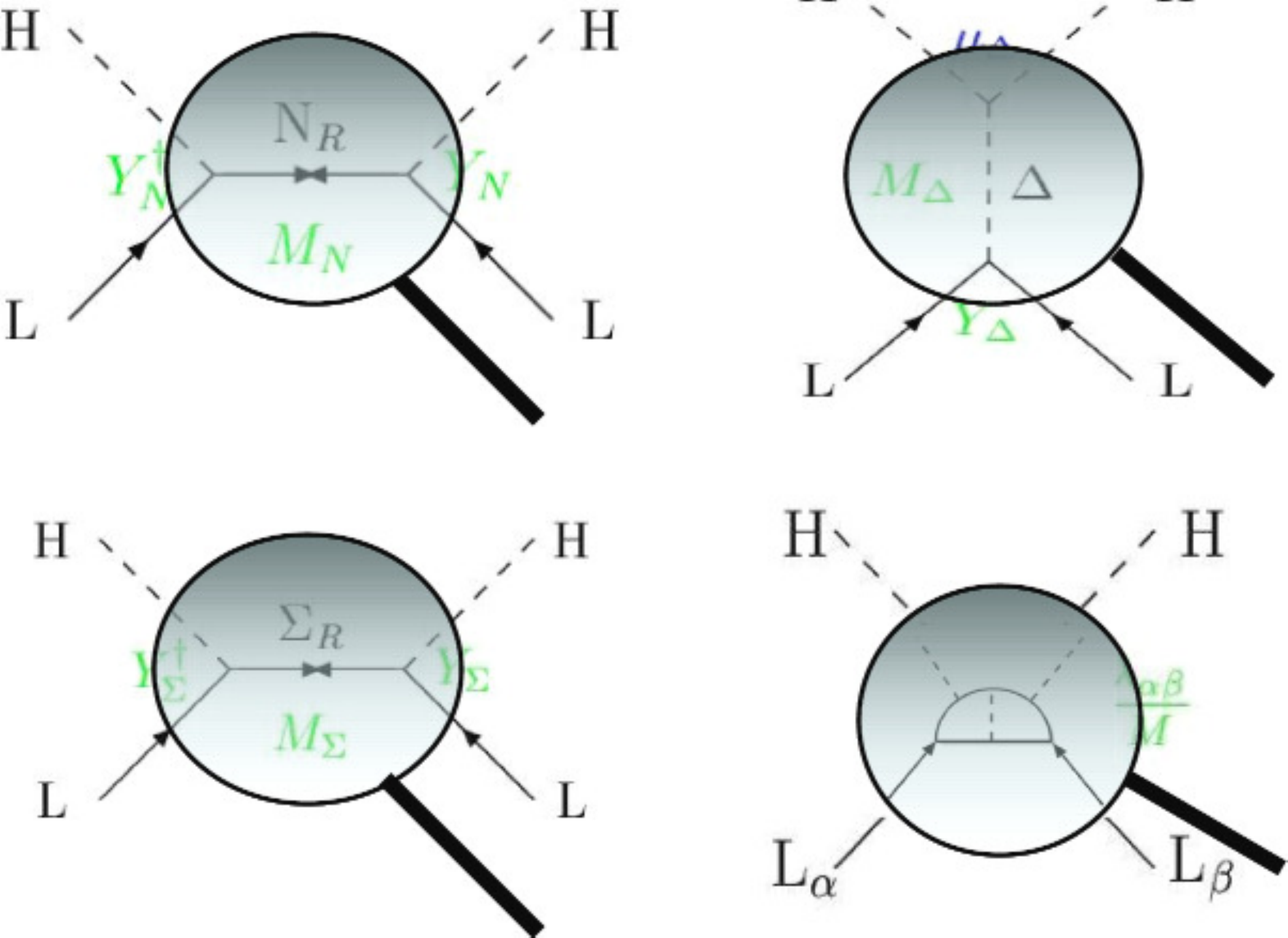}
\caption[]{Magnifying-glass view of Weinberg operator in see-saws
           Type I (top left), Type II (top right), Type III (bottom left)
           and Zee--Babu model (bottom right).}
\label{fig:weinberg}
\end{figure}

We could argue that there is not better motivated BSM physics than the one that gives rise to the Weinberg operator. 
The simplest possibility is that Weinberg's operator, like the Fermi one in Fig.\ref{fig:fermi}, arises from the exchange of a massive particle at tree level.
The classification of what particles can induce the Weinberg operator at tree level has been done, and reproduces the three types of seesaw models, as depicted in
\Fref{fig:weinberg}:
\begin{itemize}
\item type I see-saw: SM+ heavy singlet fermions \cite{Minkowski:1977sc,GellMann:1980vs,Yanagida:1979as,Mohapatra:1979ia},
\item type II see-saw: SM + heavy triplet scalar \cite{Magg:1980ut,Schechter:1980gr,Wetterich:1981bx,Lazarides:1980nt,Mohapatra:1980yp},
\item type III see-saw: SM  + heavy triple fermions \cite{Foot:1988aq,Ma:1998dn},
\end{itemize}
or combinations.  The masses of the extra states define the scale
$\Lambda$.

It is also possible that Weinberg's interaction is generated by new
physics at higher orders, such as in the famous Zee model \cite{zee}
and related ones \cite{Zee:1985rj,Babu:1988ki}. In this case, neutrino masses have an additional suppression  by loop factors $1/(16 \pi^2)$.

The $d=6$ operators induced at
tree level in see-saw models of Types I to III have been worked out
\cite{Abada:2007ux}.  They give rise to a rich phenomenology that could help 
discriminate between the models. In particular, they could induce
beyond-the-standard-model signals in $Z$ and $W$ decays, deviations in
the $\rho$ parameter or the $W$ mass, and mediate rare lepton decays,
as well as violations of universality and unitarity of the neutrino
mass matrix.  It would therefore be extremely important to search for
these effects. Whether they are large enough to be observed or not
depends strongly on how high the scale $\Lambda$ is, since all these
effects are suppressed by two powers of $\Lambda$.

\vspace{0.5cm}
Unfortunately, the measurement of Weinberg's interaction leaves behind
an unresolved ${\rm \lambda} \leftrightarrow \Lambda$ degeneracy that
makes it impossible to know what the scale of the new physics is, even
if we were to know the absolute value of neutrino masses.  

The recent discovery of the Higgs field and in particular the value of its mass 
$m_H=125\UGeV$ \cite{Aad:2015zhl} suggests that the SM is as healthy as ever. In spite of the Landau poles, 
the value of the SM couplings surprisingly conspire to make the model consistent up to arbitrarily large scales \cite{Degrassi:2012ry}. 

The most popular choice for  $\Lambda$ has traditionally been a 
grand-unification scale, given the intriguing fact that the
seesaw-type ratio $\frac{v^2}{M_\text{GUT}} \sim 0.01$--$0.1\UeV$, in the
right ballpark of the neutrino mass scale. Howeer, in the absence of any stabilizing mechanism such as 
supersymmetry, however, the electroweak scale needs to be fine-tunned \cite{Vissani:1997ys,Casas:2004gh} since the Higgs mass  receives quadratic loop corrections in 
$\Lambda$. A naturalness argument would then imply that $\Lambda <
10^7\UGeV$. The opposite is not true however, the scale
$\Lambda$ would not get corrections from the electroweak scale: any value of $\Lambda \leq v$ is technically natural.

The possibility that the scale $\Lambda$ might be of the order of the EW scale or lower has recently been studied in more detail, with 
special emphasis on establishing the existing experimental constraints, and the possibility that this new physics could explain 
other open problems in the SM such as: the LSND and reactor anomalies, dark matter, leptogenesis, etc. 
The type I seesaw model is the case better studied so we will concentrate on pinning down the scale $\Lambda$ in this context.

\subsection{Type I seesaw model}

It is arguably the minimal extension of the SM allowing for neutrino masses \cite{Minkowski:1977sc,GellMann:1980vs,Yanagida:1979as,Mohapatra:1979ia},. It involves the addition of $n_R\geq 2$ singlet Weyl fermions, $\nu_{R}$, 
to the SM. The most general renormalizable Lagrangian which satisfies Lorentz and the gauge symmetries is given by:
\begin{eqnarray}
 \mathcal{L}_{\rm Type I} &=&  \mathcal{L}_{\rm SM}- \sum_{\alpha, i} \bar{L}^\alpha Y^{\alpha i}_\nu \;\tilde{\Phi}\; \nu^i_R - \sum_{i,j}^{n_R} \frac{1}{2} \bar{\nu}_R^{ic} \;M^{ij}_N \;\nu^j_R  + \text{ h.c. },
 \label{eq:seesawI}
\end{eqnarray}
where the new parameters involved are a $3\times n_R$ neutrino Yukawa matrix and a $n_R \times n_R$ symmetric Majorana mass matrix
for the singlet fields.  Upon spontaneous symmetry
breaking these couplings become mass terms, that can be writen in the Majorana basis $(\nu_L^c, \nu_R)$ as 
\begin{eqnarray}
\mathcal{L}_{\rm Type I} &\rightarrow
  & \mathcal{L}_{\rm SM} -\frac{1}{2} \begin{pmatrix}\bar{\nu}_L & {\bar{\nu^c_R}}\end{pmatrix} 
               \begin{pmatrix}0 & m_D\\
                              m_D^T & M_N
               \end{pmatrix} 
               \begin{pmatrix}\nu_L^c \cr \nu_R\end{pmatrix}\SPp +h.c.+...
               \label{eq:typeI}
\end{eqnarray}
where 
\begin{eqnarray}
m_D= Y_\nu {v \over \sqrt{2}}.
\end{eqnarray}
Note that Dirac neutrinos are a particular case of the model. If we invoke a global lepton number symmetry to force $M_N =0$, the singlets
are exactly equivalent to the right-handed neutrinos in the Dirac case described in sec.~\ref{sec:dirac}.  In the opposite limit $M_N \gg v$, the singlets
can be integrated out and give rise to the Weinberg interaction as well as others at $d=6$, etc. For finite $M_N$, 
the spectrum of this theory contains in general $3+n_R$ Majorana neutrinos, which are  admixtures of the active ones and the extra singlets. 
It is easy to diagonalize the mass matrix in eq.~(\ref{eq:typeI}) in an expansion in $m_D/M_N$. The result to leading order in this expansion is
%When $v \ll M_R$, the diagonalization of the mass matrix can be done
%in perturbation theory:
%\begin{eqnarray}
%\mathcal{M} = \mathcal{M}^{(0)} + \mathcal{M}^{(1)}\equiv 
 %             \begin{pmatrix}0  & 0 \\ 0 & M_R \end{pmatrix} 
 %           + \begin{pmatrix}0  & {\tilde \lambda}_\nu v \\
 %                            {\tilde \lambda}^T_\nu v & 0 
 %             \end{pmatrix}\SPp.
%\end{eqnarray}
%To second order we find: 
\begin{eqnarray}
U^T  \begin{pmatrix}0 & m_D\\
                              m_D^T & M_N
               \end{pmatrix} 
                U \simeq \begin{pmatrix} 
                      - m_D {1\over M_N} m_D^T & 0 \\
                      0  & M_N
                    \end{pmatrix} + {\mathcal O}(\theta^2) ,
                    \;\;\; 
                U = \begin{pmatrix} 
                      1 & \theta \\
                      -\theta^\dagger& 1 
                    \end{pmatrix}\SPp,
\end{eqnarray}
where 
\begin{eqnarray}
\theta = m_D^* {1 \over M_N} .
\end{eqnarray}
To this order therefore the light neutrino and heavy neutrino masses are given by
\begin{eqnarray}
m_l = {\rm Diag}\left[{ - m_D {1 \over M_N} m_D^T} \right], \;\;\; M_h = {\rm Diag}[M_N].  
\end{eqnarray}
\begin{figure}[h]
\centering
\includegraphics[width=.55\linewidth]{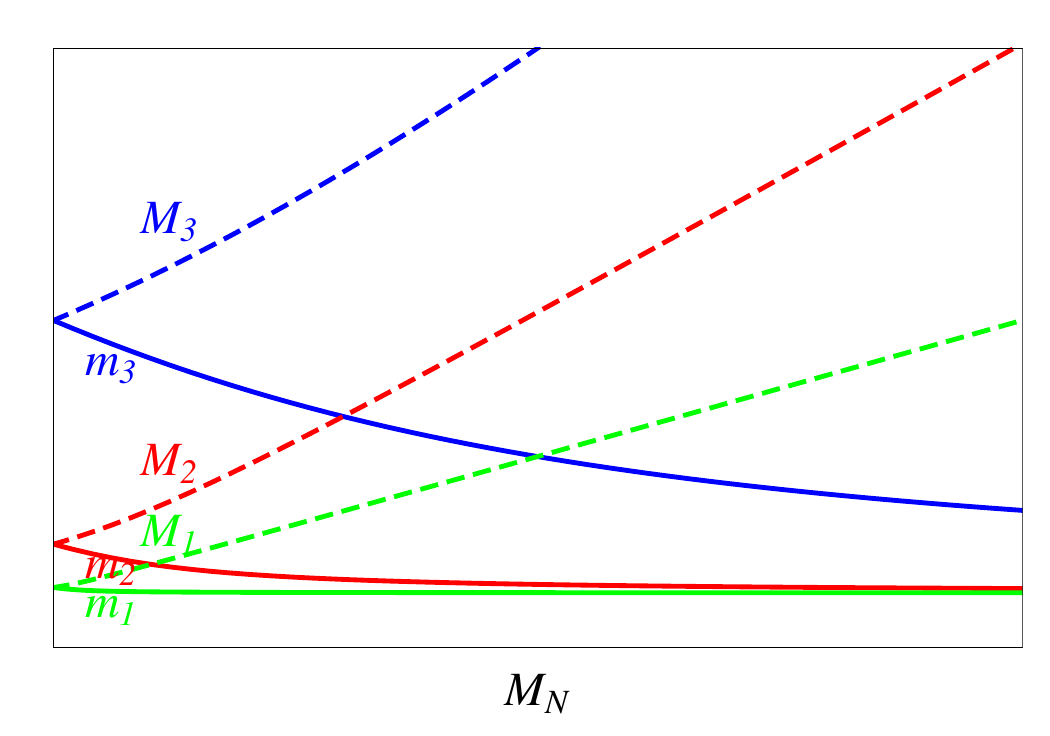}
\caption[]{Spectrum of th type I seesaw model for $n_R=3$ as a function of a common $M_N$.}
\label{fig:seesawspec}
\end{figure}

Fig.~(\ref{fig:seesawspec}) depicts the spectrum for the case of $n_R =3$ as a function of a common $M_N$. In the limit $M_N \rightarrow 0$ the states degenerate in pairs to form Dirac fermions. As $M_N$ increases three states get more massive proportional to $M_N$,  while three get lighter proportional to $M_N^{-1}$. This is why the model is called seesaw. 
The number of  new free parameters  is large. For the case $n_R=3$ there are 18 fundamental parameters in the lepton sector: six of them are masses, six mixing angles and six phases. The counting of parameters for general $n_R$ is shown in Table~\ref{tab:seesaw}. Out of these 18 parameters we have determined only 5: two mass differences and three neutrino mixing angles. 

 \begin{table}
\[
\begin{array}{l|l|l|l|l|l}                        \hline\hline
 & \text{Yukawas} 
  & \text{Field redefinitons} 
   & No.~ m 
    &No.~ \theta 
     & No.~ \phi                                   \\\hline 
 &&&&&                                           \\
\text{see-saw} 
 & Y_l, Y_\nu, M_R=M^T_R 
  & U(n)^3 
   &&&                                           \\  
 E\geq M_i 
 & 5 n^2+n   
  & \frac{3 (n^2-n)}{2}, \frac{3(n^2+n)}{2} 
   & 3 n 
    & n^2-n 
     & n^2-n                                     \\
 &&&&&                                           \\ \hline
 &&&&&                                           \\ 
\text{see-saw}
 & Y_l, \alpha_\nu^T=\alpha_\nu 
  & U(n)^2
   &&&                                           \\  
E \ll M_i
 & 3 n^2 + n 
  & n^2-n, n^2+n 
   & 2 n 
    & \frac{n^2-n}{2}
     & \frac{n^2-n}{2}                           \\ 
 &&&&&                                           \\ \hline
\end{array}
\]
\caption[]{Number of physical parameters in the see-saw model with $n$
           families and the same number of right-handed Majorana
           neutrinos at high and low energies.}
\label{tab:seesaw}
\end{table}

 A very convenient parametrization in this model is the so-called Casas-Ibarra \cite{Casas:2001sr} parametrization, which allows to write in all generality (up to corrections of ${\mathcal O}(\theta^2)$) the Lagrangian parameters in terms of the parameters that can be measured at low energies: light neutrino masses and mixings, and others that cannot. In particular the phenomenology of this model depends on the spectrum of neutrino mass eigenstates,
 that we denote by $(\nu_1, \nu_2, \nu_3, N_1,N_2,...N_{n_R})$, and their admixture in the flavour neutrino states which is given by:
 \begin{eqnarray}
 \begin{pmatrix}\nu_e \\
 \nu_\mu\\
 \nu_\tau \end{pmatrix} = U_{ll} \begin{pmatrix}\nu_1 \\
 \nu_2\\
 \nu_3 \end{pmatrix}+ U_{lh} \begin{pmatrix}N_1 \\
 N_2\\
 ..\\
 N_{n_R} \end{pmatrix}.
  \end{eqnarray}
In the Casas-Ibarra parametrization we have
\begin{eqnarray}
U_{ll} &=& U_{\rm PMNS} + {\mathcal O}(\theta^2),\nonumber\\
U_{lh} &=& i U_{\rm PMNS}  \sqrt{m_l} R {1\over \sqrt{M_h}} + {\mathcal O}(\theta^2).
\label{eq:ci}
\end{eqnarray}
where $R$ is a general complex orthogonal matrix, 
$R^T R = 1$, which together with the heavy neutrino masses, $M_h$, parametrizes the parameter space inaccessible to neutrino oscillation experiments. Note that $U_{ll}$ is the mixing matrix that we measure in neutrino oscillation experiments, assuming the heavy states are too 
heavy to play a role. This matrix is however no longer unitary, but the unitarity violations are parametrically of ${\mathcal O}(\theta^2) \sim m_l/M_h$. 

The Casas-Ibarra parametrization needs to be modified in the presence of large unitarity violations. A similar parametrization valid to all orders
in $\theta$ is given in \cite{Donini:2012tt}. 

Eqs.~(\ref{eq:ci}) indicate that in this model there is a strong correlation between  flavour mixings of the heavy states, $U_{lh}$, and the ratio of light-to-heavy neutrino masses. However the presence of the unknown matrix $R$, which is not bounded, implies that the naive seesaw
scaling, $|U_{lh}|^2 \sim m_l/M_h$, that would hold exactly for one neutrino family, is far too naive for $n_R > 1$. In fact there are regions of parameter space where these mixings can be much larger than suggested by the naive scaling, and these are precisely the regions with more phenomenological interest, as we will see below. 

In this model we can ask the question. What is the value of the $M_N$ scale to avoid the hierarchy between neutrinos and the remaining fermions. If we plot the distribution of Yukawa couplings instead of the masses, we find that neutrino masses can be explained with a scale $M_N \simeq$ GUT, if the neutrino yukawa couplings are of $O(1)$ like the top. However if the yukawas are of the order of the electron yukawa, a scale $M_N \sim$ TeV can also explain neutrino masses. Clearly, in both cases we have avoided making neutrinos especial, and the flavour puzzle is no worse than in the charged fermion sector. Note that this wide range of scales between TeV-GUT is the result of the quadratic dependence of the light neutrino masses on the yukawas, as opposed to the linear dependence in the Dirac case. 

Let us discuss some phenomenological implications of the different choices of the scale $M_N$. 

{\bf Neutrinoless double-beta  decay}

For $M_N \geq 100$ MeV, the model implies the  presence of neutrinoless double beta decay at some level. The amplitude for this process gets contribution from the light and heavy states:
\begin{eqnarray}
m_{\beta\beta} \equiv \sum_{i=1}^3 (U_{\rm PMNS})_{ei}^2 m_i + \sum_{j=1}^{n_R} (U_{lh})_{ej}^2 M_j {{\mathcal M}^{\beta\beta 0\nu} (M_j) \over{\mathcal M}^{\beta\beta 0\nu}(0)},
\end{eqnarray}
where the ratio of matrix elements ${\mathcal M}^{\beta\beta 0\nu}$ for heavy and light mediators satisfy \cite{Blennow:2010th}:
\begin{eqnarray}
{{\mathcal M}^{\beta\beta 0\nu} (M_j) \over{\mathcal M}^{\beta\beta 0\nu}(0)} \propto \left({100 {\rm MeV} \over M_j }\right)^2,\;\;\; M_j \rightarrow \infty.
\end{eqnarray}
If all the heavy state masses $\gg 100 \UMeV$, the second term is suppressed and the amplitude contains only the light neutrino masses and mixings:
\begin{equation}
m_{\beta\beta} \simeq  
|c_{13}^2 ( m_1 c_{12}^2 + m_2 e^{i \alpha_1} s_{12}^2 ) 
 + m_3 e^{i \alpha_2} s_{13}^2|\SPp,
\end{equation}
and  is quite well constrained from  neutrino oscillation experiments.
 \Fref[b]{fig:lisibb} shows the present allowed regions for $m_{\beta\beta}$ neglecting the heavy state contributions as a function of 
the sum of the light neutrino masses, that can be constrained from cosmology:
 \begin{equation}
\Sigma \equiv m_1+ m_2+ m_3. \, 
\end{equation}
%The present experimental bounds from cosmology (vertical lines) and 
 %the experimental $\beta\beta 0\nu$ searches(horizontal lines) are also shown.
  It the neutrino ordering would be inverted or $\Sigma$ not much smaller than $0.1\UeV$, there is a good chance that the next generation of $\beta\beta 0\nu$ experiments will see a signal. A plethora of experiments using different technologies 
 have been proposed to reach a sensitivity in $m_{\beta\beta}$ in the range of $10^{-2}\UeV$
, which could be sufficient to explore the full parameter space
in the case of the IO. The importance of this measurement can hardly be overstated. A non-zero $m_{\beta\beta}$ will imply that 
neutrinos are Majorana and therefore a new physics scale must exist,  that lepton number is violated, and might give very valuable information on  the lightest neutrino mass, and even help establishing the
neutrino mass ordering.
\begin{figure}[h]
\centering
\includegraphics[width=.5\linewidth]{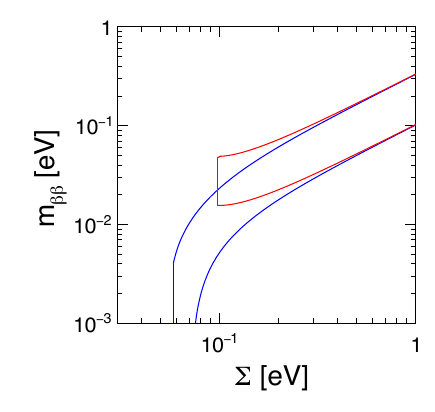}
\caption[]{Allowed region for $m_{\beta\beta}$ for NO (blue contour) and IO (red contour) from a global analysis of neutrino data (from Ref.~\cite{Capozzi:2016rtj}).}
\label{fig:lisibb}
\end{figure}
If the heavy states are not too heavy, within $100\UMeV$-few$\UGeV$, they could also  contribute to the process significantly and even dominate over the light neutrino contribution for both orderings \cite{Ibarra:2010xw,Mitra:2011qr,Lopez-Pavon:2015cga}. 

{\bf Mini-seesaw and oscillations }

If the scale $M_N \leq  \UeV$ (mini-seesaw models \cite{deGouvea:2005er}), the extra heavy states could affect neutrino oscillations significantly. Strong constraints can be derived
therefore from neutrino oscillation experiments which essentially exclude the possibility that $M_N \in[10^{-9}(10^{-11})\UeV, 1\UeV]$ for NO(IO) \cite{deGouvea:2009fp,deGouvea:2011zz,Donini:2011jh}. The possibility that $M_N \sim1\UeV$ could explain the LSND and reactor anomalies has also been studied. It is intriguing
that the best fit to the LSND and reactor anomalies \cite{Giunti:2011gz,Kopp:2013vaa} point to mixings and masses of the extra neutrino(s) that nicely match the naive seesaw scaling.  In fact mini-seesaw models provide similar fits to data  \cite{Donini:2012tt}, with much less parameters,  than general phenomenological models with $3+N$ neutrino mixing.  Both are affected however by  the tension in data between the anomalies and the non-observation of $\nu_\mu$ disappearance. 

{\bf Cosmology and the seesaw scale}

For $M_N \leq 100 \UMeV$, the heavy states in seesaw models can sizeably modify the history of the Universe: the abundance of light elements, the fluctuations in the CMB and the galaxy distribution at large scales. This is the case because these extra states contribute to the expansion either as a significant extra component of dark matter ($\Omega_m$) or radiation ($\Delta N_{\rm eff}$). 

The singlet states in this mass range are produced at $T$ below the electroweak phase transition via mixing. A simple estimate of their production rate
is
\begin{eqnarray}
\Gamma_{s_i}(T) \simeq \sum_\alpha |(\tilde{U}_{lh})_{\alpha i}|^2\times \Gamma_{\nu_\alpha}(T), 
\end{eqnarray}
where $\Gamma_{\nu_\alpha}$ is the interaction rate of the active neutrinos and the ${\tilde U}_{lh}$ is the light-heavy mixing at $T$, 
strongly modified by forward scattering on the plasma particles  \cite{Notzold:1987ik}. 
The state $i$ will reach thermal equilibrium if $\Gamma_{s_i}(T)$ is larger than the Hubble parameter at some $T$. If this is the case, the extra species will contribute like one extra neutrino for $T>M_i$ or like an extra component of dark matter for $T < M_i$. 
The latest results from Planck strongly constrain an extra radiation component at CMB:
\begin{eqnarray}
N_{\rm eff}({\rm CMB}) = 3.2\pm 0.5.
\end{eqnarray}
and also measures the dark matter component to be $\Omega_m=0.308\pm 0.012$. 
Similar bounds are obtained from the abundance of light elements, BBN. 
These bounds exclude the possibility of having essentially any extra fully thermalized neutrino that is sufficiently long-lived to 
survive BBN. It can be shown that 
the ratio ${\Gamma_{s_i}(T)\over H(T)}$ reaches a maximum at $T_{max}$ \cite{Barbieri:1989ti,Kainulainen:1990ds} and
\begin{eqnarray}
{\Gamma_{s_i}(T_{max})\over H(T_{max})} \sim {\sum_\alpha |(U_{lh})_{\alpha i}|^2 M_i \over \sqrt{g_*(T_{max})}}.
\end{eqnarray}
The naive seesaw scaling $U_{lh}^2 M_h \sim m_l $, would seem to imply that the thermalization condition depends only on the light neutrino masses and is independent on the seesaw scale. In fact a detailed study  shows that indeed this naive expectation holds. 

For $n_R=2$, the heavy states must be $M_i \geq 100\UMeV$ \cite{Hernandez:2013lza}, so that they might decay before BBN. 
For $n_R=3$ two things can happen\cite{Hernandez:2014fha}. If the lightest neutrino mass
$m_{\rm lightest} \geq 3 \times 10^{-3}\UeV$, all the three heavy states thermalize so $M_i \geq 100\UMeV$. If $m_{\rm lightest} \leq 3 \times 10^{-3}\UeV$ two states must be above this limit, but 
one of the states with mass $M_1$ might not thermalize and therefore be sufficiently diluted.  $M_1$  may take any value  provided  $m_{\rm lightest}$, which is presently unconstrained,  is tuned accordingly. 

The states that could explain the LSND and reactor anomalies will imply $\Delta N_{\rm eff} \geq 1$ which is essentially excluded by cosmology. For a recent detailed analysis see \cite{Hannestad:2015tea}. Exotic extensions involving hidden interactions of the extra singlet states would be needed to make them compatible.

{\bf Warm dark matter}

For $m_{\rm lightest} \leq 10^{-5} \UeV$, $M_1$ might be ${\mathcal O}(\UkeV)$, and  a viable warm dark matter candidate \cite{Dodelson:1993je,Asaka:2005pn}. This scenario is the so-called $\nu$MSM model \cite{Asaka:2005pn}. The most spectacular signal of this type of Dark Matter is a monocromatic X-ray line. Two recent analyses \cite{Bulbul:2014sua,Boyarsky:2014jta} have recently shown some evidence for an unexplained X-ray line in galaxy clusters that might be compatible with a 7$\UkeV$ neutrino. These results are under intense debate. If interpreted in terms of a $\UkeV$ neutrino, the mixing however is too small and some extra mechanism is needed to enhance the production so that it matches the required dark matter density, such as the presence of large primordial lepton asymmetries \cite{Shi:1998km}. 

{\bf Direct searches}

\begin{figure}[h]
\centering
\includegraphics[width=.45\linewidth]{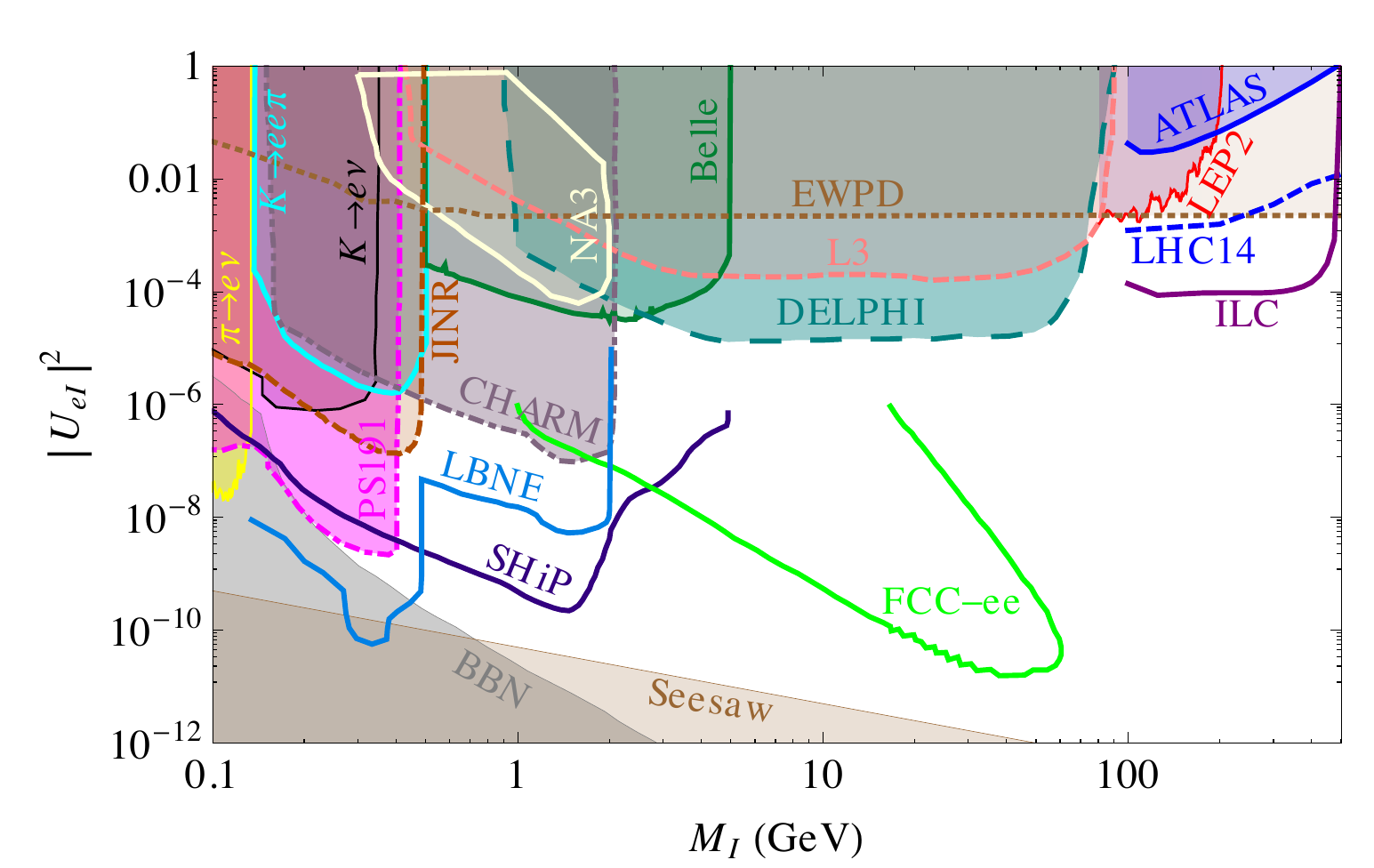}
\includegraphics[width=.45\linewidth]{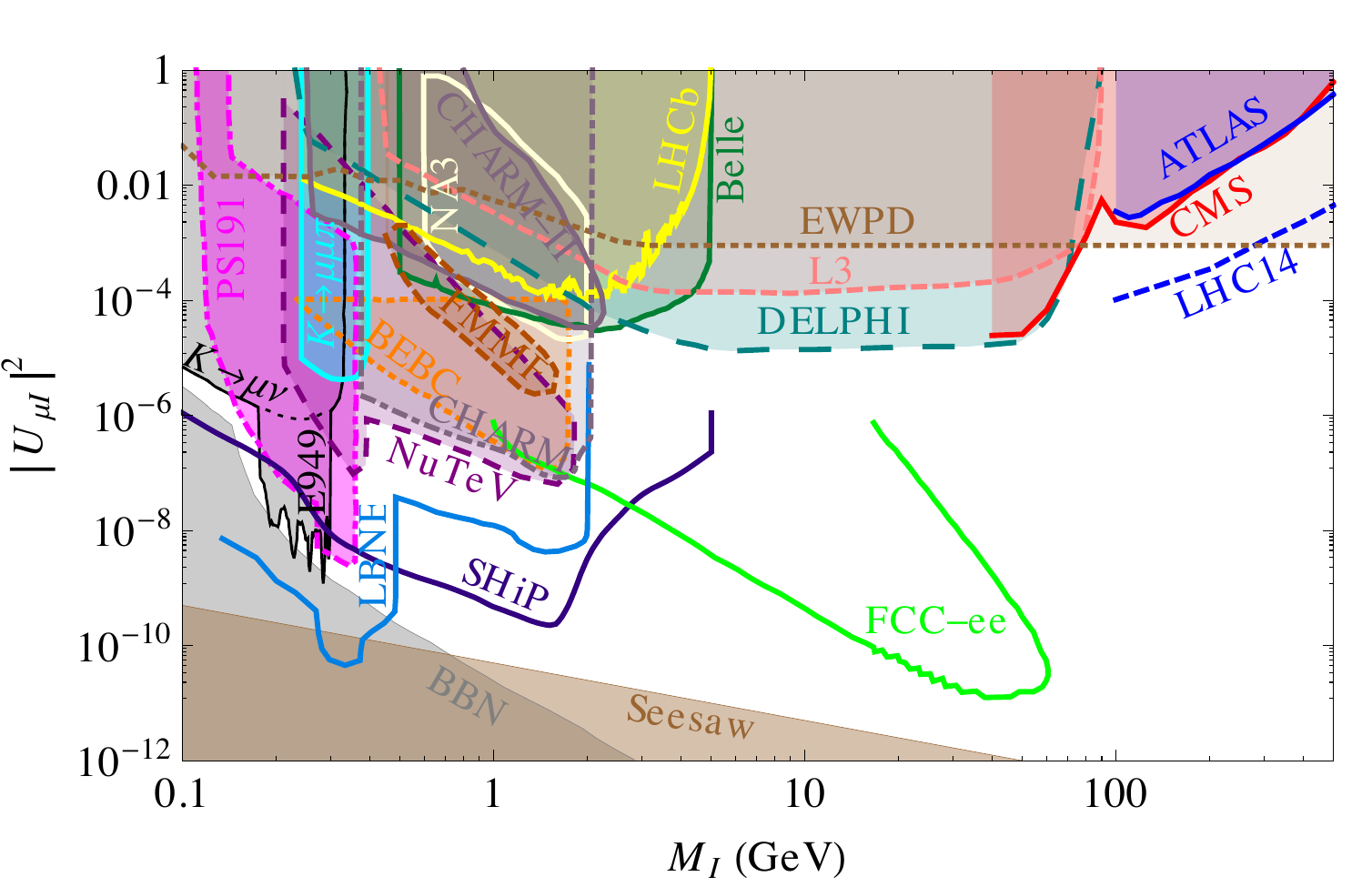}
\caption[]{Constraints from present and future experiments on a heavy neutrino with mixing to the electron (left) and muon (right). Shaded
regions are existing bounds and the empty ones are prospects (from Ref.~\cite{ship}).}
\label{fig:ship}
\end{figure}

\begin{figure}[h]
\centering
\includegraphics[width=.45\linewidth]{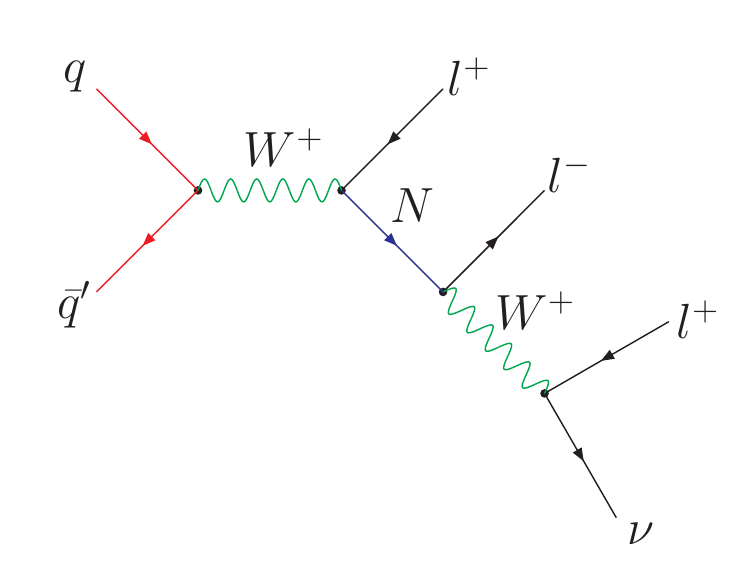}
\caption[]{Process to search for heavy Majoranas at LHC.}
\label{fig:3lep}
\end{figure}

In summary, cosmology and neutrino oscillations restrict a huge range of $M_N\in [10^{-17} - 10^2]\UMeV$. Naturalness arguments on the other hand point to a scale $M_N \leq 10^{10}\UMeV$, suggesting that maybe the scale of $M_N$ is not far from the electroweak scale. States with masses in this range could be produced on the lab and searched for as peaks in meson decays,  in beam dump experiments, colliders, etc. \cite{Atre:2009rg}. The present experimental bounds on the $e,\mu$ mixings of these heavy states are shown in Figs.~\ref{fig:ship}. The shaded regions correspond to existing constraints and the unshaded ones to prospects of various new experiments. 
For masses below  a few GeV, the best constraints come from peak searches in meson decays. In particular the new beam dump experiment SHiP \cite{ship} can improve considerably the sensitivity in the region above kaon decays. For the lighter hadrons, improvements can be achieved with the more intense beams expected in long-baseline accelerator neutrino projects such as DUNE \cite{Adams:2013qkq}.
For masses above the $W, Z$ masses the best constraints are presently coming from LHC searches. Processes with three leptons in the final state as in Fig.~\ref{fig:3lep} seem most promising \cite{delAguila:2008cj}, although other production mechanisms like $W\gamma$ fusion can dominate at higher masses \cite{Dev:2013wba}. For a recent review  and further references see \cite{ship}. 
For masses below the $W$ mass, it has been pointed out recently that LHC might also improve the present constraints by looking for displaced vertices in the range 1mm-1m \cite{Helo:2013esa,Izaguirre:2015pga,Gago:2015vma}.   

Note that present sensitivities are very far from the naive seesaw scaling 
$|U_{l h}|^2 \sim m_l/M_h$, so they are only exploring a relatively small corner of parameter space. 

{\bf Lepton flavour violating processes}

\begin{figure}[h]
\centering
\includegraphics[width=.5\linewidth]{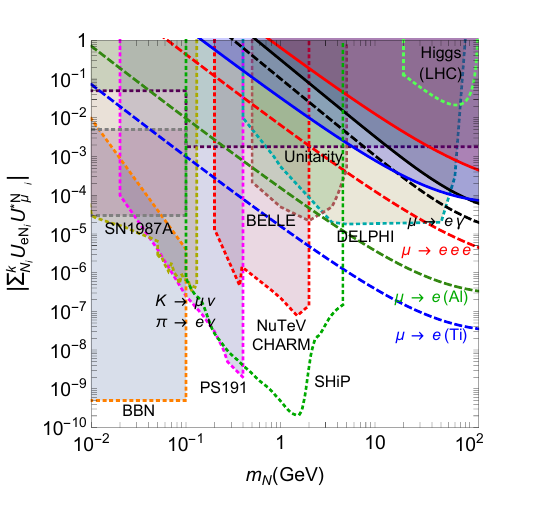}
\caption[]{Present bounds and prospects from lepton $\mu\rightarrow e \gamma, \mu \rightarrow eee$ and $\mu-e$ conversion searches (from Ref.\cite{ship}).}
\label{fig:lfv}
\end{figure}

 Massive neutrinos imply that lepton flavour violating processes, such as $\mu\rightarrow e \gamma$, $eee$ or $\mu-e$ conversion in atoms, must exist at some level. Heavy Majorana neutrinos around the EW scale can significantly enhance these rates. The constraints on the mixing and mass coming from these searches  cannot be included in Figs.~\ref{fig:ship} without further assumptions, since they depend on the different combination, $|\sum_i U_{ei} U_{\mu i}^*|$. They are shown in Fig.~\ref{fig:lfv} and compared with other present constraints.  Future searches will significantly improve present constraints for $M_N \in [1,100]\UGeV$. 

\subsection{Approximate Lepton Number}
Type I seesaw models  with a scale around the electroweak scale are very hard to test unless $|U_{lh}|^2 \gg m_l/M_h$. Although this is possible in some corners of parameter space for $n_R \geq 2$, being in such corners might be enforced and technically natural by an approximate lepton number symmetry \cite{Branco:1988ex,Kersten:2007vk}.

Let us consider the simplest case $n_R=2$ \cite{Gavela:2009cd}. If the two singlet states have opposite  lepton charges and we impose an exact $U(1)$ global symmetry, the $3\times 2$ Yukawa matrix, and the Majorana mass matrix have the following structures:
\begin{eqnarray}
Y_\nu =               \begin{pmatrix}Y_{e1} &  0\\
                              Y_{\mu 1} & 0\\
                              Y_{\tau1} & 0
               \end{pmatrix}, ~~M_N = \begin{pmatrix}0 & M\\
               M & 0\end{pmatrix}.
\end{eqnarray}
For this texture, the heavy states form a Dirac pair, while the light neutrino masses vanish identically.  The global symmetry can be only approximate if the zero entries in these matrices are small compared to the non-zero ones, but non vanishing.  For example if we lift the zero in the 22 element  of the $M_N$ matrix to be $\mu \ll M$,  we get the type of texture found in the so-called inverse seesaw models\footnote{In order to get at least two non-zero light neutrino masses by lifting the zeros of $M_N$ only, it is necessary to have two pairs of singlets, each pair with +1 or -1 lepton charge, ie. $n_R =4$. For $n_R=2$, the zero's in the Yukawa matrix must be lifted aswell.} \cite{Wyler:1982dd,Mohapatra:1986bd}. In this case the light neutrino mass satisfies $m_l \propto {\mu\over M^2}$. If we integrate out the scale $M$, the effective field theory describing this type of models is of the form
\begin{equation}
\mathcal{L} = \mathcal{L}_\text{SM} 
            + \sum_i \frac{\alpha_i}{\Lambda_\text{LN}}   \mathcal{O}_i^{d=5} 
            + \sum_i \frac{\beta_i}{\Lambda^2_\text{LFV}} \mathcal{O}_i^{d=6} 
            + ..., 
\label{eq:eft2}
\end{equation} 
where the operators that break lepton number  ($d=5$) and those that preserve
this symmetry ($d=6$) are generically suppressed by different scales:  $\Lambda_\text{LN} \simeq {M^2 \over \mu} \gg \Lambda_\text{LFV}\simeq M$.
These models therefore have a richer phenomenology if $M$ is at the EW scale, since yukawa's need not be suppressed. Future searches such as those mentioned in the previous section will be particularly important to constraint this subclass of seesaw models. 

\vspace{0.5cm}

We have discussed the phenomenological implications of the minimal Type I seesaw model, which will be the hardest to test. The other types of models leading to the
Weinberg operator have a richer phenomenology since the extra states couple to gauge fields (e.g the triplet scalar in type II or the
fermion in type III),  and therefore can be more copiously produced at colliders. In particular lepton number violation could give rise to spectacular signals at LHC, like same-sign lepton resonances in the type II seesaw model \cite{Keung:1983uu}:
\begin{eqnarray}
pp \rightarrow H^{++} H^{--} \rightarrow l^+ l^+ l^- l^-. 
\end{eqnarray}
Searches for triplet scalar and fermions are now standard LHC analyses.

\section{Leptogenesis}

The Universe is made of matter. The matter--antimatter asymmetry is
measured to be
\begin{equation}
\eta_B \equiv \frac{N_b - N_{\bar{b}}}{N_\gamma} \sim 6.21(16) \times 10^{-10}\SPp.
\end{equation}
One generic implication of neutrino mass models is that they provide a new mechanism
to explain this asymmetry dynamically. 

It has been known for a long time that all the ingredients to generate 
such an asymmetry from a symmetric initial state are present in the 
laws of particle physics. These ingredients were first put forward by Sakharov \cite{sakharov}: 

\emph{Baryon number violation}

\begin{figure}
\centering
\includegraphics[width=.4\linewidth]{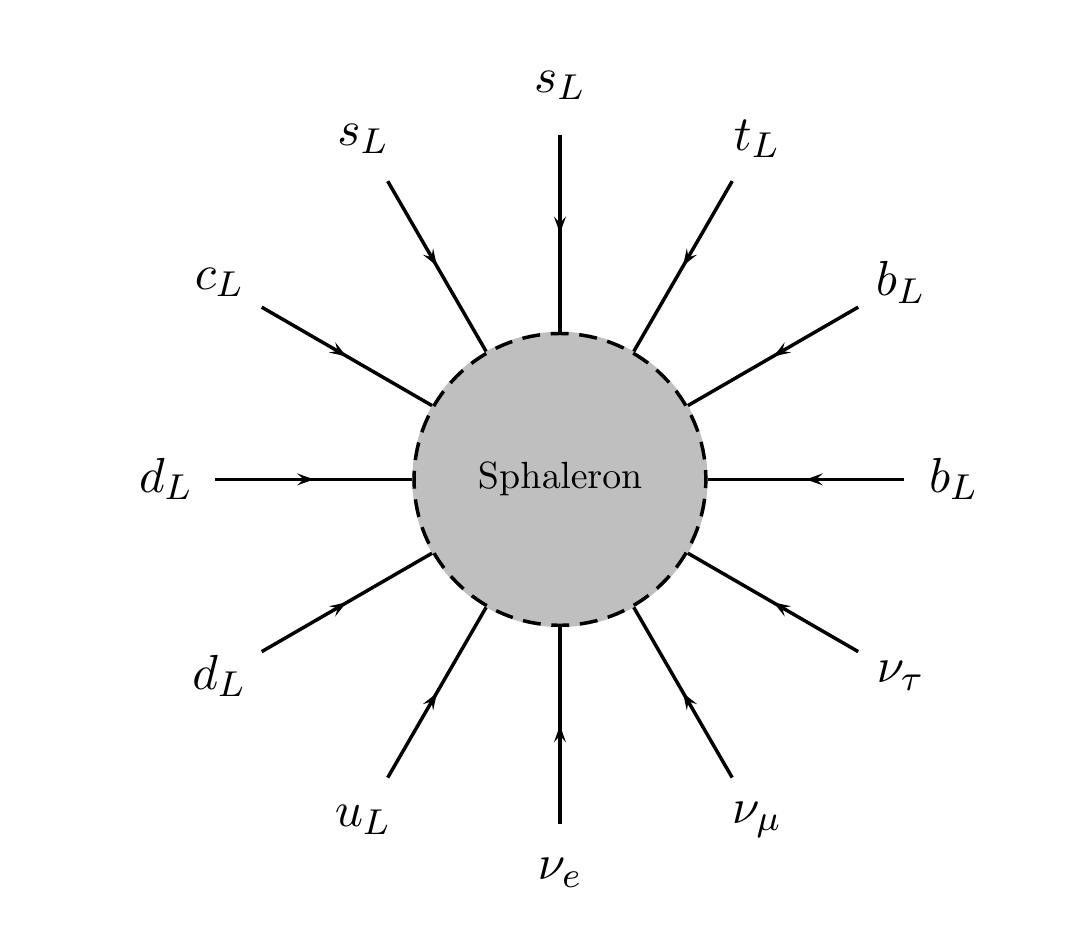}
\caption[]{Artistic view of a sphaleron.}
\label{fig:spha}
\end{figure}

$B+L$ is anomalous in the SM \cite{thooft} both with and without
massive neutrinos.  At high T in the early Universe, $B+L$ violating
transitions are in thermal equilibrium \cite{krs} due to the
thermal excitation of configurations with topological charge called
sphalerons, see \Fref{fig:spha}.

These processes violate baryon and lepton numbers by the same amount:
\begin{eqnarray}
\Delta B = \Delta L. 
\end{eqnarray}
In  seesaw models, there
is generically an additional source of $L$ violation (and $B-L$). If a lepton
charge is generated at temperatures where the sphalerons are still in
thermal equilibrium, a baryon charge can be generated.

\emph{$C$ and $CP$ violation}

Any lepton or baryon asymmetry can only be generated if there is $C$ and $CP$ violation.
Seesaw models generically include new sources of $CP$ violation. As we have seen in type I 
seesaw model with $n_R =3$ there are six new CP phases in the lepton sector. They can be 
absorbed in the Yukawa matrix, $Y_\nu$ of eq.~(\ref{eq:seesawI}). For example, 
in the Casas-Ibarra parametrization, this matrix is writen as 
\begin{eqnarray}
Y_\nu = U^*_{\rm PMNS} \sqrt{m_l} R \sqrt{M_h} {\sqrt{2} \over v}.
\end{eqnarray}
Three phases can be chosen as those in the PMNS matrix, and therefore accessible via neutrino oscillations and neutrinoless double-beta decay. The other three are the parameters of the general complex matrix $R$, that we cannot access at low-energies. Note that the combination $Y_\nu^\dagger Y_\nu$ only depends on the latter.

\emph{Departure from thermal equilibrium}

\begin{figure}
\centering
\includegraphics[width=.6\linewidth]{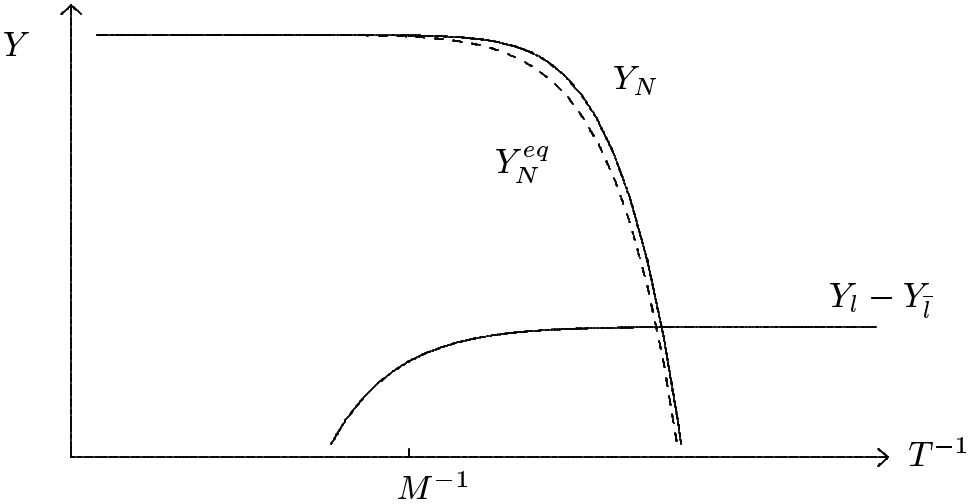}
\caption[]{Abundance of the heavy Majorana singlets at the decoupling
           temperature and the lepton number generated in the
           decay.}
\label{fig:outofeq}
\end{figure}

\begin{figure}
\centering
\includegraphics[width=.6\linewidth]{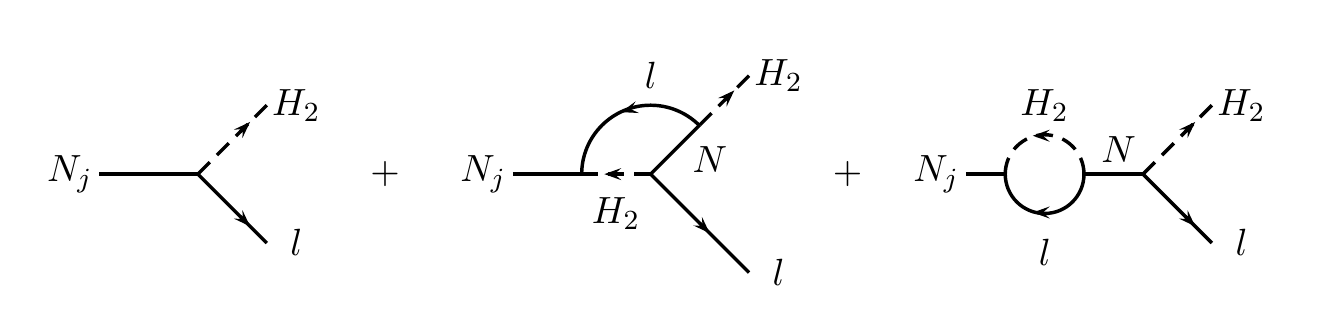}
\caption[]{Tree-level and one-loop diagrams contributing to heavy
           neutrino decays.}
\label{fig:decay}
\end{figure}

In seesaw models, $B-L$ violating processes can be out-of-equilibrium  at $T\gg T_{\rm EW}$ where
the sphalerons are still in thermal equilibrium. 
In the type I seesaw model two possibilities of non-equilibrium $L$ violation can be realised. In the high
scale scenario $M_i \gg v$, the non-equilibrium condition is met at freeze out. The heavy states are 
 thermally produced and freeze out 
at a temperatures similar to their masses  \cite{lepto0}. 
A net lepton asymmetry can be produced if  the decay rate is slower than the expansion of the
Universe close to the decoupling temperature, so that the 
distribution functions of these states differ slightly from the thermal
ones, as shown in 
 \Fref{fig:outofeq}. 
It is necessary however that CP and C be violated in the
out-of-equilibrium decays:
\begin{equation}
\epsilon_1 = \frac{\Gamma(N\rightarrow \Phi l)- \Gamma(N\rightarrow \Phi \bar{l})}
                  {\Gamma(N\rightarrow \Phi l)+ \Gamma(N\rightarrow \Phi \bar{l})} 
\neq 0\SPp.
\end{equation}
The 
new CP phases in the Yukawa matrix induce
an asymmetry, $\epsilon_1$,  at the one-loop level (see \Fref{fig:decay}). 
The final asymmetry is given by
\begin{equation}
Y_B = 10^{-2} \;\overbrace{\epsilon_1}^\text{CP-asym} \; 
                \overbrace{\kappa}^\text{eff.~factor}\SPp,
\end{equation}
where $\kappa$ is an efficiency factor which depends on the
non-equilibrium dynamics. Therefore a relation between the baryon
number of the Universe and the neutrino flavour parameters in
$\epsilon_1$ exists.

In the low-scale scenario, for $M_i < v$, the  out-of-equilibrium condition is met at freeze-in \cite{Akhmedov:1998qx}\cite{Asaka:2005pn,Canetti:2012kh}. It is 
possible that not  all the massive states reach thermal equilibrium before $T_{EW}$. A non-vanishing lepton and baryon asymmetry 
can survive at $T_{EW}$  and, if this is the case, sphaleron transitions that decouple at this point, can no longer 
wash it out. It turns out that these conditions can be met naturally in type I seesaw model for masses in the range $[0.1, 100]$ GeV. 
The relevant CP asymmetries arise in the production of the heavy seesaw states via the interference
of CP-odd phases from the Yukawa's with CP-even phases from propagation.   A quantum treatment of the corresponding kinetic equations is mandatory in this case.

An interesting question is whether the baryon asymmetry can be
predicted quantitatively from the measurements at low energies of the
neutrino mass matrix.  Unfortunately this is not the case generically, 
because the asymmetry depends on more parameters than
those that are observable at low energies.

For example, in the high-scale scenario, $\epsilon_1$   can be approximated by\cite{Davidson:2008bu}
\begin{equation}
\epsilon_1 = - \frac{3}{16 \pi} 
  \sum_i \frac{{\rm Im}[(Y_\nu^\dagger Y_\nu)^2_{i1}]}
              {(Y_\nu^\dagger Y_\nu)_{11}}
         \frac{M_1}{M_i}\SPp,
\end{equation}
in the minimal model with $M_{2,3} \gg M_1$. It  depends only on the CP phases of $R$, but not those in $U_{\rm PMNS}$.

If the prediction of the lepton asymmetry is not possible, 
it is possible to constrain the neutrino mass matrix, assuming
that the lepton asymmetry explains the measured baryon asymmetry.
Indeed, various upper bounds can be derived on the generated
asymmetry. In
particular $\epsilon_1$ has been shown \cite{Davidson:2002qv} to satisfy
\begin{equation}
|\epsilon_1| \leq \frac{8}{16 \pi} \frac{M_1}{v^2} |\Delta m^2_\text{atm}|^{1/2}\SPp,
\end{equation}
and therefore leptogenesis in this model requires that 
the lightest heavy neutrino is rather heavy:
\begin{equation}
M_1 \geq \mathcal{O}(10^9\UGeV)\SPp.
\end{equation}
%A sufficiently large $\kappa$ implies an upper bound on the lightest neutrino mass:
%\begin{equation}
%m_i \leq \mathcal{O}(\UeVZ).
%\end{equation} 
For further details and references see Ref.~\cite{Davidson:2008bu}.
\begin{figure}[h]
\centering
\includegraphics[width=.6\linewidth]{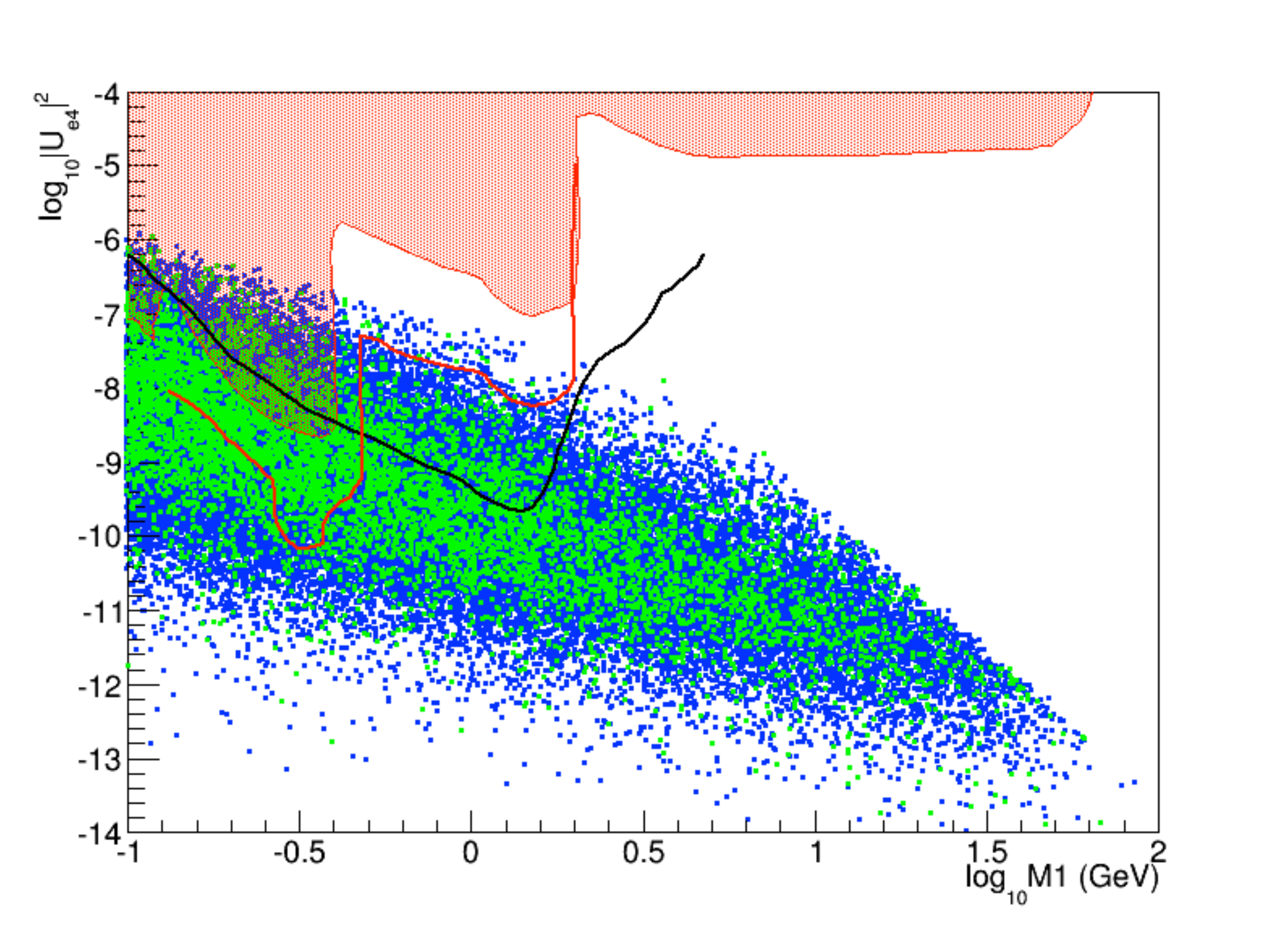}
\caption[]{Points on the plane $|U_{e4}|^2$ vs $M_1$   for which the baryon asymmetry, $Y_B$,  is in the range $[1/5-1]\times Y_B^{\rm exp}$ (blue) and $[1-5] \times Y_B^{\rm exp}$ (green)  for IO (from Ref.~\cite{Hernandez:2015wna}). The red band are the present constraints \cite{Atre:2009rg}, the solid black line shows the reach of the SHiP experiment \cite{ship} and the solid red line is the reach of LBNE near detector \cite{Adams:2013qkq}. }
\label{fig:leptovsship}
\end{figure}

Interestingly, in the low-scale scenario, the states responsible for generating the baryon asymmetry might be accessible experimentally. For example, Fig.~\ref{fig:leptovsship} shows the values of the mixing $|(U_{lh})_{e1}|^2$ and mass $M_1$ for which the baryon asymmetry can be explained within the type I seesaw model, compared to the sensitivity of future experiments such as SHiP and DUNE.

\section{Conclusions}

The results of many beautiful experiments in the last decade have
demonstrated beyond doubt that  neutrinos are massive and mix. The
standard $3\nu$ scenario can explain all available data, except that of the unconfirmed signal
of LSND. The lepton flavour sector of the Standard Model is expected
to be at least as complex as the quark one, even though we know it
only partially.

The structure of the neutrino spectrum and mixing is quite different
from the one that has been observed for the quarks: there are large
leptonic mixing angles and the neutrino masses are much smaller than
those of the remaining leptons. These peculiar features of the lepton
sector strongly suggest that leptons and quarks constitute two
complementary approaches to understanding the origin of flavour in the
Standard Model.  In fact, the smallness of neutrino masses can be
naturally understood if there is new physics beyond the electroweak
scale.

Many fundamental questions remain to be answered in future neutrino
experiments, and these can have very important implications for our
understanding of the Standard Model and of what lies beyond: Are
neutrinos Majorana particles?  Are neutrino masses the result of a new
physics scale? Is CP violated in the lepton sector? Could neutrinos
be the seed of the matter--antimatter asymmetry in the Universe?

A rich experimental programme lies ahead where fundamental physics
discoveries are very likely (almost warrantied). We can only hope that
neutrinos will keep up with their old tradition and  provide a
window to what lies beyond the Standard Model.

\bibliography{biblio}

\begin{thebibliography}{100}

\bibitem{Fermi:1934sk}
E.~Fermi, ``{Trends to a Theory of beta Radiation. (In Italian)},'' {\em Nuovo
  Cim.}, vol.~11, pp.~1--19, 1934.
\newblock [,535(1934)].

\bibitem{Bethe:1934qn}
H.~Bethe and R.~Peierls, ``{The 'neutrino'},'' {\em Nature}, vol.~133, p.~532,
  1934.

\bibitem{Pontecorvo:1946mv}
B.~Pontecorvo, ``{Inverse beta process},'' {\em Camb. Monogr. Part. Phys. Nucl.
  Phys. Cosmol.}, vol.~1, pp.~25--31, 1991.

\bibitem{Reines:1956rs}
F.~Reines and C.~L. Cowan, ``{The neutrino},'' {\em Nature}, vol.~178,
  pp.~446--449, 1956.

\bibitem{Cowan:1992xc}
C.~L. Cowan, F.~Reines, F.~B. Harrison, H.~W. Kruse, and A.~D. McGuire,
  ``{Detection of the free neutrino: A Confirmation},'' {\em Science},
  vol.~124, pp.~103--104, 1956.

\bibitem{Danby:1962nd}
G.~Danby, J.~M. Gaillard, K.~A. Goulianos, L.~M. Lederman, N.~B. Mistry,
  M.~Schwartz, and J.~Steinberger, ``{Observation of High-Energy Neutrino
  Reactions and the Existence of Two Kinds of Neutrinos},'' {\em Phys. Rev.
  Lett.}, vol.~9, pp.~36--44, 1962.

\bibitem{Agashe:2014kda}
K.~A. Olive {\em et~al.}, ``{Review of Particle Physics},'' {\em Chin. Phys.},
  vol.~C38, p.~090001, 2014.

\bibitem{Assamagan:1995wb}
K.~Assamagan {\em et~al.}, ``{Upper limit of the muon-neutrino mass and charged
  pion mass from momentum analysis of a surface muon beam},'' {\em Phys. Rev.},
  vol.~D53, pp.~6065--6077, 1996.

\bibitem{Barate:1997zg}
R.~Barate {\em et~al.}, ``{An Upper limit on the tau-neutrino mass from
  three-prong and five-prong tau decays},'' {\em Eur. Phys. J.}, vol.~C2,
  pp.~395--406, 1998.

\bibitem{Weinberg:1979sa}
S.~Weinberg, ``{Baryon and Lepton Nonconserving Processes},'' {\em Phys. Rev.
  Lett.}, vol.~43, pp.~1566--1570, 1979.

\bibitem{Weinberg:1978kz}
S.~Weinberg, ``{Phenomenological Lagrangians},'' {\em Physica}, vol.~A96,
  pp.~327--340, 1979.

\bibitem{Buchmuller:1985jz}
W.~Buchmuller and D.~Wyler, ``{Effective Lagrangian Analysis of New
  Interactions and Flavor Conservation},'' {\em Nucl. Phys.}, vol.~B268,
  pp.~621--653, 1986.

\bibitem{Maki:1962mu}
Z.~Maki, M.~Nakagawa, and S.~Sakata, ``{Remarks on the unified model of
  elementary particles},'' {\em Prog. Theor. Phys.}, vol.~28, pp.~870--880,
  1962.

\bibitem{Pontecorvo:1967fh}
B.~Pontecorvo, ``{Neutrino Experiments and the Problem of Conservation of
  Leptonic Charge},'' {\em Sov. Phys. JETP}, vol.~26, pp.~984--988, 1968.
\newblock [Zh. Eksp. Teor. Fiz.53,1717(1967)].

\bibitem{Pontecorvo:1957cp}
B.~Pontecorvo, ``{Mesonium and anti-mesonium},'' {\em Sov. Phys. JETP}, vol.~6,
  p.~429, 1957.
\newblock [Zh. Eksp. Teor. Fiz.33,549(1957)].

\bibitem{Akhmedov:2009rb}
E.~K. Akhmedov and A.~{\relax Yu}. Smirnov, ``{Paradoxes of neutrino
  oscillations},'' {\em Phys. Atom. Nucl.}, vol.~72, pp.~1363--1381, 2009,
  0905.1903.

\bibitem{Akhmedov:2010ms}
E.~K. Akhmedov and J.~Kopp, ``{Neutrino oscillations: Quantum mechanics vs.
  quantum field theory},'' {\em JHEP}, vol.~04, p.~008, 2010, 1001.4815.
\newblock [Erratum: JHEP10,052(2013)].

\bibitem{Cerda}
M.~Cerd\'a, ``{},'' {\em Master Thesis University of Valencia}, 2011,
  0905.1903.

\bibitem{wolf}
L.~Wolfenstein, ``{Neutrino Oscillations in Matter},'' {\em Phys. Rev.},
  vol.~D17, pp.~2369--2374, 1978.

\bibitem{Mikheev:1986gs}
S.~P. Mikheev and A.~{\relax Yu}. Smirnov, ``{Resonance Amplification of
  Oscillations in Matter and Spectroscopy of Solar Neutrinos},'' {\em Sov. J.
  Nucl. Phys.}, vol.~42, pp.~913--917, 1985.
\newblock [Yad. Fiz.42,1441(1985)].

\bibitem{Bethe:1939bt}
H.~A. Bethe, ``{Energy production in stars},'' {\em Phys. Rev.}, vol.~55,
  pp.~434--456, 1939.

\bibitem{Bahcall:2000nu}
J.~N. Bahcall, M.~H. Pinsonneault, and S.~Basu, ``{Solar models: Current epoch
  and time dependences, neutrinos, and helioseismological properties},'' {\em
  Astrophys. J.}, vol.~555, pp.~990--1012, 2001, astro-ph/0010346.

\bibitem{homestake}
B.~T. Cleveland, T.~Daily, R.~Davis, Jr., J.~R. Distel, K.~Lande, C.~K. Lee,
  P.~S. Wildenhain, and J.~Ullman, ``{Measurement of the solar electron
  neutrino flux with the Homestake chlorine detector},'' {\em Astrophys. J.},
  vol.~496, pp.~505--526, 1998.

\bibitem{gallex-gno}
W.~Hampel {\em et~al.}, ``{GALLEX solar neutrino observations: Results for
  GALLEX IV},'' {\em Phys. Lett.}, vol.~B447, pp.~127--133, 1999.

\bibitem{sage}
J.~N. Abdurashitov {\em et~al.}, ``{Solar neutrino flux measurements by the
  Soviet-American Gallium Experiment (SAGE) for half the 22 year solar
  cycle},'' {\em J. Exp. Theor. Phys.}, vol.~95, pp.~181--193, 2002,
  astro-ph/0204245.
\newblock [Zh. Eksp. Teor. Fiz.122,211(2002)].

\bibitem{kamio}
Y.~Fukuda {\em et~al.}, ``{Solar neutrino data covering solar cycle 22},'' {\em
  Phys. Rev. Lett.}, vol.~77, pp.~1683--1686, 1996.

\bibitem{Fukuda:1998ua}
Y.~Fukuda {\em et~al.}, ``{Measurement of the solar neutrino energy spectrum
  using neutrino electron scattering},'' {\em Phys. Rev. Lett.}, vol.~82,
  pp.~2430--2434, 1999, hep-ex/9812011.

\bibitem{Hosaka:2005um}
J.~Hosaka {\em et~al.}, ``{Solar neutrino measurements in
  super-Kamiokande-I},'' {\em Phys. Rev.}, vol.~D73, p.~112001, 2006,
  hep-ex/0508053.

\bibitem{sno}
Q.~R. Ahmad {\em et~al.}, ``{Measurement of the rate of $\nu_e+d \to p+p+e^-$
  interactions produced by $^8B$ solar neutrinos at the Sudbury Neutrino
  Observatory},'' {\em Phys. Rev. Lett.}, vol.~87, p.~071301, 2001,
  nucl-ex/0106015.

\bibitem{Ahmad:2002jz}
Q.~R. Ahmad {\em et~al.}, ``{Direct evidence for neutrino flavor transformation
  from neutral current interactions in the Sudbury Neutrino Observatory},''
  {\em Phys. Rev. Lett.}, vol.~89, p.~011301, 2002, nucl-ex/0204008.

\bibitem{kamland}
K.~Eguchi {\em et~al.}, ``{First results from KamLAND: Evidence for reactor
  anti-neutrino disappearance},'' {\em Phys. Rev. Lett.}, vol.~90, p.~021802,
  2003, hep-ex/0212021.

\bibitem{Abe:2008aa}
S.~Abe {\em et~al.}, ``{Precision Measurement of Neutrino Oscillation
  Parameters with KamLAND},'' {\em Phys. Rev. Lett.}, vol.~100, p.~221803,
  2008, 0801.4589.

\bibitem{Bellini:2013lnn}
G.~Bellini {\em et~al.}, ``{Final results of Borexino Phase-I on low energy
  solar neutrino spectroscopy},'' {\em Phys. Rev.}, vol.~D89, no.~11,
  p.~112007, 2014, 1308.0443.

\bibitem{Honda:2004yz}
M.~Honda, T.~Kajita, K.~Kasahara, and S.~Midorikawa, ``{A New calculation of
  the atmospheric neutrino flux in a 3-dimensional scheme},'' {\em Phys. Rev.},
  vol.~D70, p.~043008, 2004, astro-ph/0404457.

\bibitem{sk-atmos}
Y.~Fukuda {\em et~al.}, ``{Evidence for oscillation of atmospheric
  neutrinos},'' {\em Phys. Rev. Lett.}, vol.~81, pp.~1562--1567, 1998,
  hep-ex/9807003.

\bibitem{sk-zen}
Y.~Ashie {\em et~al.}, ``{A Measurement of atmospheric neutrino oscillation
  parameters by SUPER-KAMIOKANDE I},'' {\em Phys. Rev.}, vol.~D71, p.~112005,
  2005, hep-ex/0501064.

\bibitem{sk_atmos_el}
Y.~Ashie {\em et~al.}, ``{Evidence for an oscillatory signature in atmospheric
  neutrino oscillation},'' {\em Phys. Rev. Lett.}, vol.~93, p.~101801, 2004,
  hep-ex/0404034.

\bibitem{minos}
A.~Holin, ``{Results from the MINOS Experiment and New MINOS+ Data},'' {\em
  PoS}, vol.~NUFACT2014, p.~028, 2014, 1507.08564.

\bibitem{An:2012eh}
F.~P. An {\em et~al.}, ``{Observation of electron-antineutrino disappearance at
  Daya Bay},'' {\em Phys. Rev. Lett.}, vol.~108, p.~171803, 2012, 1203.1669.

\bibitem{Ahn:2012nd}
J.~K. Ahn {\em et~al.}, ``{Observation of Reactor Electron Antineutrino
  Disappearance in the RENO Experiment},'' {\em Phys. Rev. Lett.}, vol.~108,
  p.~191802, 2012, 1204.0626.

\bibitem{Abe:2011fz}
Y.~Abe {\em et~al.}, ``{Indication for the disappearance of reactor electron
  antineutrinos in the Double Chooz experiment},'' {\em Phys. Rev. Lett.},
  vol.~108, p.~131801, 2012, 1112.6353.

\bibitem{t2k}
K.~Abe {\em et~al.}, ``{Observation of Electron Neutrino Appearance in a Muon
  Neutrino Beam},'' {\em Phys. Rev. Lett.}, vol.~112, p.~061802, 2014,
  1311.4750.

\bibitem{Gonzalez-Garcia:2014bfa}
M.~C. Gonzalez-Garcia, M.~Maltoni, and T.~Schwetz, ``{Updated fit to three
  neutrino mixing: status of leptonic CP violation},'' {\em JHEP}, vol.~11,
  p.~052, 2014, 1409.5439.

\bibitem{Ade:2015xua}
P.~A.~R. Ade {\em et~al.}, ``{Planck 2015 results. XIII. Cosmological
  parameters},'' {\em arXiv:1502.01589[astro-ph.CO]}, 2015, 1502.01589.

\bibitem{sanjib}
S.~K. Agarwalla and P.~Hernandez, ``{Probing the Neutrino Mass Hierarchy with
  Super-Kamiokande},'' {\em JHEP}, vol.~10, p.~086, 2012, 1204.4217.

\bibitem{Petcov:2001sy}
S.~T. Petcov and M.~Piai, ``{The LMA MSW solution of the solar neutrino
  problem, inverted neutrino mass hierarchy and reactor neutrino
  experiments},'' {\em Phys. Lett.}, vol.~B533, pp.~94--106, 2002,
  hep-ph/0112074.

\bibitem{Choubey:2003qx}
S.~Choubey, S.~T. Petcov, and M.~Piai, ``{Precision neutrino oscillation
  physics with an intermediate baseline reactor neutrino experiment},'' {\em
  Phys. Rev.}, vol.~D68, p.~113006, 2003, hep-ph/0306017.

\bibitem{juno}
F.~An {\em et~al.}, ``{Neutrino Physics with JUNO},'' {\em J. Phys.}, vol.~G43,
  no.~3, p.~030401, 2016, 1507.05613.

\bibitem{golden}
A.~Cervera, A.~Donini, M.~B. Gavela, J.~J. Gomez~Cadenas, P.~Hernandez,
  O.~Mena, and S.~Rigolin, ``{Golden measurements at a neutrino factory},''
  {\em Nucl. Phys.}, vol.~B579, pp.~17--55, 2000, hep-ph/0002108.
\newblock [Erratum: Nucl. Phys.B593,731(2001)].

\bibitem{hyperk}
K.~Abe {\em et~al.}, ``{Physics potential of a long-baseline neutrino
  oscillation experiment using a J-PARC neutrino beam and Hyper-Kamiokande},''
  {\em PTEP}, vol.~2015, p.~053C02, 2015, 1502.05199.

\bibitem{dune}
R.~Acciarri {\em et~al.}, ``{Long-Baseline Neutrino Facility (LBNF) and Deep
  Underground Neutrino Experiment (DUNE) Conceptual Design Report Volume 2: The
  Physics Program for DUNE at LBNF},'' {\em arXiv:1512.06148 [hep-ex]}, 2015,
  1512.06148.

\bibitem{lsnd}
A.~Aguilar-Arevalo {\em et~al.}, ``{Evidence for neutrino oscillations from the
  observation of anti-neutrino(electron) appearance in a anti-neutrino(muon)
  beam},'' {\em Phys. Rev.}, vol.~D64, p.~112007, 2001, hep-ex/0104049.

\bibitem{karmen}
B.~Armbruster {\em et~al.}, ``{Upper limits for neutrino oscillations
  muon-anti-neutrino ---> electron-anti-neutrino from muon decay at rest},''
  {\em Phys. Rev.}, vol.~D65, p.~112001, 2002, hep-ex/0203021.

\bibitem{miniboone}
A.~A. Aguilar-Arevalo {\em et~al.}, ``{Improved Search for $\bar \nu_\mu
  \rightarrow \bar \nu_e$ Oscillations in the MiniBooNE Experiment},'' {\em
  Phys. Rev. Lett.}, vol.~110, p.~161801, 2013, 1207.4809.

\bibitem{Mueller:2011nm}
T.~A. Mueller {\em et~al.}, ``{Improved Predictions of Reactor Antineutrino
  Spectra},'' {\em Phys. Rev.}, vol.~C83, p.~054615, 2011, 1101.2663.

\bibitem{Mention:2011rk}
G.~Mention, M.~Fechner, T.~Lasserre, T.~A. Mueller, D.~Lhuillier, M.~Cribier,
  and A.~Letourneau, ``{The Reactor Antineutrino Anomaly},'' {\em Phys. Rev.},
  vol.~D83, p.~073006, 2011, 1101.2755.

\bibitem{Huber:2011wv}
P.~Huber, ``{On the determination of anti-neutrino spectra from nuclear
  reactors},'' {\em Phys. Rev.}, vol.~C84, p.~024617, 2011, 1106.0687.
\newblock [Erratum: Phys. Rev.C85,029901(2012)].

\bibitem{dayabayflux}
F.~P. An {\em et~al.}, ``{Measurement of the Reactor Antineutrino Flux and
  Spectrum at Daya Bay},'' {\em Phys. Rev. Lett.}, vol.~116, no.~6, p.~061801,
  2016, 1508.04233.

\bibitem{minosplus}
A.~B. Sousa, ``{First MINOS+ Data and New Results from MINOS},'' {\em AIP Conf.
  Proc.}, vol.~1666, p.~110004, 2015, 1502.07715.

\bibitem{Harrison:2002er}
P.~F. Harrison, D.~H. Perkins, and W.~G. Scott, ``{Tri-bimaximal mixing and the
  neutrino oscillation data},'' {\em Phys. Lett.}, vol.~B530, p.~167, 2002,
  hep-ph/0202074.

\bibitem{King:2013eh}
S.~F. King and C.~Luhn, ``{Neutrino Mass and Mixing with Discrete Symmetry},''
  {\em Rept. Prog. Phys.}, vol.~76, p.~056201, 2013, 1301.1340.

\bibitem{Minkowski:1977sc}
P.~Minkowski, ``{$\mu \to e\gamma$ at a Rate of One Out of $10^{9}$ Muon
  Decays?},'' {\em Phys. Lett.}, vol.~B67, pp.~421--428, 1977.

\bibitem{GellMann:1980vs}
M.~Gell-Mann, P.~Ramond, and R.~Slansky, ``{Complex Spinors and Unified
  Theories},'' {\em Conf. Proc.}, vol.~C790927, pp.~315--321, 1979, 1306.4669.

\bibitem{Yanagida:1979as}
T.~Yanagida, ``{HORIZONTAL SYMMETRY AND MASSES OF NEUTRINOS},'' {\em Conf.
  Proc.}, vol.~C7902131, pp.~95--99, 1979.

\bibitem{Mohapatra:1979ia}
R.~N. Mohapatra and G.~Senjanovic, ``{Neutrino Mass and Spontaneous Parity
  Violation},'' {\em Phys. Rev. Lett.}, vol.~44, p.~912, 1980.

\bibitem{Magg:1980ut}
M.~Magg and C.~Wetterich, ``{Neutrino Mass Problem and Gauge Hierarchy},'' {\em
  Phys. Lett.}, vol.~B94, p.~61, 1980.

\bibitem{Schechter:1980gr}
J.~Schechter and J.~W.~F. Valle, ``{Neutrino Masses in SU(2) x U(1)
  Theories},'' {\em Phys. Rev.}, vol.~D22, p.~2227, 1980.

\bibitem{Wetterich:1981bx}
C.~Wetterich, ``{Neutrino Masses and the Scale of B-L Violation},'' {\em Nucl.
  Phys.}, vol.~B187, p.~343, 1981.

\bibitem{Lazarides:1980nt}
G.~Lazarides, Q.~Shafi, and C.~Wetterich, ``{Proton Lifetime and Fermion Masses
  in an SO(10) Model},'' {\em Nucl. Phys.}, vol.~B181, pp.~287--300, 1981.

\bibitem{Mohapatra:1980yp}
R.~N. Mohapatra and G.~Senjanovic, ``{Neutrino Masses and Mixings in Gauge
  Models with Spontaneous Parity Violation},'' {\em Phys. Rev.}, vol.~D23,
  p.~165, 1981.

\bibitem{Foot:1988aq}
R.~Foot, H.~Lew, X.~G. He, and G.~C. Joshi, ``{Seesaw Neutrino Masses Induced
  by a Triplet of Leptons},'' {\em Z. Phys.}, vol.~C44, p.~441, 1989.

\bibitem{Ma:1998dn}
E.~Ma, ``{Pathways to naturally small neutrino masses},'' {\em Phys. Rev.
  Lett.}, vol.~81, pp.~1171--1174, 1998, hep-ph/9805219.

\bibitem{zee}
A.~Zee, ``{A Theory of Lepton Number Violation, Neutrino Majorana Mass, and
  Oscillation},'' {\em Phys. Lett.}, vol.~B93, p.~389, 1980.
\newblock [Erratum: Phys. Lett.B95,461(1980)].

\bibitem{Zee:1985rj}
A.~Zee, ``{Charged Scalar Field and Quantum Number Violations},'' {\em Phys.
  Lett.}, vol.~B161, p.~141, 1985.

\bibitem{Babu:1988ki}
K.~S. Babu, ``{Model of 'Calculable' Majorana Neutrino Masses},'' {\em Phys.
  Lett.}, vol.~B203, p.~132, 1988.

\bibitem{Abada:2007ux}
A.~Abada, C.~Biggio, F.~Bonnet, M.~B. Gavela, and T.~Hambye, ``{Low energy
  effects of neutrino masses},'' {\em JHEP}, vol.~12, p.~061, 2007, 0707.4058.

\bibitem{Aad:2015zhl}
G.~Aad {\em et~al.}, ``{Combined Measurement of the Higgs Boson Mass in $pp$
  Collisions at $\sqrt{s}=7$ and 8 TeV with the ATLAS and CMS Experiments},''
  {\em Phys. Rev. Lett.}, vol.~114, p.~191803, 2015, 1503.07589.

\bibitem{Degrassi:2012ry}
G.~Degrassi, S.~Di~Vita, J.~Elias-Miro, J.~R. Espinosa, G.~F. Giudice,
  G.~Isidori, and A.~Strumia, ``{Higgs mass and vacuum stability in the
  Standard Model at NNLO},'' {\em JHEP}, vol.~08, p.~098, 2012, 1205.6497.

\bibitem{Vissani:1997ys}
F.~Vissani, ``{Do experiments suggest a hierarchy problem?},'' {\em Phys.
  Rev.}, vol.~D57, pp.~7027--7030, 1998, hep-ph/9709409.

\bibitem{Casas:2004gh}
J.~A. Casas, J.~R. Espinosa, and I.~Hidalgo, ``{Implications for new physics
  from fine-tuning arguments. 1. Application to SUSY and seesaw cases},'' {\em
  JHEP}, vol.~11, p.~057, 2004, hep-ph/0410298.

\bibitem{Casas:2001sr}
J.~A. Casas and A.~Ibarra, ``{Oscillating neutrinos and muon ---> e, gamma},''
  {\em Nucl. Phys.}, vol.~B618, pp.~171--204, 2001, hep-ph/0103065.

\bibitem{Donini:2012tt}
A.~Donini, P.~Hernandez, J.~Lopez-Pavon, M.~Maltoni, and T.~Schwetz, ``{The
  minimal 3+2 neutrino model versus oscillation anomalies},'' {\em JHEP},
  vol.~07, p.~161, 2012, 1205.5230.

\bibitem{Blennow:2010th}
M.~Blennow, E.~Fernandez-Martinez, J.~Lopez-Pavon, and J.~Menendez,
  ``{Neutrinoless double beta decay in seesaw models},'' {\em JHEP}, vol.~07,
  p.~096, 2010, 1005.3240.

\bibitem{Capozzi:2016rtj}
F.~Capozzi, E.~Lisi, A.~Marrone, D.~Montanino, and A.~Palazzo, ``{Neutrino
  masses and mixings: Status of known and unknown $3\nu$ parameters},'' {\em
  arXiv:1601.07777 [hep-ph]}, 2016, 1601.07777.

\bibitem{Ibarra:2010xw}
A.~Ibarra, E.~Molinaro, and S.~T. Petcov, ``{TeV Scale See-Saw Mechanisms of
  Neutrino Mass Generation, the Majorana Nature of the Heavy Singlet Neutrinos
  and $(\beta\beta)_{0\nu}$-Decay},'' {\em JHEP}, vol.~09, p.~108, 2010,
  1007.2378.

\bibitem{Mitra:2011qr}
M.~Mitra, G.~Senjanovic, and F.~Vissani, ``{Neutrinoless Double Beta Decay and
  Heavy Sterile Neutrinos},'' {\em Nucl. Phys.}, vol.~B856, pp.~26--73, 2012,
  1108.0004.

\bibitem{Lopez-Pavon:2015cga}
J.~Lopez-Pavon, E.~Molinaro, and S.~T. Petcov, ``{Radiative Corrections to
  Light Neutrino Masses in Low Scale Type I Seesaw Scenarios and Neutrinoless
  Double Beta Decay},'' {\em JHEP}, vol.~11, p.~030, 2015, 1506.05296.

\bibitem{deGouvea:2005er}
A.~de~Gouvea, ``{See-saw energy scale and the LSND anomaly},'' {\em Phys.
  Rev.}, vol.~D72, p.~033005, 2005, hep-ph/0501039.

\bibitem{deGouvea:2009fp}
A.~de~Gouvea, W.-C. Huang, and J.~Jenkins, ``{Pseudo-Dirac Neutrinos in the New
  Standard Model},'' {\em Phys.Rev.}, vol.~D80, p.~073007, 2009, 0906.1611.

\bibitem{deGouvea:2011zz}
A.~de~Gouvea and W.-C. Huang, ``{Constraining the (Low-Energy) Type-I
  Seesaw},'' {\em Phys.Rev.}, vol.~D85, p.~053006, 2012, 1110.6122.

\bibitem{Donini:2011jh}
A.~Donini, P.~Hernandez, J.~Lopez-Pavon, and M.~Maltoni, ``{Minimal models with
  light sterile neutrinos},'' {\em JHEP}, vol.~07, p.~105, 2011, 1106.0064.

\bibitem{Giunti:2011gz}
C.~Giunti and M.~Laveder, ``{3+1 and 3+2 Sterile Neutrino Fits},'' {\em Phys.
  Rev.}, vol.~D84, p.~073008, 2011, 1107.1452.

\bibitem{Kopp:2013vaa}
J.~Kopp, P.~A.~N. Machado, M.~Maltoni, and T.~Schwetz, ``{Sterile Neutrino
  Oscillations: The Global Picture},'' {\em JHEP}, vol.~05, p.~050, 2013,
  1303.3011.

\bibitem{Notzold:1987ik}
D.~Notzold and G.~Raffelt, ``{Neutrino Dispersion at Finite Temperature and
  Density},'' {\em Nucl. Phys.}, vol.~B307, p.~924, 1988.

\bibitem{Barbieri:1989ti}
R.~Barbieri and A.~Dolgov, ``{Bounds on Sterile-neutrinos from
  Nucleosynthesis},'' {\em Phys. Lett.}, vol.~B237, p.~440, 1990.

\bibitem{Kainulainen:1990ds}
K.~Kainulainen, ``{Light Singlet Neutrinos and the Primordial
  Nucleosynthesis},'' {\em Phys. Lett.}, vol.~B244, pp.~191--195, 1990.

\bibitem{Hernandez:2013lza}
P.~Hernandez, M.~Kekic, and J.~Lopez-Pavon, ``{Low-scale seesaw models versus
  $N_{eff}$},'' {\em Phys.Rev.}, vol.~D89, no.~7, p.~073009, 2014, 1311.2614.

\bibitem{Hernandez:2014fha}
P.~Hernandez, M.~Kekic, and J.~Lopez-Pavon, ``{$N_{\rm eff}$ in low-scale
  seesaw models versus the lightest neutrino mass},'' {\em Phys.Rev.},
  vol.~D90, no.~6, p.~065033, 2014, 1406.2961.

\bibitem{Hannestad:2015tea}
S.~Hannestad, R.~S. Hansen, T.~Tram, and Y.~Y.~Y. Wong, ``{Active-sterile
  neutrino oscillations in the early Universe with full collision terms},''
  {\em JCAP}, vol.~1508, no.~08, p.~019, 2015, 1506.05266.

\bibitem{Dodelson:1993je}
S.~Dodelson and L.~M. Widrow, ``{Sterile-neutrinos as dark matter},'' {\em
  Phys. Rev. Lett.}, vol.~72, pp.~17--20, 1994, hep-ph/9303287.

\bibitem{Asaka:2005pn}
T.~Asaka and M.~Shaposhnikov, ``{The nuMSM, dark matter and baryon asymmetry of
  the universe},'' {\em Phys.Lett.}, vol.~B620, pp.~17--26, 2005,
  hep-ph/0505013.

\bibitem{Bulbul:2014sua}
E.~Bulbul, M.~Markevitch, A.~Foster, R.~K. Smith, M.~Loewenstein, and S.~W.
  Randall, ``{Detection of An Unidentified Emission Line in the Stacked X-ray
  spectrum of Galaxy Clusters},'' {\em Astrophys. J.}, vol.~789, p.~13, 2014,
  1402.2301.

\bibitem{Boyarsky:2014jta}
A.~Boyarsky, O.~Ruchayskiy, D.~Iakubovskyi, and J.~Franse, ``{Unidentified Line
  in X-Ray Spectra of the Andromeda Galaxy and Perseus Galaxy Cluster},'' {\em
  Phys. Rev. Lett.}, vol.~113, p.~251301, 2014, 1402.4119.

\bibitem{Shi:1998km}
X.-D. Shi and G.~M. Fuller, ``{A New dark matter candidate: Nonthermal sterile
  neutrinos},'' {\em Phys. Rev. Lett.}, vol.~82, pp.~2832--2835, 1999,
  astro-ph/9810076.

\bibitem{Atre:2009rg}
A.~Atre, T.~Han, S.~Pascoli, and B.~Zhang, ``{The Search for Heavy Majorana
  Neutrinos},'' {\em JHEP}, vol.~05, p.~030, 2009, 0901.3589.

\bibitem{ship}
S.~Alekhin {\em et~al.}, ``{A facility to Search for Hidden Particles at the
  CERN SPS: the SHiP physics case},'' {\em arXiv:1504.04855[hep-ph]}, 2015,
  1504.04855.

\bibitem{Adams:2013qkq}
C.~Adams {\em et~al.}, ``{The Long-Baseline Neutrino Experiment: Exploring
  Fundamental Symmetries of the Universe},'' 2013, 1307.7335.

\bibitem{delAguila:2008cj}
F.~del Aguila and J.~A. Aguilar-Saavedra, ``{Distinguishing seesaw models at
  LHC with multi-lepton signals},'' {\em Nucl. Phys.}, vol.~B813, pp.~22--90,
  2009, 0808.2468.

\bibitem{Dev:2013wba}
P.~S.~B. Dev, A.~Pilaftsis, and U.-k. Yang, ``{New Production Mechanism for
  Heavy Neutrinos at the LHC},'' {\em Phys. Rev. Lett.}, vol.~112, no.~8,
  p.~081801, 2014, 1308.2209.

\bibitem{Helo:2013esa}
J.~C. Helo, M.~Hirsch, and S.~Kovalenko, ``{Heavy neutrino searches at the LHC
  with displaced vertices},'' {\em Phys. Rev.}, vol.~D89, p.~073005, 2014,
  1312.2900.

\bibitem{Izaguirre:2015pga}
E.~Izaguirre and B.~Shuve, ``{Multilepton and Lepton Jet Probes of
  Sub-Weak-Scale Right-Handed Neutrinos},'' {\em Phys. Rev.}, vol.~D91, no.~9,
  p.~093010, 2015, 1504.02470.

\bibitem{Gago:2015vma}
A.~M. Gago, P.~Hernandez, J.~Jones-Perez, M.~Losada, and A.~Moreno Brice\~no,
  ``{Probing the Type I Seesaw Mechanism with Displaced Vertices at the LHC},''
  {\em Eur. Phys. J.}, vol.~C75, no.~10, p.~470, 2015, 1505.05880.

\bibitem{Branco:1988ex}
G.~C. Branco, W.~Grimus, and L.~Lavoura, ``{The Seesaw Mechanism in the
  Presence of a Conserved Lepton Number},'' {\em Nucl. Phys.}, vol.~B312,
  p.~492, 1989.

\bibitem{Kersten:2007vk}
J.~Kersten and A.~{\relax Yu}. Smirnov, ``{Right-Handed Neutrinos at CERN LHC
  and the Mechanism of Neutrino Mass Generation},'' {\em Phys. Rev.}, vol.~D76,
  p.~073005, 2007, 0705.3221.

\bibitem{Gavela:2009cd}
M.~B. Gavela, T.~Hambye, D.~Hernandez, and P.~Hernandez, ``{Minimal Flavour
  Seesaw Models},'' {\em JHEP}, vol.~09, p.~038, 2009, 0906.1461.

\bibitem{Wyler:1982dd}
D.~Wyler and L.~Wolfenstein, ``{Massless Neutrinos in Left-Right Symmetric
  Models},'' {\em Nucl. Phys.}, vol.~B218, p.~205, 1983.

\bibitem{Mohapatra:1986bd}
R.~N. Mohapatra and J.~W.~F. Valle, ``{Neutrino Mass and Baryon Number
  Nonconservation in Superstring Models},'' {\em Phys. Rev.}, vol.~D34,
  p.~1642, 1986.

\bibitem{Keung:1983uu}
W.-Y. Keung and G.~Senjanovic, ``{Majorana Neutrinos and the Production of the
  Right-handed Charged Gauge Boson},'' {\em Phys. Rev. Lett.}, vol.~50,
  p.~1427, 1983.

\bibitem{sakharov}
A.~D. Sakharov, ``{Violation of CP Invariance, c Asymmetry, and Baryon
  Asymmetry of the Universe},'' {\em Pisma Zh. Eksp. Teor. Fiz.}, vol.~5,
  pp.~32--35, 1967.
\newblock [Usp. Fiz. Nauk161,61(1991)].

\bibitem{thooft}
G.~'t~Hooft, ``{Symmetry Breaking Through Bell-Jackiw Anomalies},'' {\em Phys.
  Rev. Lett.}, vol.~37, pp.~8--11, 1976.

\bibitem{krs}
V.~A. Kuzmin, V.~A. Rubakov, and M.~E. Shaposhnikov, ``{On the Anomalous
  Electroweak Baryon Number Nonconservation in the Early Universe},'' {\em
  Phys. Lett.}, vol.~B155, p.~36, 1985.

\bibitem{lepto0}
M.~Fukugita and T.~Yanagida, ``{Baryogenesis Without Grand Unification},'' {\em
  Phys. Lett.}, vol.~B174, p.~45, 1986.

\bibitem{Akhmedov:1998qx}
E.~K. Akhmedov, V.~Rubakov, and A.~Y. Smirnov, ``{Baryogenesis via neutrino
  oscillations},'' {\em Phys.Rev.Lett.}, vol.~81, pp.~1359--1362, 1998,
  hep-ph/9803255.

\bibitem{Canetti:2012kh}
L.~Canetti, M.~Drewes, T.~Frossard, and M.~Shaposhnikov, ``{Dark Matter,
  Baryogenesis and Neutrino Oscillations from Right Handed Neutrinos},'' {\em
  Phys. Rev.}, vol.~D87, p.~093006, 2013, 1208.4607.

\bibitem{Davidson:2008bu}
S.~Davidson, E.~Nardi, and Y.~Nir, ``{Leptogenesis},'' {\em Phys. Rept.},
  vol.~466, pp.~105--177, 2008, 0802.2962.

\bibitem{Davidson:2002qv}
S.~Davidson and A.~Ibarra, ``{A Lower bound on the right-handed neutrino mass
  from leptogenesis},'' {\em Phys. Lett.}, vol.~B535, pp.~25--32, 2002,
  hep-ph/0202239.

\bibitem{Hernandez:2015wna}
P.~Hernandez, M.~Kekic, J.~Lopez-Pavon, J.~Racker, and N.~Rius, ``{Leptogenesis
  in GeV scale seesaw models},'' {\em JHEP}, vol.~10, p.~067, 2015, 1508.03676.

\end{thebibliography}
\bibliographystyle{hieeetr}
\end{document}